\documentclass[twocolumn, showpacs, amsmath]{revtex4}
\usepackage{graphicx,epsfig}
\usepackage{amssymb,amsmath,amsfonts,multirow,rotate,color}

\DeclareGraphicsExtensions{.  jpg,.  pdf, .  mps, . png, . eps, .  ps, .  EPS}

\usepackage[framed,numbered,autolinebreaks,useliterate]{mcode}

\usepackage{url,textcomp}
\begin{document}

\newcommand{\bea}{\begin{eqnarray}}
\newcommand{\eea}{\end{eqnarray}}
\newcommand{\avg}[1]{\langle{#1}\rangle}
\newcommand{\ket}[1]{\left |{#1}\right \rangle}
\newcommand{\beq}{\begin{equation}}
\newcommand{\eneq}{\end{equation}}
\newcommand{\beqnn}{\begin{equation*}}
\newcommand{\eneqnn}{\end{equation*}}
\newcommand{\beqy}{\begin{eqnarray}}
\newcommand{\eneqy}{\end{eqnarray}}
\newcommand{\beqynn}{\begin{eqnarray*}}
\newcommand{\eneqynn}{\end{eqnarray*}}
\newcommand{\half}{\mbox{$\textstyle \frac{1}{2}$}}
\newcommand{\proj}[1]{\ket{#1}\bra{#1}}
\newcommand{\av}[1]{\langle #1\rangle}
\newcommand{\braket}[2]{\langle #1 | #2\rangle}
\newcommand{\bra}[1]{\langle #1 | }
\newcommand{\Avg}[1]{\left\langle{#1}\right\rangle}
\newcommand{\inprod}[2]{\braket{#1}{#2}}
\newcommand{\upket}{\ket{\uparrow}}
\newcommand{\downket}{\ket{\downarrow}}
\newcommand{\Tr}{\mathrm{Tr}}
\newcommand{\hcontrol}{*!<0em,. 025em>-=-{\Diamond}}
\newcommand{\hctrl}[1]{\hcontrol \qwx[#1] \qw}
\newenvironment{proof}[1][Proof]{\noindent\textbf{#1. } }{\ \rule{0.5em}{0.5em}}
\newtheorem{mytheorem}{Theorem}
\newtheorem{mylemma}{Lemma}
\newtheorem{mycorollary}{Corollary}
\newtheorem{myproposition}{Proposition}
\newcommand{\vp}{\vec{p}}
\newcommand{\Or}{\mathcal{O}}
\newcommand{\so}[1]{{\ignore{#1}}}

\newcommand{\red}[1]{\textcolor{red}{#1}}
\newcommand{\blue}[1]{\textcolor{blue}{#1}}

\title{Network  geometry with flavor: from complexity to quantum geometry}
\author{Ginestra Bianconi$^1$ and Christoph Rahmede$^2$ }

\affiliation{$^1$School of Mathematical Sciences, Queen Mary University of London, London E1 4NS, United Kingdom\\
$^2$ Rome International Centre for  Material Science Superstripes RICMASS, via dei Sabelli 119A, 
00185 Roma, Italy}

\begin{abstract}
Network geometry is attracting increasing attention  because it has a wide range of applications, ranging from data mining to routing protocols in the Internet. At the same time advances in the understanding of the geometrical properties of networks are essential for further progress in quantum gravity. In network geometry, simplicial complexes  describing the interaction between two or more nodes play a special role. In fact these structures can be used to  discretize a geometrical $d$ dimensional space,  and for this reason they have  already been widely used in quantum gravity.     
Here we introduce the  Network Geometry with Flavor $s=-1,0,1$ (NGF) describing simplicial complexes defined in  arbitrary dimension $d$ and evolving by a non-equilibrium dynamics.  The NGF can generate discrete geometries of different nature, ranging from chains and higher dimensional manifolds to scale-free networks with small-world properties, scale-free degree distribution and non-trivial community structure. The NGF admits as limiting cases both  the Bianconi-Barab\'asi model for complex networks the stochastic Apollonian network,  and the recently introduced model for Complex Quantum Network Manifolds. 
The thermodynamic properties of NGF reveal that NGF  obeys a generalized  area law opening a new scenario for formulating its coarse-grained limit. The  structure of NGF  is strongly dependent on the dimensionality  $d$.  In $d=1$ NGF are growing complex networks for which  the preferential attachment mechanism is necessary in order to obtain a scale-free degree distribution. Instead, for NGF with dimension $d>1$  it is not necessary to have  an explicit preferential attachment rule to generate scale-free topologies. 
We also  show that NGF admits  a quantum mechanical description in terms of associated quantum network states. Quantum network states  are evolving by a Markovian dynamics and  a quantum network state at time $t$ encodes all possible NGF evolutions up to time $t$. Interestingly the NGF remains fully classical but its statistical properties reveal the relation to its quantum mechanical description. In fact the $\delta$-dimensional faces of the NGF  have generalized degrees that follow either  the Fermi-Dirac,  Boltzmann or Bose-Einstein statistics depending on the flavor $s$ and the dimensions $d$ and $\delta$.
\end{abstract}
\maketitle
\section{Introduction}
Recently, network geometry  \cite{interdisciplinary} is gaining increasing interest. Progress in this field is expected to have relevance for a number of applications, including routing protocols   \cite{Kleinberg,Boguna_navigability,Boguna_Internet}, data mining  \cite{Aste_filtering,Reka,Vaccarino2,Mason,Caldarelli}, and advances in the theoretical foundations of network clustering  \cite{Mukherjee}. In this context, several  theoretical questions have been recently approached including the formulation of models for  emergent geometry  \cite{Emergent,PRE,CQNM},  the  characterization of hyperbolic networks  \cite{Aste,Hyperbolic,Boguna_growing,Saniee}, the modelling of   complex networks embedded in the plane or in surfaces  \cite{Apollonian1,Apollonian4,Apollonian2,Apollonian3,det,Aste2}  and finally the development of a geometric information theory of networks  \cite{Franzosi1}. \\
 It  is also   believed that network geometry  \cite{Yau1,Yau2,Jost,Ollivier,Gromov,Majid} could provide a theoretical framework for establishing cross-fertilization between the field of network theory and quantum gravity. 
In fact most quantum gravity approaches rely on a discretization of space-time that takes a network-like structure.  These approaches include causal sets  \cite{Bombelli,Dowker}, causal dynamical triangulations  \cite{CDT1,CDT2,Burda1,Burda2},  group field theory  \cite{Oriti,Oriti_PRL}, loop quantum gravity  \cite{Smolin1,Smolin2,Rovelli}, energetic causal sets  \cite{Energetic1,Energetic2}, quantum gravity as an information network \cite{Trugenberger} and quantum graphity  \cite{graphity_rg,graphity1,graphity2}.  
Already several works explore the frontier territory between complex networks and quantum gravity. 
The relation between complex hyperbolic networks and causal sets  has  been  exploited by building a "network cosmology"  \cite{Cosmology}.  Moreover  causal sets  have been used  to analyze citation networks and measuring their effective dimension  \cite{Evans}. 
Recently  Complex Quantum Network Manifolds (CQNMs)  \cite{CQNM} have been introduced as  models of discrete manifolds that show the relation between quantum statistics and emergent network geometry.  

When faced with the problem of describing a network geometry, simplicial complexes of dimension $d$ become very useful.  These are discrete structures formed by the simplices of dimension $\delta$, with $0 \leq\delta\leq d$,  i.e.  nodes ($\delta=0$), links ($\delta=1$), triangles ($\delta=2$), tetrahedra ($\delta=3$) and so on. Simplicial complexes are widely used  in the  quantum gravity literature.  For example in the context of causal dynamical triangulations  \cite{CDT1,CDT2,Burda1,Burda2} and  group field theory  \cite{Oriti,Oriti_PRL} space-time is described using these discrete structures. 
In network theory, large attention  \cite{RMP,Doro_Book,Newman_Book} has been devoted to complex networks described as sets of nodes and links, i.e.  forming simplicial complexes of dimension $d=1$. Only recently additional attention has been addressed to simplicial complexes of higher dimension also called hypergraphs in the network science community.  These structures are important to capture relations existing between more than two nodes, such as the one existing in  collaboration networks (where each paper might result from a collaboration of more than two individuals, or a movie might have a large cast of actors), protein interaction networks (where  proteins form complexes consisting in general  of more than two types of proteins)  or in Twitter  (where one tweet might include several hashtags). 
Therefore equilibrium and non-equilibrium models of random simplicial complexes  and hypergraphs have been recently proposed by physicists and mathematicians  \cite{Emergent, PRE, CQNM, Dima_SC,Farber1,Farber2, Kahle,Newman1,Newman2}.  

Modeling complex networks has been the subject of intense research in network theory over the years.  In particular attention has been focusing on the minimal models able to generate network structure with the universal properties observed in real complex network datasets: the small-world property  \cite{WS}, the scale-free degree distribution  \cite{BA} and a non-trivial community structure  \cite{Santo}. 
In this context, non-equilibrium growing network models generating scale-free networks  \cite{BA,Fitness,Bose,Chayes,Weight,Doro_link,RMP,Doro_Book,Newman_Book} have been widely studied.  Scale-free  networks have highly inhomogeneous degree distribution $P(k)$ decaying as a power-law for large value of $k$, i.e.  $P(k)\simeq k^{-\gamma}$, with the power-law exponent $\gamma\leq 3$. The scale-free network distribution   affects the properties of dynamical processes defined on networks  \cite{crit,Dynamics,interdisciplinary} such as the Ising model, percolation, epidemic spreading, and quantum phase transitions.   In  growing network models formed by nodes and links, the so-called preferential attachment mechanism has been identified as a key element for obtaining scale-free networks as shown in the framework of the famous Barab\'asi-Albert model  \cite{BA}.  The preferential attachment rule determines that the probability that a node acquires new links is proportional to its degree.  Additional heterogeneity of the nodes, capturing intrinsic characteristics of the nodes that are different from the node degree, have been  modeled by associating an energy $\epsilon$ to the nodes of the network.   The energy $\epsilon$ of a node determines its fitness $\eta=e^{-\beta \epsilon}$, measuring the ability of the node to attract new links compared to the ability of other nodes with the same degree.  The first growing scale-free network model introducing this heterogeneity of the nodes is the Bianconi-Barab\'asi model  \cite{Fitness, Bose,Chayes} that has been used to model the Internet and the World-Wide-Web.  This model captures the competition existing between nodes to attract new links. In fact, nodes acquire new links with a generalized preferential attachment rule which assigns  to high degree and high fitness nodes  higher probability to acquire new links than  to  lower degree or lower fitness nodes. \\
The characterization of the Bianconi-Barab\'asi model has unveiled an important relation between complex networks and quantum statistics. 
In fact, the Bianconi-Barab\'asi model  \cite{Fitness,Bose,Chayes} can be mapped to a quantum Bose gas and, under the same circumstances in which the Bose gas  undergoes a Bose-Einstein condensation, a structural phase transition is observed in the network structure in which one node grabs a finite fraction of all the links  \cite{Bose,Chayes}.  Interestingly, the Fermi-Dirac statistics characterizes growing Cayley trees with energy of the nodes  \cite{Fermi}, and these results have been extended in different directions  \cite{Complex,Weight,Multiplex}, including weighted networks and multiplex networks.  It is to note that not only growing network models but also equilibrium network models have been shown to be related to quantum statistics  \cite{Garlaschelli}.  \\
Recently the new results obtained in  \cite{CQNM} for CQNMs show that also growing network manifolds describing a complex network geometry are related to  quantum statistics. In fact, in  Complex Quantum Network Manifolds the Fermi-Dirac, the Boltzmann and the Bose-Einstein statistics  coexist in the same network geometry describing the statistical properties of the $\delta$-dimensional faces of the CQNM.
   
Here our goal is to introduce  Network Geometry with flavor $s=-1,0,1$ (for short NGF) showing the strong effect of dimensionality $d$  on the geometry emergent from these models and the relation between NGF and quantum statistics. The NGFs describe growing simplicial complexes with energies associated to all their simplices, ( i.e. to their nodes, links, triangular faces, etc.) and evolving with (case $s=1$) or without (cases $s=-1,0$) explicit preferential attachment, forming either manifolds (case $s=-1$) or more general simplicial complexes (cases $s=0,1$). The NGF generalizes the  CQNM introduced in  Ref. \cite{CQNM} which constitutes the NGF with flavor  $s=-1$. For $s=-1,$ $d=3$  and $\beta=0$ the model reduces to the random Apollonian network \cite{Apollonian1,Apollonian4,Apollonian2,Apollonian3}. Moreover the NGF with flavor $s=1$ and dimension $d=1$ reduces to  the Bianconi-Barab\'asi model.
 
We will focus specifically  on the  thermodynamic properties of  NGF, on the relation of NGF to  complexity theory, and on the relation between these geometrical network structures and    their quantum mechanical description. 
In particular we will characterize the thermodynamic relations satisfied by the  NGF evolving by a non-equilibrium dynamics and obeying a generalized area law; we will identify  in which  dimension $d$ and for which flavor $s$ NGF are scale-free networks; and finally  we will provide a quantum mechanical description of NGF, constructing quantum network states characterizing the evolution of these models, and showing how quantum statistics emerges from the statistical properties of these networks.

In order to determine the thermodynamics of NGF, we  define its  total energy $E$,  total entropy $S$ and  area $A$. 
The thermodynamic properties of the NGFs reveal that these structures follow a generalized area law.  Since in quantum gravity the celebrated Jacobson  \cite{Jacobson,Chirco_liberati,Chirco_Rovelli} result relates the area law  to the Einstein equations as equation of state, this result could play a crucial role in determining the dynamics of NGFs at the macroscopic, coarse-grained level.   

Our results  highlight the strong effect of the dimensionality $d$ on the structure of the  NGF.  For NGF in $d=1$, like in the Barab\'asi-Albert model, preferential attachment is a necessary element for obtaining scale-free networks. 
Here we show that for NGF formed by simplicial complexes of dimension $d>1$ an explicit preferential attachment is not necessary to obtain scale-free networks, as an effective preferential attachment can emerge in simplicial complexes of dimension $d>1$ by dynamical rules that do not include an explicit preferential attachment. 
Therefore in dimension $d=2$ also Network Geometry with flavor $s=0$ that is not driven by an explicit preferential attachment generates scale-free networks.  In dimension $d\geq 3$ all the NGFs are  scale-free, independently of their flavor $s$.

The NGF can  be mapped to quantum network states evolving by a Markovian dynamics.  The relation between the NGF and their quantum mechanical description is also emerging from their statistical properties.  In fact,  NGFs in dimension $d$ have the generalized degree of their $\delta-$faces that as a function of the flavor $s$ and the dimensions $d,\delta$ follows Fermi-Dirac, Boltzmann or Bose Einstein  statistics. 
The dimension $d=3$ again plays a special role because it is the lowest dimension for   observing the coexistence of the Fermi-Dirac, Boltzmann and  Bose-Einstein statistics describing the statistical properties of the faces of the NGF  of dimensions $d$.

 \section{ Network Geometry with flavor $s$}
 \subsection{Network  Geometry with Flavor $s$  (NGF) and simplicial complexes}

Here we define  NGFs in a constructive way by characterizing  their non-equilibrium dynamical evolution.  

By $d$-dimensional simplex here we indicate a fully connected graph (a clique) of $(d+1)$ nodes. Its $\delta$-faces are all the $\delta$-dimensional simplices that can be built by a subset of $(\delta+1)$ of its nodes. In general, a   simplicial complex of dimension $d$ is formed by a set of simplices of dimension $d'\leq d$.

A  NGF of dimension $d\geq 1$ is a simplicial complex formed by  $d$-dimensional simplices glued along their $(d-1)$-dimensional faces also called $(d-1)$-faces.  
For example, a NGF of $d=1$  is formed by links glued at their end nodes, a NGF of  $d=2$  is formed by  triangles glued along their links, and a NGF of $d=3$  is formed by  tetrahedra glued along their triangular faces. The set of all possible $\delta$-dimensional faces (or $\delta$-faces) belonging to the $d$-dimensional NGF with $N$ nodes is here indicated by ${\cal Q}_{d,\delta}(N)$.
The set of all $\delta$-dimensional faces  belonging to the $d$-dimensional NGF with $\delta<d$ is indicated by ${\cal S}_{d, \delta}$.  

\subsection{Energies and Generalized degrees of NGF}
To each node $i$ of the NGF we assign an  {\em energy } of the node $\epsilon_i$  from a distribution $g(\epsilon)$.   The energy of the node is quenched and does not change during the evolution of the network.  This parameter describes  the intrinsic and heterogeneous properties of the nodes. 
To every  $\delta$-face  $\alpha\in {\cal S}_{d, \delta}$ we associate an {\em energy} $\epsilon_{\alpha}$ given by the sum of the energy of the nodes that belong to the face  $\alpha$,
\bea
\epsilon_{\alpha}=\sum_{i \in \alpha}\epsilon_i.  
\label{ea}
\eea
Therefore, each link will be associated to an energy of the link given by the sum of energies of the two nodes incident to it, and each triangular face will be associated to the sum of the energy of the three nodes incident to it and so on. 
The energy $\epsilon_{(i,j)}$ of the  links $\alpha=(i,j)$ belonging to any given  triangle  of the NGF formed by the  nodes $i$, $j$ and $r$   satisfy the triangular inequality 
\bea
|\epsilon_{(i,r)}-\epsilon_{(j,r)}|\leq \epsilon_{(i,j)}\leq \epsilon_{(i,r)}+\epsilon_{(j,r)}.
\label{triangulard}
\eea 
This result remains valid for any permutation of the order of the nodes $i,j$ and $r$ belonging to the triangle.
The energy of the links can  therefore be interpreted as length of the links and related to the use of spins in spin-networks and loop quantum gravity \cite{Rovelli}.\\
Although most of the derivations shown in this paper  can be performed similarly for either continuous or discrete energy of the nodes and of the higher dimensional $\delta$-faces, here we consider  the case in which the energies of the nodes $\{\epsilon_i\}$ and consequently the energy of the $\delta$-faces  $\{\epsilon_{\alpha}\}$ are integers. 

The {\em generalized degrees} $k_{d,\delta}(\alpha)$ of the $\delta$-face $\alpha$ (i.~e. $\alpha\in {\cal S}_{d,\delta}$) in a $d$-dimensional NGF is defined as the number of $d$-dimensional simplices  incident to it. 
Let us define the adjacency indicator function ${\bf a}$ of elements $a_{\alpha'}$ with $\alpha'\in{\cal Q}_{d,d-1}(N)$ taking value  $a_{\alpha'}=1$ if the $d$-dimensional complex $\alpha'$ is part of the NGF and otherwise taking value zero, $a_{\alpha'}=0$.
Using the adjacency indicator function, we can define the  generalized degree $k_{d,\delta}$ of a $\delta$-face $\alpha$ as
\bea
k_{d,\delta}(\alpha)=\sum_{\alpha' | \alpha \subset \alpha'}a_{\alpha'}. 
\eea
Therefore, in a NGF of dimension $d=1$ the generalized degree $k_{1,0}(\alpha)$ is the number of links incident to a node $\alpha$, i.e.  its degree. In $d=2$, the generalized degree $k_{2,1}(\alpha)$ is the number of triangles incident to a link $\alpha$ while the generalized degree $k_{2,0}(\alpha)$ indicates the number of triangles incident to a node $\alpha$. 
Similarly in a NGF of dimension $d=3$, the generalized degrees $k_{3,2}$, $k_{3,1}$ and $k_{3,0}$ indicate  the number of tetrahedra incident respectively to a triangular face, a link or a node. 

\subsection{NGF evolution}
\begin{figure}
\begin{center}
{\includegraphics[width=0.9\columnwidth]{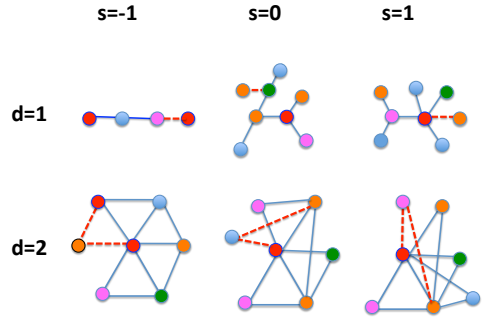}}
\end{center}
\caption{(Color online) The figure schematically illustrates the temporal evolution of the NGF of flavor $s$ in dimension $d=1,2$.  In dimension $d=1$ at each time step a new node is added  to the network  and is connected to the existing network by a single link.  In dimension $d=2$, at each time a new node is added to the network. This node is connected to the existing network by a triangle, i.e.  it is linked to two adjacent nodes of the network.  When NGF has flavor $s=-1$, each $(d-1)-$simplex can be connected at most to two $d-$dimensional simplices.  This implies that in $d=1$ each node can have at most degree two and in $d=2$ each link can be adjacent to at most two triangles.  Therefore NGF with flavor $s=-1$ are manifolds, and are also called CQNM  \cite{CQNM}.  On the contrary, NGF with flavor $s=0,1$ does not have this constraint and each $(d-1)$-dimensional simplex can be connected to an arbitrarily large number of $d$-dimensional simplices.  Therefore in $d=1$ each node can have an arbitrarily large degree and in $d=2$ each link can be incident to an arbitrarly large number of triangles.  \\
For $s=1$ the NGF evolution includes an explicit preferential attachment rule implying that each new $d-$dimensional simplex is linked to a $(d-1)-$face $\alpha$ with a probability that  increases linearly with its generalized degree $k_{d,d-1}(\alpha)$.  
Therefore the NGF with $d=1$, $s=1$ for  $\beta=0$ reduces to  the Barab\'asi-Albert model \cite{BA} and for $\beta>0$ it reduces to  the Bianconi-Barab\'asi model  \cite{Fitness,Bose}.   }
\label{fig1}
\end{figure}

The NGF comes in three {\em flavors} indicated by the variable $s=-1,0,1$. 
In order to define the non-equilibrium dynamics  of NGF we associate to each   $(d-1)$-face $\alpha$  the number $n_{\alpha}$ given by the sum of the $d$-dimensional simplices incident to $\alpha$ minus one, i.e. 
\bea
n_{\alpha}=k_{d,d-1}(\alpha)-1. 
\label{na}
\eea
If the variable $n_{\alpha}$ can only take values $0,1$ the NGF is a manifold also called CQNM. If instead the variable $n_{\alpha}$ can  also take values greater than two we have a NGF which is not a manifold.  
As we will see in the following, NGFs with flavor $s=-1$ describe manifolds, the CQNMs, while NGFs with flavor $s=0,1$ do not generate manifolds. \\
 The NGFs in dimension $d$ are evolving according to a non-equilibrium dynamics enforcing that at each time  the NGF is growing by the addition of a new $d$-dimensional simplex.   
 Here we describe the NGF evolution for NGF with every type of flavor $s=-1,0,1$ (see Supplementary Material  \cite{SM} for the MATLAB code generating NGF in dimensions $d=1,2,3$). \\
At time $t=1$ the NGF is formed by a single $d$-dimensional simplex.  
At  each time $t>1$ we add a simplex of dimension $d$ to a $(d-1)$-face $\alpha\in{\cal S}_{d, d-1}$  which is chosen with   probability  $\Pi_{\alpha}^{[s]}$ given by  
\bea
\Pi_{\alpha}^{[s]}&=&\frac{1}{Z^{[s]}(t)}e^{-\beta \epsilon_{\alpha}}(1+sn_{\alpha}), 
\label{P1}
\eea
where $\beta\geq 0$ is a parameter of the model called {\em inverse temperature}, $s=-1,0,1$ and $Z^{[s]}(t)$ is a normalization sum given by 
\bea
Z^{[s]}(t)=\sum_{\alpha\in {\cal S}_{d, d-1}}e^{-\beta \epsilon_{\alpha}}(1+s n_{\alpha}).  
\eea
Having chosen the $(d-1)$-face $\alpha$,  we glue to it a new $d$-dimensional simplex containing  all the nodes of the $(d-1)-$face $\alpha$  plus the new node $i$.   It follows that the new node $i$ of the new simplex is linked to each node $j$ belonging to $\alpha$.  
Finally we note here that the number of nodes $N$ at time $t$ is given by $N=t+d$. In fact for $t=1$ the NGF is formed by a single $d$-dimensional simplex, and has $N=d+1$ nodes.
At each time $t>1$, a new $d$ dimensional simplex is added to the NGF. This simplex has a single new node. Therefore the number of nodes grows at each time step by one, and is given by $N=t+d$.
\\
In Figure $\ref{fig1}$ we show the first few steps of  the  NGF evolution for the cases $d=1,2$ and  $s=-1,0,1$. 
In Figure $\ref{fig2}$ we show a visualization of NGF with $s=-1,0,1$, $d=1,2,3$ and $\beta=0.1$.  These NGFs for $d=1$ are trees, for $d>1$ they have at the same time large clustering and small average distance between the nodes, i.e.  they are small world and they have a  non-trivial community structure.  \\
\subsection{The NGF of different flavor $s$ have  significantly different structure and dynamics}

The NGF of different flavor $s$ have significantly different geometry and statistical properties. In fact, depending on the flavor $s$ either manifolds ($s=-1$) or more general simplicial complexes are generated. The dynamical properties of NGF of different flavor $s$ are also very different, with NGF of flavor $s=1$ including an explicit preferential attachment while NGF with flavor $s$ are driven by an homogeneous attachment dynamics. 
In the following we will discuss the properties of NGF  as a function of their flavor $s$ and their dimension $d$. Moreover we will relate specific limiting cases of NGFs with existing models of complex networks.\\
 
 The dynamical rules of the NGF imply that only for the case $s=-1$ NGF are actually manifolds, also called CQNMs  \cite{CQNM}.  In fact, for $s=-1$ the probability $\Pi_{\alpha}^{[-1]}$ defined in Eq.  $(\ref{P1})$ is zero, (i.e.  $\Pi_{\alpha}^{[-1]}=0$) for every $(d-1)-$face $\alpha$ with $n_{\alpha}=1$.  If a $(d-1)-$face has $n_{\alpha}=1$ it is  already incident to two $d-$dimensional simplices, as its generalized degree is $k_{d,d-1}(\alpha)=n_{\alpha}+1=2$.  Such a face $\alpha$ cannot be glued to any additional $d-$dimensional simplex because the probability that we glue an additional $d-$simplex to this face is $\Pi_{\alpha}^{[-1]}=0$. In particular the NGF of $d=1$ and flavor $s=-1$ is a chain.\\
For $s=1$, we observe that the probability to attach a new simplex to the $(d-1)-$face $\alpha$, $\Pi_{\alpha}^{[1]}$, is proportional to its generalized degree $k_{d,d-1}(\alpha)=1+n_{\alpha}$ providing a generalization of the so-called preferential attachment mechanism, known to be  necessary  for generating scale-free networks in simplicial complexes of dimension $d=1$. \\
The evolution of NGF is  related to existing complex network models with fitness of the nodes  \cite{BA,Fitness,Bose,Fermi,Weight,Chayes,Complex,Multiplex}. 
In particular the NGF with $d=1,\beta=0$ and $s=1$ is the  Barab\'asi-Albert model  \cite{BA} (with the number of initial links of each node given by one), while for $d=1,\beta>0$ and $s=1$  it is the Bianconi-Barab\'asi model  \cite{Fitness,Bose} (always with the number of initial links given by one). 
Moreover, the NGF of $d=2$ with flavor $s=0$ and $\beta=0$ has been first proposed as a scale-free network model in Ref.   \cite{Doro_link}. 
The  NGF in $d=2$ is related to models proposed in the recent literature on emergent network geometry  \cite{Emergent,PRE}. \\
Finally the NGF for $s=-1$, $d=3$  and $\beta=0$ is a stacked polytope model and as such reduces to the stochastic Apollonian network \cite{Apollonian1,Apollonian4,Apollonian2,Apollonian3}.
We note here that it  is  possible to define NGF allowing also for a $\Pi_{\alpha}^{[s]}$ given by Eq. $(\ref{P1})$ with real  values of $s$, as long as    $s>0$. These models will include energy of the $\delta$-faces and preferential attachment with an initial additive constant \cite{Doro_add}. These models will qualitatively behave like the NGF with $s=1$.
Also it is  possible to consider negative  values $s\neq-1$. Nevertheless, to avoid having negative probabilities  $\Pi_{\alpha}^{[s]}$ given by Eq. (\ref{P1}), we should impose that $s$ takes negative rational values   $s=-1/m$ with  $m\geq 1$. This model allows  the generalized degree of $(d-1)$-faces to be  at most $m$ and therefore $n_{\alpha}\leq m$.  These models are related to the ones recently proposed in Ref. \cite{Emergent} for simplicial complexes in $d=2$. For simplicity here  we restrict our study only to NGF with flavor $s=-1,0,1$ that display a significant change in their structural properties. 

\subsection{Area and volume of NGFs}
The {\em boundary} of the NGF is defined as the set of $(d-1)-$faces with $n_{\alpha}=0$, i.e.  incident to exactly one $d-$dimensional simplex.  We will call the {\it area} $A$ of the NGF the  number of $(d-1)-$faces in the boundary, i.e. 
\bea
A=\sum_{\alpha\in  {\cal S}_{d, d-1}}\delta(n_{\alpha},0). 
\eea
At each time step of the NGF dynamical evolution,  a  $(d-1)$-face is chosen and a new  simplex is attached to it. If this face is initially  at the boundary of the NGF,  after the addition of the simplex it will leave  the boundary, contributing to a negative change of  $A$  of  one.
At the same time the new simplex adds $d$ new $(d-1)-$faces to the boundary,  contributing to an increase of  $A$ by $d$.
For  NGF with flavor $s=-1$ (i.e.  for CQNMs),  the new $d$ dimensional complex is attached exclusively to a $(d-1)$-face at the boundary. Moreover at time $t=1$ the area is the area of a single $d$-dimensional simplex, and is given by $A=d+1$ . Therefore we have
\bea 
A=(d-1)t+2.
\eea
In  general for NGF with every flavor $s=-1,0,1$, and sufficiently low values of $\beta,$ we have 
\bea
A\simeq \lambda t
\label{At}
\eea
for $t\gg 1$ with $\lambda\in [d-1,d)$.  
The {\em volume} $V$ of the NGF is given by the total number of $d-$dimensional simplices that form the NGF. 
The volume $V$ of the NGF at time $t$ is equal to the time, i.e. 
\bea
V=t,
\eea
since at each timestep one $d$-dimensional simplex is added to the NGF.  Therefore in NGF the area $A$ is proportional to the volume $V$, i.e.  $A\propto V$. 
This property of the NGF is  crucial to determine the NGF small-world diameter, i.e.  a diameter at most increasing like the logarithm of time $t$, for sufficiently low values of the inverse temperature $\beta$. 

\begin{figure*}
\begin{center}
$\begin{array}{ccc}
{\includegraphics[width=0.55\columnwidth]{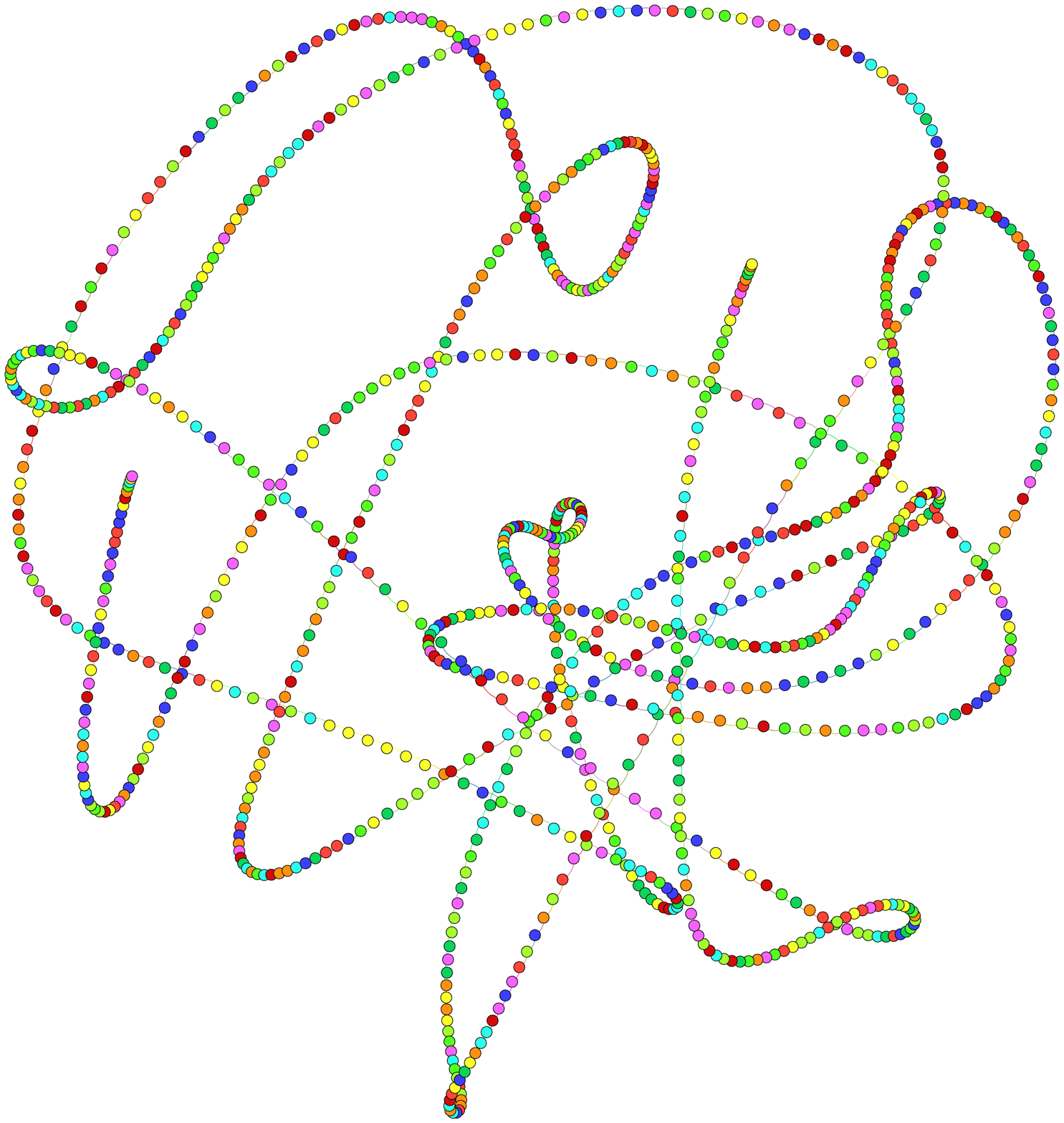}}&
{\includegraphics[width=0.55\columnwidth]{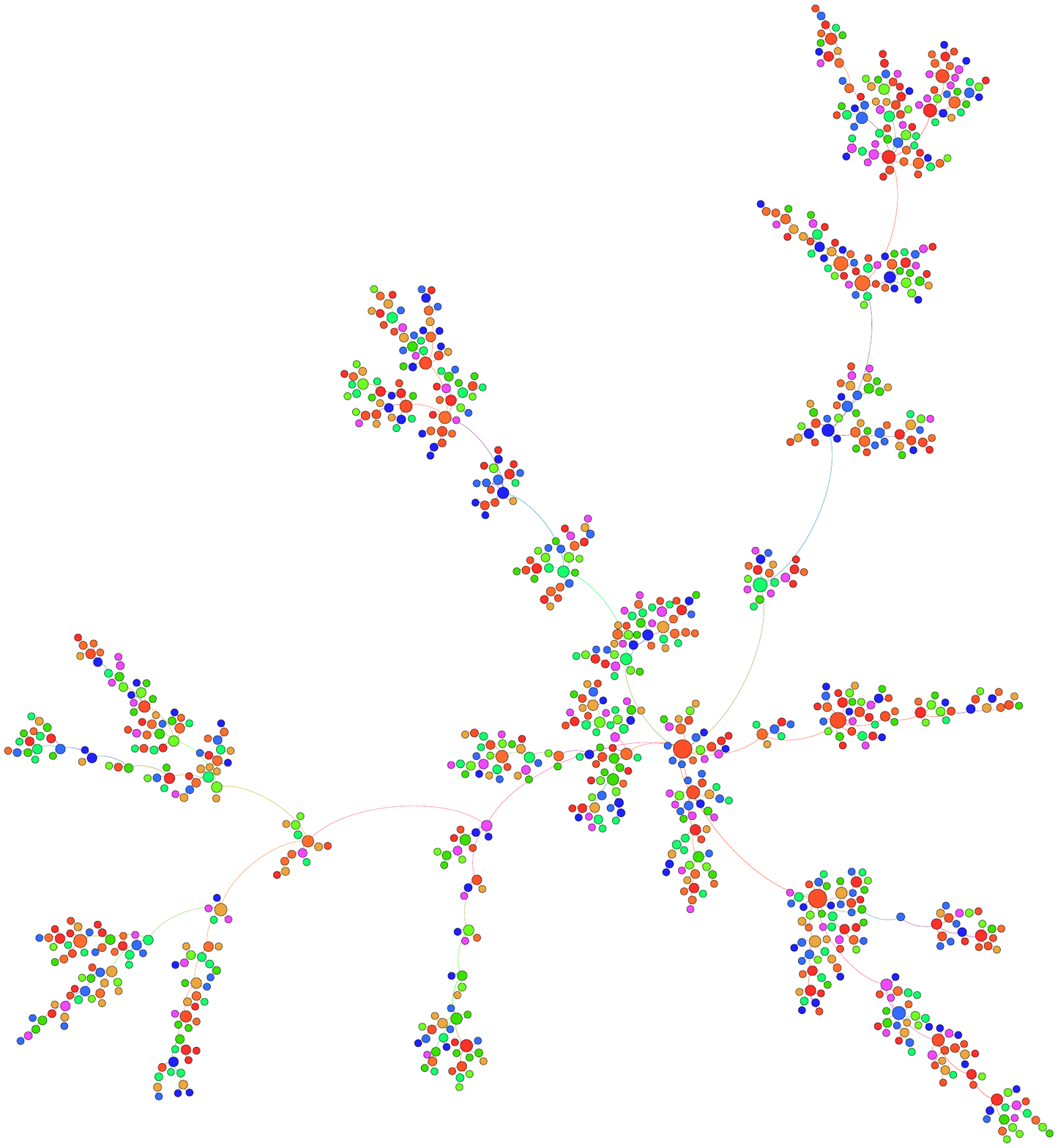}}&
{\includegraphics[width=0.55\columnwidth]{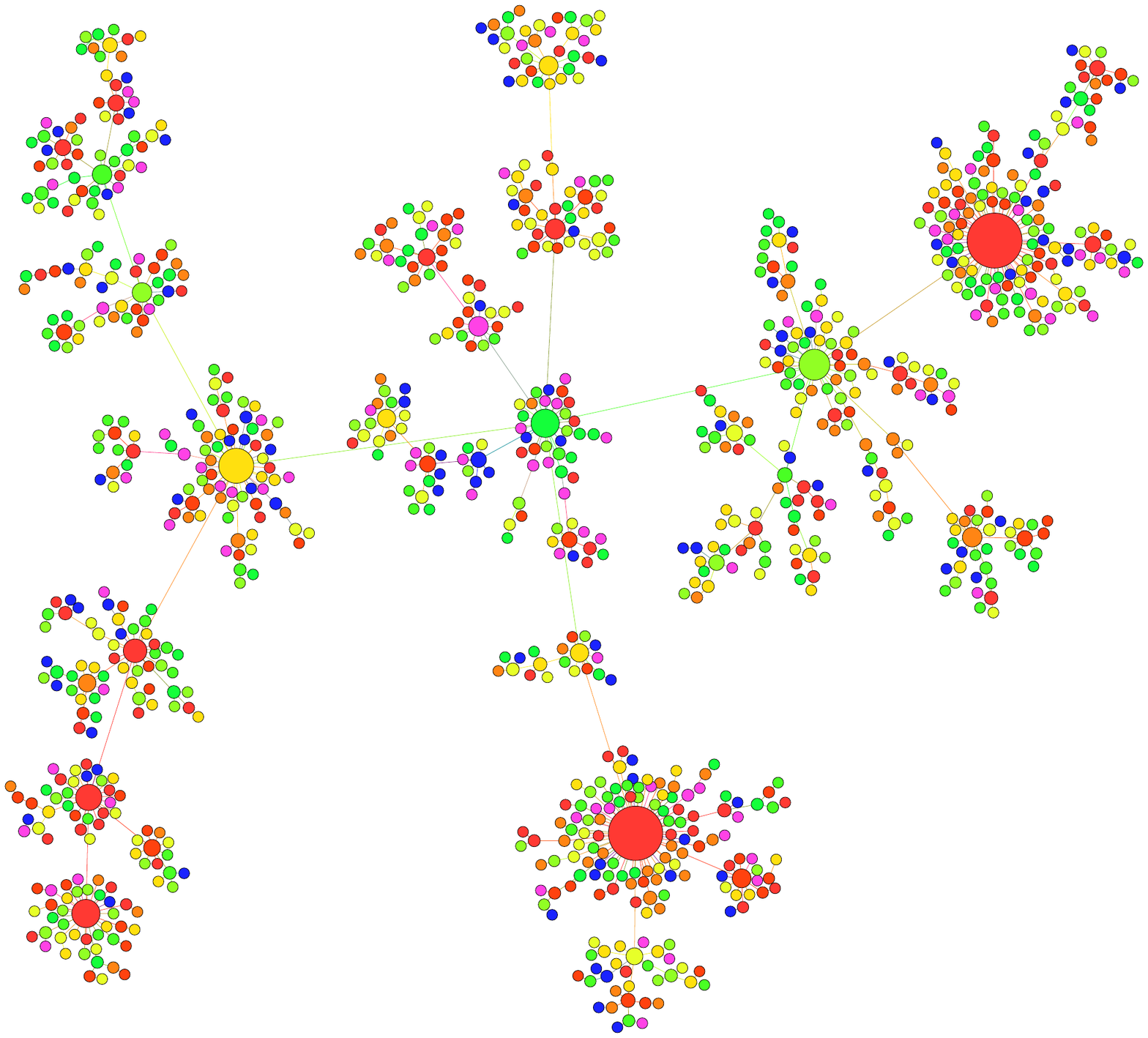}}\\
$(a) $&
$(b) $&
$(c) $\\
$d = 1, s=-1$&
$d = 1, s = 0$&
$d = 1, s = 1$\\
{\includegraphics[width=0.55\columnwidth]{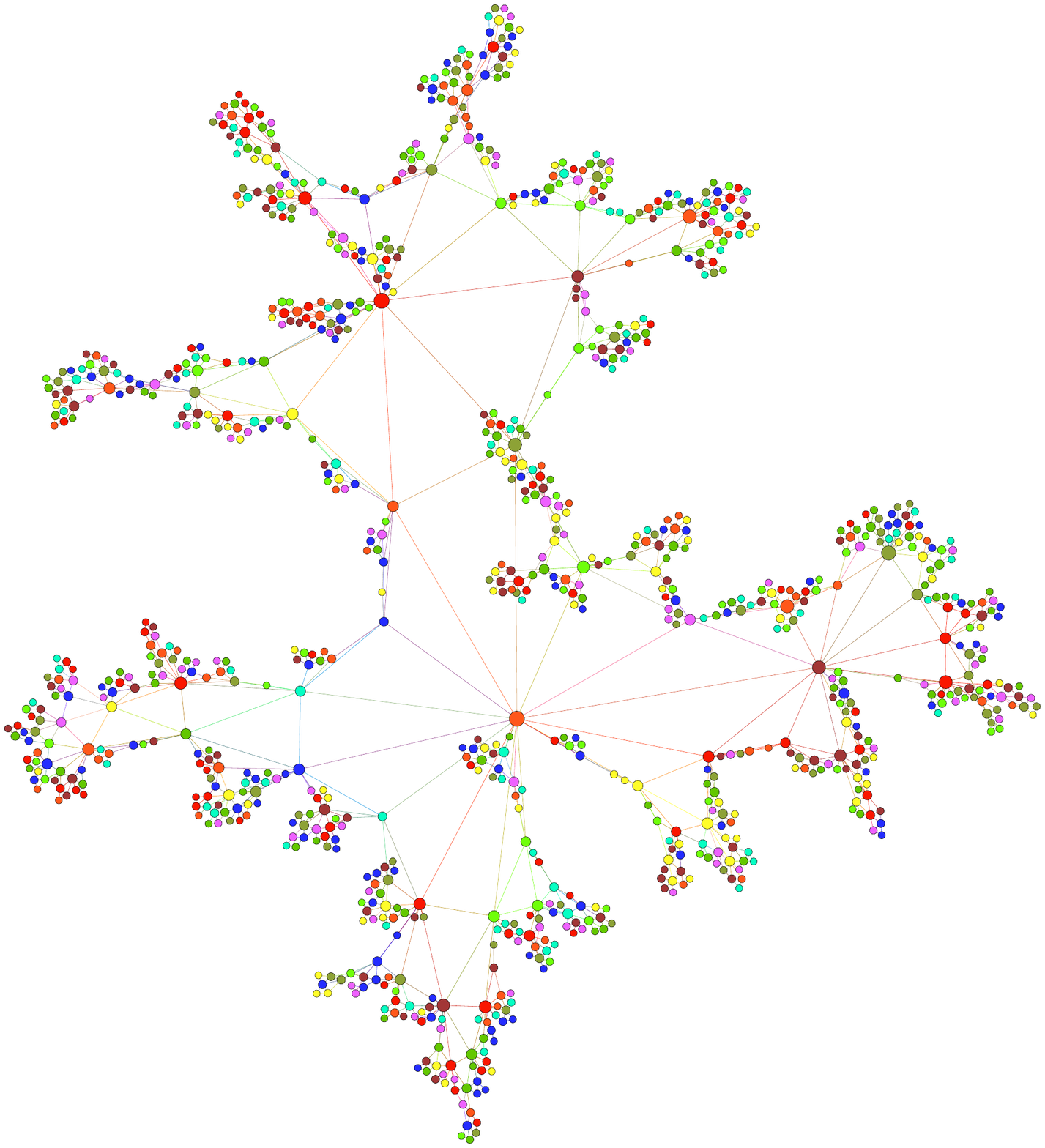}}	&
{\includegraphics[width=0.55\columnwidth]{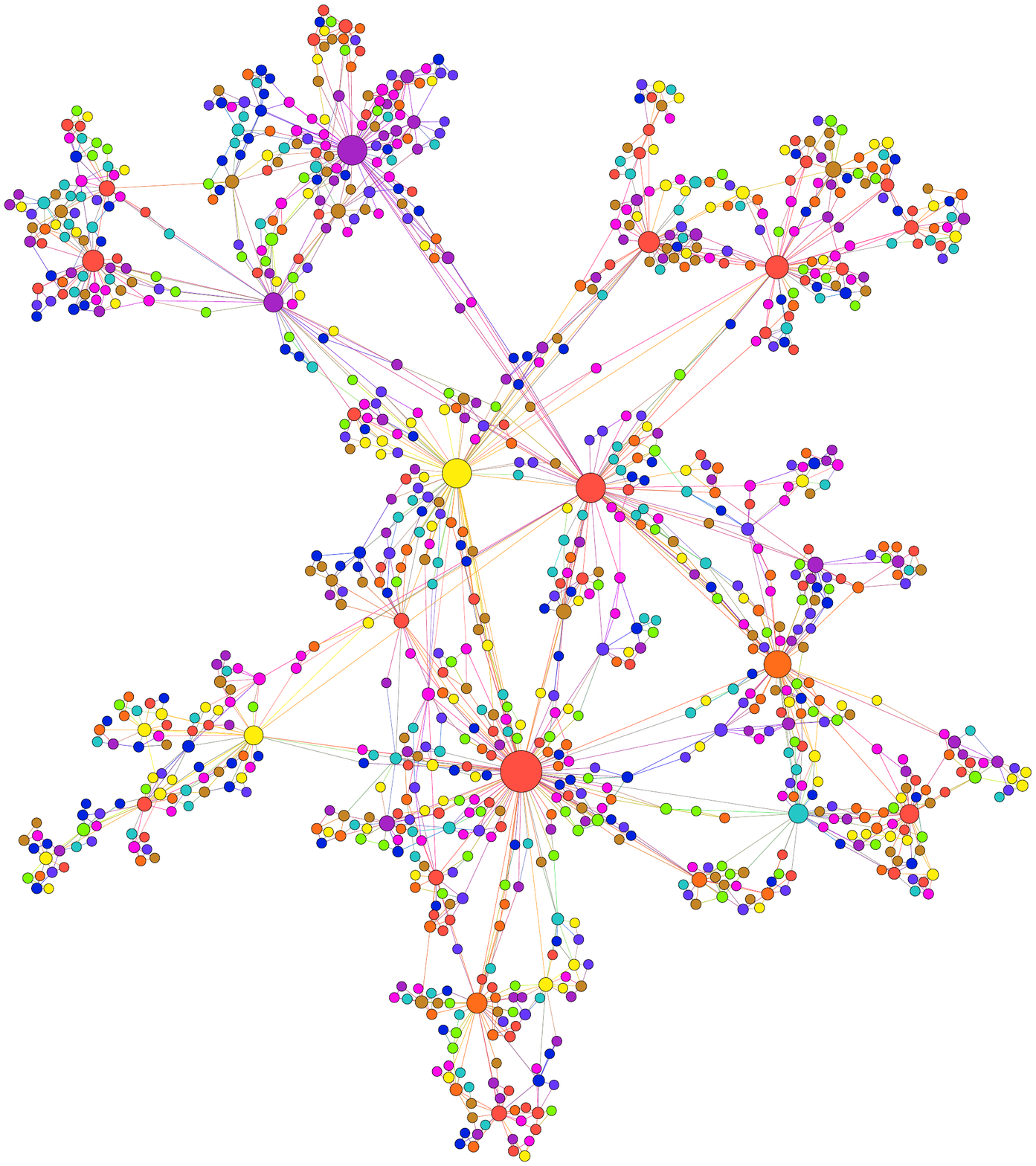}}&
{\includegraphics[width=0.55\columnwidth]{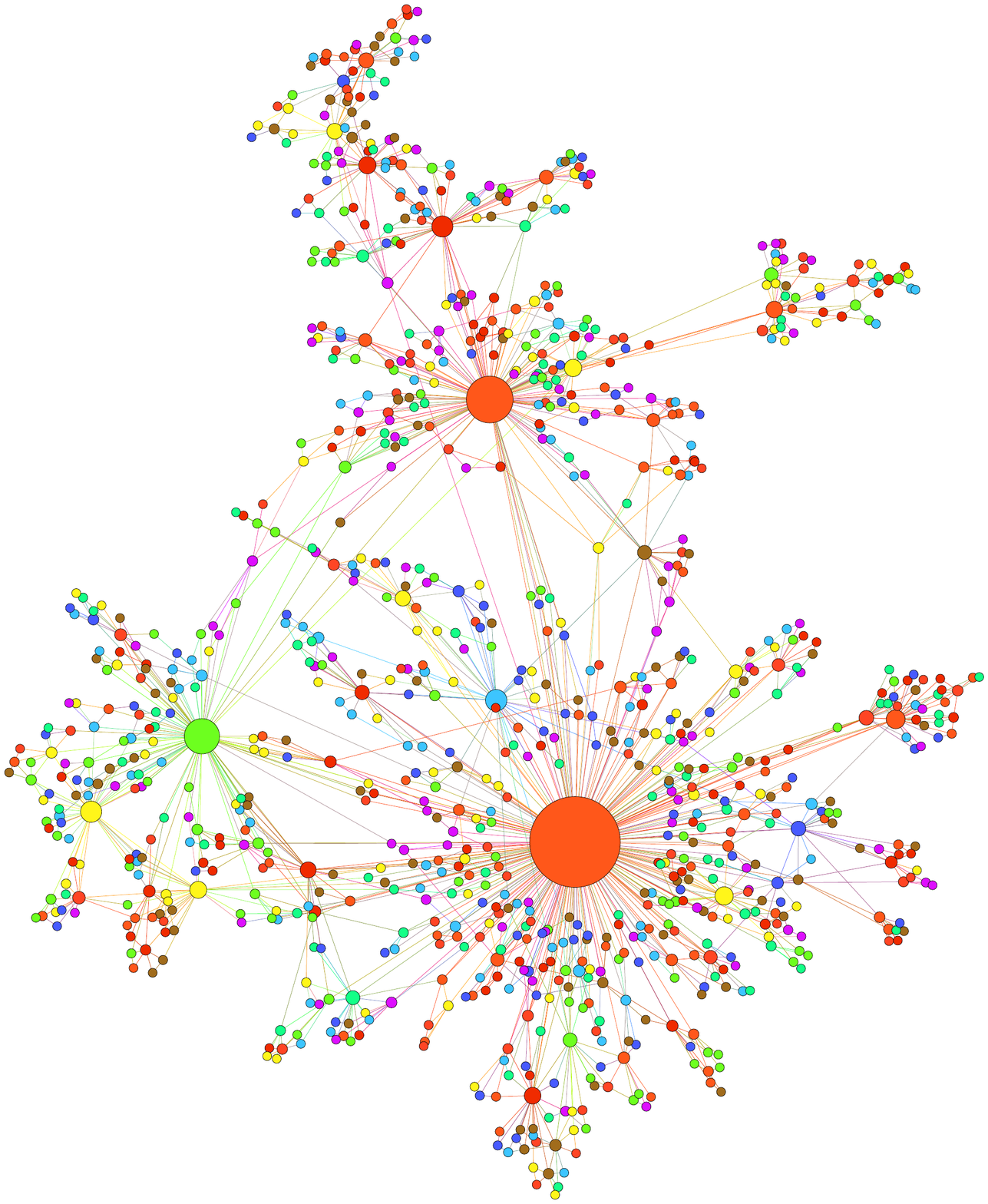}}\\
$(d) $&
$(e)$&
$(f) $\\
$d = 2, s = -1$&
$d = 2, s = 0$&
$d = 2, s = 1$\\
{\includegraphics[width=0.55\columnwidth]{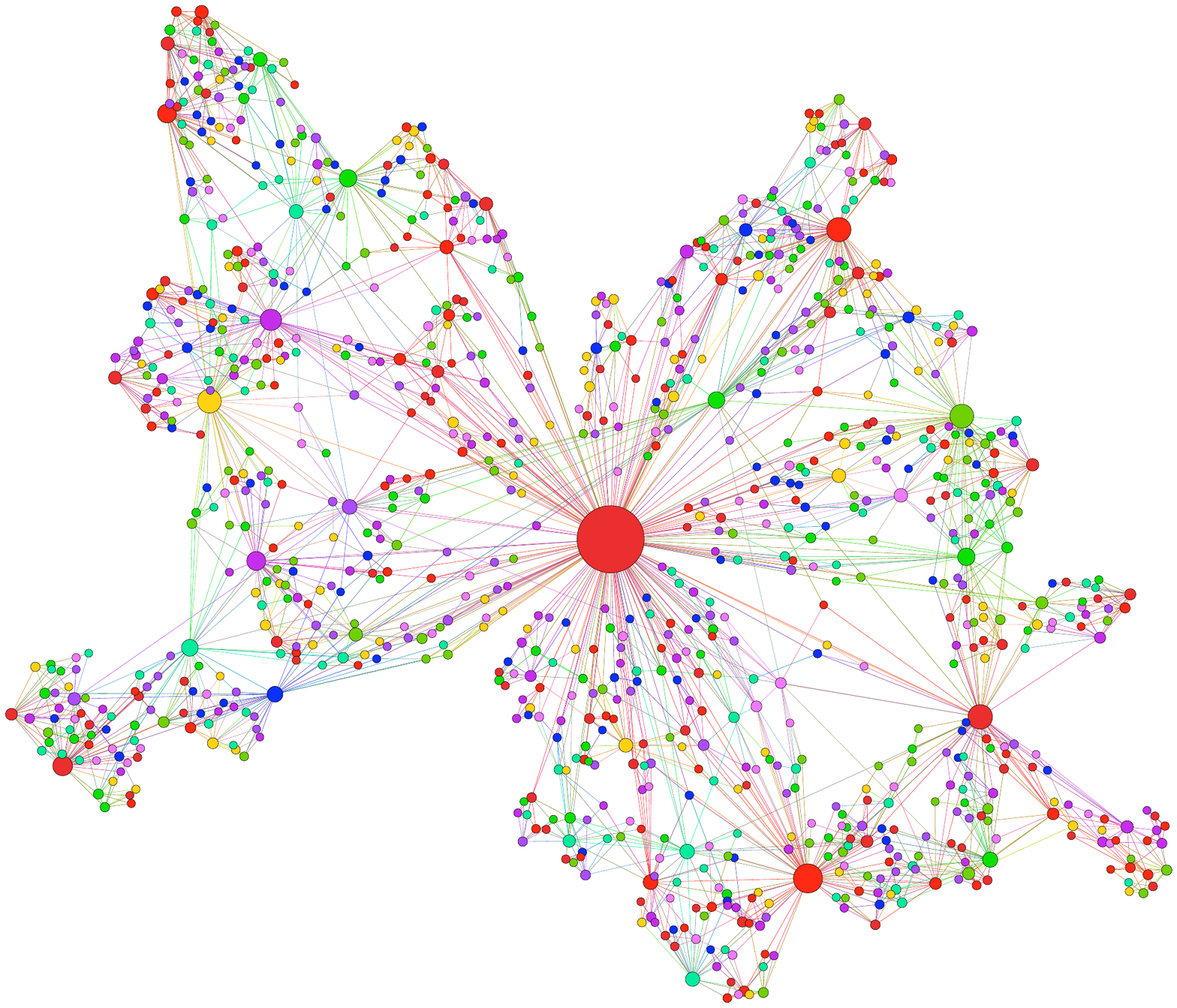}}&
{\includegraphics[width=0.55\columnwidth]{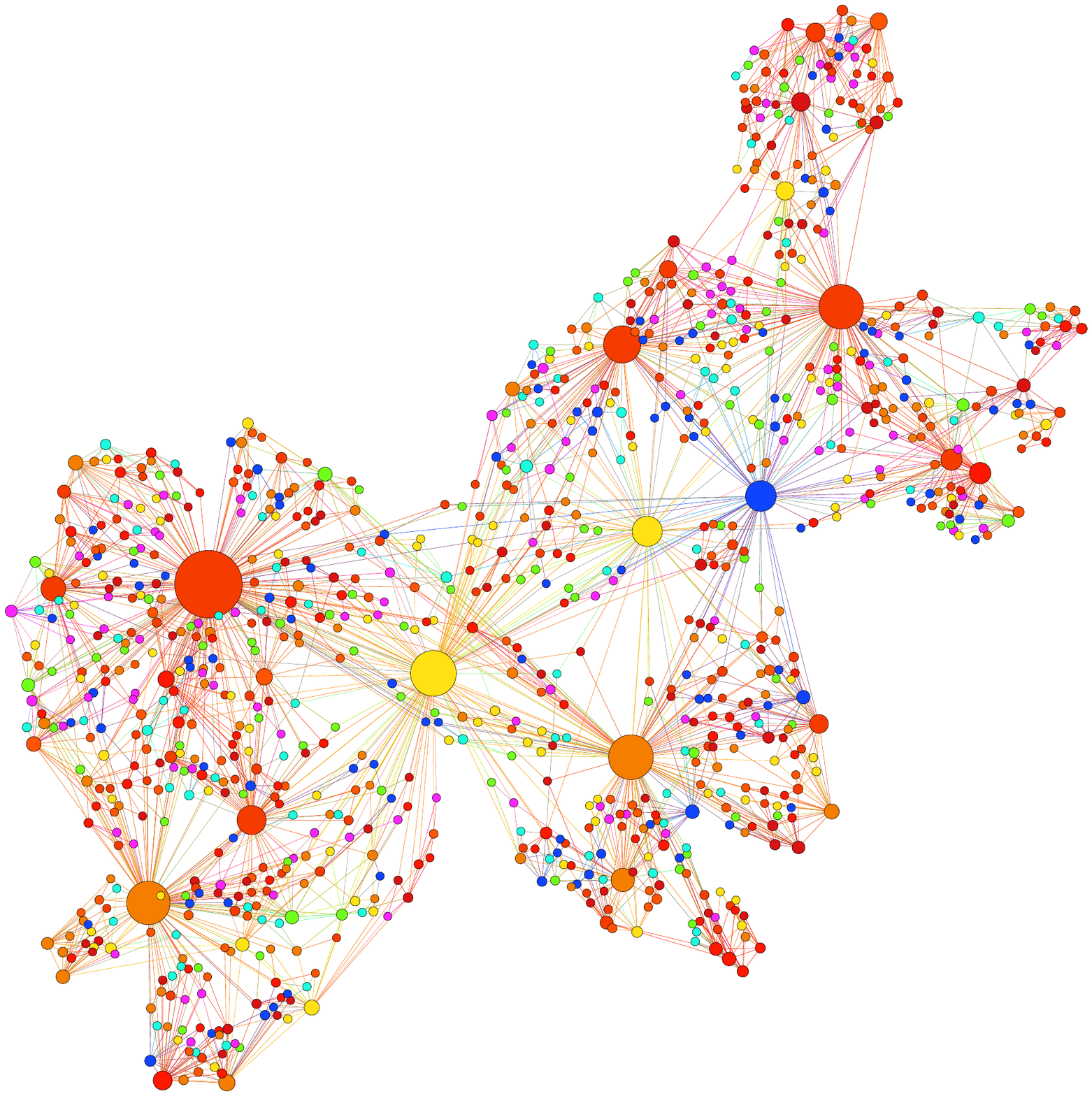}}&
{\includegraphics[width=0.55\columnwidth]{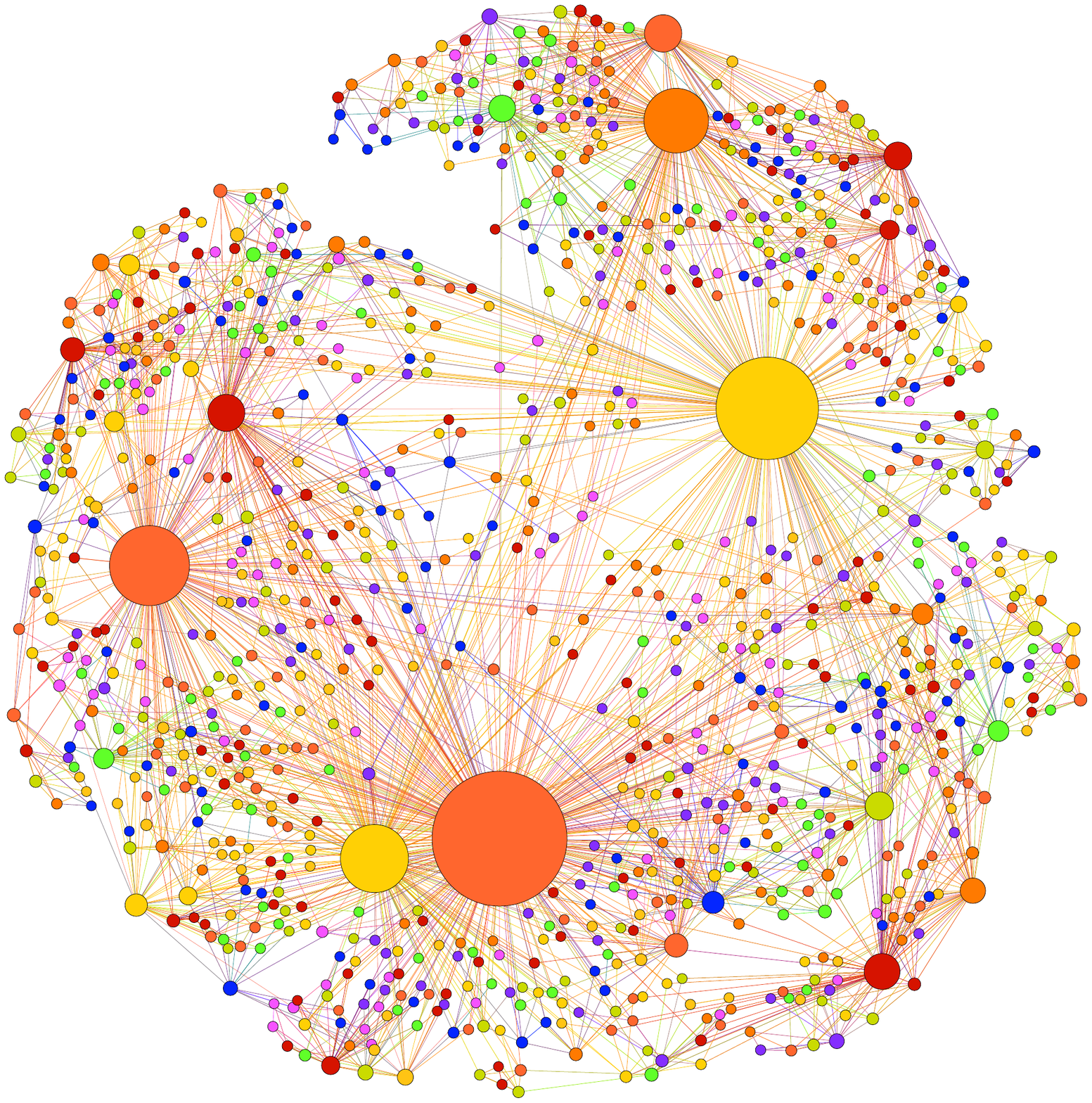}}\\
$(g)$&
$(h)$&
$(i)$\\
$d = 3, s = -1$&
$d = 3, s = 0$&
$d = 3, s = 1$\\
\end{array}$
\end{center}
\caption{(Color online) Network Geometry with flavor $s=-1,0,1$ and dimension $d=1,2,3$.  The NGFs  have $N=10^3$ nodes, $\beta=0.1$ and uniform distribution of the energy of the nodes $g(\epsilon)=1/10$ for $0\leq \epsilon<10$.  The color of the nodes indicates their energy,  the color code keeps the same order of the frequency of light (in order of increasing energy we have red,orange, yellow,green, blue, violet) the size of the nodes is proportional to their degree.}
\label{fig2}
\end{figure*}
\subsection{The dual of the NGFs}

The NGF  have a particularly simple dual network structure. The dual network is formed by considering nodes indicating the $d$-dimensional simplices and links connecting $d$-dimensional simplices that share a $(d-1)$-face.
For NGF with flavor $s=-1$, i.e. for the CQNMs, the dual is a tree with degree bounded by  $d+1$. In fact each $(d-1)$-face connects at most two $d$-dimensional simplices and each $d$-dimensional simplex has exactly $d+1$ $(d-1)$-dimensional faces.  Interestingly, as it is possible to see in Figure $\ref{fig2}$,  the CQNMs, also if they have  very  homogeneous dual networks, can display very complex structure, and as we will see in the next section they are scale-free for  $d\geq 3$. This shows  a clear example in which  the  relation between simplicial complexes and their dual networks might not preserve the same complexity properties. \\
For Network Geometry with flavor $s=0,1$ the dual network remains a tree but the degree of its nodes is no longer bounded.\\
The tree like nature of the dual network of the NGF allows for relevant simplifications in the analytical calculations.

\section{Thermodynamics of  NGFs}
\subsection{Probability of a given NGF evolution and total energy of a given NGF}
Given the evolutionary dynamics of the NGFs, the  evolution of the NGF  up to  time $t$ is fully determined  by the sequence  $\{\alpha_{t'}\}_{t' \leq  t}$, where    $\alpha_{t'}$ indicates  the $(d-1)$-face to  which the new $d$-dimensional simplex is added at time $t'>1$.  
Moreover the NGF is associated with the sequence of the energies of its $N=t+d$ nodes $\{\epsilon(t')\}_{t'\leq t+d}$. Of those only the energy of the nodes arrived in the NGF before time $t$, i.e. the sequence $\{\epsilon(t')\}_{t'< t+d}$ determines the probabilities of choosing a particular sequence of $\{\alpha_{t'}\}_{t' \leq \ldots t}$.
Finally   it is possible  to evaluate the probability  $P(\{\alpha(t')\}_{t' \leq t}|\{\epsilon(t')\}_{t'< t+d},s)$  that  the temporal evolution  until time $t$ of the NGF with flavor $s$ is described by the subsequent addition of $d$-simplices to the $(d-1)$-faces   $\{\alpha(t')\}_{t' \leq t}$  given that the energies 
of the nodes until time $t-1$ are  $\{\epsilon(t')\}_{t'<t+d}$.
  In fact $P(\{\alpha(t')\}_{t' \leq t}|\{\epsilon(t')\}_{t'< t+d},s)$ is given by  the product of the probability of each subsequent addition of the new simplex to the $\alpha(t)$ face, i.e. 
\bea
P(\{\alpha(t')\}_{t' \leq t}|\{\epsilon(t')\}_{t'< t+d},s)=\prod_{t'\leq t}\Pi_{\alpha(t')}^{[s]},
\label{Pat}
\eea
where $\Pi_{\alpha(t')}$ is given by Eq.  $(\ref{P1})$. 
Inserting the explicit expression of $\Pi_{\alpha(t')}$ in Eq.  $(\ref{Pat})$, we obtain
\bea 
 P(\{\alpha(t')\}_{t' \leq t}|\{\epsilon(t')\}_{t'< t+d},s)&=&\frac{1}{{\cal Z}^{[s]}(t)}e^{-\beta E}\nonumber \\
 &&\hspace{-20mm}\times\prod_{\alpha\in {\cal S}_{d,d-1}(t)}(1+sn_{\alpha}(t))!
\eea
Here we have indicated by $E$ the {\em total energy} of the NGF, given by 
\bea
E(t)=\sum_{\alpha \in {\cal S}_{d, d-1}}\epsilon_{\alpha} n_{\alpha}(t),
\label{E}
\eea
and with ${\cal Z}^{[s]}(t)$ the normalization constant,
\bea
{\cal Z}^{[s]}(t)&=&\sum_{\{\alpha(t')\}_{t' \leq t}} e^{-\beta E}\prod_{\alpha\in {\cal S}_{d,d-1}(t)}(1+sn_{\alpha}(t))!\nonumber \\
&&=\prod_{t'\leq t}Z^{[s]}(t').
\label{calZ}
\eea
Moreover ${\cal S}_{d,d-1}(t)$ is the set of $(d-1)-$faces in the NGF formed by the subsequent addition of $d-$dimensional simplices to the faces $\{\alpha(t')\}_{t'\leq t}.$

For sufficiently low values of $\beta$ we have that for large times, i.e. for $t\gg 1$, the ratio $Z^{[s]}/t$ is a self-averaging quantity and   $\lim_{t\to \infty}Z^{[s]}/t= e^{-\beta \mu_{d,d-1}^{[s]}}$, with $\mu_{d,d-1}^{[s]}$ indicating the {\em chemical potential} associated to the $(d-1)$-faces in NGF of flavor $s$.  Therefore we can approximate ${\cal Z}^{[s]}$ as 
\bea
{\cal Z}^{[s]}(t)&\simeq& e^{-\beta \mu_{d,d-1}^{[s]}} t! \simeq e^{-\beta \mu_{d,d-1}^{[s]}} N!
\eea 
for large times $t\gg 1$ and $t\simeq N$.
Finally t $P(\{\alpha(t')\}_{t' \leq t}|\{\epsilon(t')\}_{t'< t+d},s)$ can be expressed  as 
\bea
\hspace*{-3mm}P(\{\alpha(t')\}_{t' \leq t}|\{\epsilon(t')\}_{t'< t+d},s)=\frac{1}{N!}e^{-\beta (E-\mu_{d,d-1}^{[s]}N-F)} 
\eea
where $F$ is given by 
\bea
e^{\beta F}=\prod_{\alpha\in {\cal S}_{d,d-1}(t)}(1+sn_{\alpha}(t))!
\eea
\subsection{The entropy of the NGF and the generalized area law}
We   note that  different histories of the NGF up to time $t$ can give rise to the same network structure. This network structure is indicated by $G_N$ where $N=N(t)$ is the number of nodes of the network and $\{\epsilon_i\}_{i\leq N}$ are the energies of the nodes. All the possible temporal evolutions $\{\alpha(t')\}_{t'\leq t}$ of the NGF corresponding to the same network $G_N$ have the same probability  $P(\{\alpha(t')\}_{t' \leq t}|\{\epsilon_i\}_{i\leq N},s)=P(\{\alpha(t')\}_{t' \leq t}|\{\epsilon(t')\}_{t'< t+d},s),$ and they can be obtained from a given history by considering all causal relabelings of the  nodes.
We define the  probability $P(G_N|\{\epsilon_i\}_{i\leq N},s)$ that the NGF of flavor $s$ at time $t$ results in a  given network structure $G_N$,  independently of its temporal evolution, given the  energy of the nodes $\{\epsilon_i\}_{i\leq N}$. Using  the fact that the dual of the NGF is a tree, this probability can be calculated with methods already developed in \cite{Burda,Ambjorn} by evaluating the number of possible causal relabelings of the dual tree.
Specifically we have 
\bea
P(G_N| \{\epsilon_i\}_{i\leq N},s)=e^{-\beta (E-\mu_{d,d-1}^{[s]}N-F)}z_N^{[s]}
\eea
where 
\bea
z_N^{[s]}=\frac{1}{N!}L^{[s]}(T)
\label{treb}
\eea
and where $L^{[s]}(T)$ indicates the number of different NGF temporal evolutions giving rise to the same network $G_N$. It can easily be realized that $L^{[s]}(T)$ indicates also 
the number of different labelings of the tree $T$ that is the dual network of the NGF. 
The introduced quantity $z_N^{[s]}$ can be calculated by following the derivation given in Ref. \cite{Burda} as long as  the NGF is in a stationary state and the degree distribution of the tree describing the dual network of the NGF is known.
In fact, it is possible to evaluate the scaling of $z_N^{[s]}$ by writing a recursive equation for $L^{[s]}(T)$  for a tree $T$ given by a  root node connected to $p$ subtrees $T_1,T_2\ldots T_p$ of $N_1,N_2\ldots N_p$ nodes respectively. The recursive equation is  given by 
\bea
L^{[s]}(T)=\frac{(N-1)!}{N_1!N_2! \ldots N_p!}\delta_{\sum_i N_i+1,N}\prod_{i=1}^pL^{[s]}(T_i).
\label{lst1}
\eea
Here, differently from the case analyzed in \cite{Burda}, the different branches $T_1,T_2,\ldots, T_p$ of the tree $T$ are not exchangeable since the tree $T$ is a dual of a labelled NGF where the labels indicate the different energies of the nodes. 
Using Eq. $(\ref{lst1})$,  it is found (see Supplementary Material for details) that $z_N^{[s]}$ scales with the number of nodes as 
\bea
z_N^{[s]}=C^{[s]}e^{\beta \nu^{[s]} N}
\label{zns}
\eea
as long as the NGF is not a chain (it is different from the case $s=-1, d=1$), and the NGF reaches a stationary state (low enough values of $\beta$).
In fact,  the prefactor ${1}/{N!}$ in Eq. (\ref{treb}) is compensated by the number of terms in the summand. Therefore,  in Eq. $(\ref{zns})$,  $C^{[s]}$ is a subleading factor, and $\nu^{[s]}$ depends on the degree distribution of the dual of the NGF, and therefore depends on its flavor $s$. 

Finally  the probability $P(G_N,s)$ scales exponentially with the number of nodes and can be written for large networks  $ N\gg1$ as
\bea
P(G_N|\{\epsilon_i\}_{i\leq N},s)=C^{[s]}e^{-\beta (E-\mu_{d,d-1}^{s}N -\nu^{[s]} N-F)}.
\label{PG}
\eea
The entropy $S(N)$ of the NGF has the natural definition 
\bea
S(N)&=-\sum_{G_N}&P(G_N|\{\epsilon_i\}_{i\leq N},s) \ln P(G_N|\{\epsilon_i\}_{i\leq N},s). \nonumber
\label{S}
\eea
The   {\em total energy} $E$ and the {\em entropy} $S$ of NGF satisfy  thermodynamics relations. In order to derive them, let us evaluate  the variation in entropy of the network $\Delta S$ given by 
\bea \Delta S(N)=S(N)-S(N-1). \eea
It  can be easily shown, using the definition of the total energy $E$ in Eq.  $(\ref{E})$ and the rules determining the NGF evolution,  that  
\bea
\Avg{\epsilon_{\alpha}}_{\Pi^{[s]}}=&\Avg{\Delta E}_{\Pi^{[s]}}. 
\label{DE}
\eea
Finally, since the dynamics of the NGF reaches stationarity for sufficiently low values of $\beta$,  both $\Avg{\epsilon_{\alpha}}_{\Pi}$ and $\Avg{\ln(1+sn_{\alpha})}_{\Pi}$ are independent  of time for sufficiently large times $t\gg 1$. 
Therefore the  relation between $\Delta S$ and $\Avg{\Delta E}$  calculated over the interval $\Delta t=1$ can be found using Eqs.  $(\ref{PG})$, $(\ref{DE})$ and is given by 
\bea
\hspace*{-4mm}\Delta S= \left\{\beta \left( \Avg{\epsilon_{\alpha}}_{\Pi^{[s]}}-\mu_{d,d-1}^{[s]}-\nu^{[s]} \right)-\Avg{\ln (1+sn_{\alpha})}_{\Pi^{[s]}} \right\}\Delta t.\nonumber
\eea
Using the scaling of the area $A$ with time given by Eq.~$(\ref{At})$, it follows that the change in entropy $\Delta S$ can then be  expressed as
\bea
\hspace*{-4mm}\Delta S= \left\{\beta \left[\Avg{\epsilon_{\alpha}}_{\Pi^{[s]}}-\mu_{d,d-1}^{[s]}-\nu^{[s]}\right]-\Avg{\ln (1+sn_{\alpha})}_{\Pi^{[s]}}\right\}\frac{\Delta A}{\lambda}.\nonumber
\eea
 This relation  provides a special type of area law because for NGF the area $A$ scales like the volume $V=N$, i.e.   $A\propto V$.  Nevertheless,  we believe that this result opens new avenues for formulating the macroscopic description of NGF at the coarse-grained level, in the light of the results obtained in Refs.  \cite{Jacobson,Chirco_liberati,Chirco_Rovelli}. 
 
\subsection{Relation between the Regge curvature and the total energy $E$ of NGF with flavor $s=-1$}

We note here that  the NGF with flavor $s=-1$ are manifolds, specifically they are  the CQNM. For these manifolds,  one may wish to characterize their geometry  using Regge's definition of curvature  \cite{Regge,Rovelli,Dittrich}. The Regge curvature is  localized on $(d-2)$-faces and is given by the excess angle formed by the $d-$dimensional simplices incident to a given $(d-2)$-face.  Therefore in the case in which the $d-$dimensional simplices are assumed all equilateral the curvature $R_{\alpha}$ associated to the $(d-2)-$face $\alpha$ is  uniquely determined by the generalized degree $k_{d,d-2}(\alpha)$, i.e.  
\bea
R_{\alpha}=a_{\alpha}\pi-\theta_d\  k_{d,d-2}(\alpha)
\label{RG}
\eea
where $\theta_d>0$ indicates the angle between any two $(d-1)$-faces of the  $d$-dimensional simplex and where $a_{\alpha}=1$  \cite{Dittrich} for all $\alpha\in S_{d,d-2}(N)$ because for the NGF all $(d-2)$-faces are at the boundary.
 
The total energy $E$ of the NGF with flavor $s$ is defined in  Eq. $(\ref{E})$ 
as
\bea
E=\sum_{\alpha \in {\cal S}_{d, d-1}}\epsilon_{\alpha} n_{\alpha},
\eea
where $n_{\alpha}$ is related to the generalized degree of the $(d-1)$-face $\alpha$ by $n_{\alpha}=k_{d,d-1}(\alpha)-1$ (Eq. $(\ref{na})$), and where the energy of the face $\alpha$ is given by the sum of the energy of the nodes belonging to that face (Eq. $(\ref{ea})$).
We note now that it is   possible to show (see Supplementary Material for details), using simple combinatorial calculations, 
that 
\bea
 \sum_{\alpha \in {\cal S}_{d, d-1}}\epsilon_{\alpha} k_{d,d-1}({\alpha})=B_d \sum_{\alpha' \in {\cal S}_{d, d-2}} \epsilon_{\alpha'}k_{d,d-2}(\alpha'),
\eea
with $B_d={2}/{(d-1)}$.
Using this expression we can express the total energy $E$ and the total energy of the boundary $\hat{E}$ of the NGF in terms of  the Regge curvature $R_{\alpha}$ of the $(d-2)$-faces.
The total energy $E$ of the NGF can  then be written as
\bea
E=\frac{B_d}{\theta_d}\left(\Lambda-\sum_{\alpha' \in {\cal S}_{d, d-2}}\epsilon_{\alpha'}R_{\alpha'}\right)
\label{ER}
\eea
with $\Lambda$ being independent of the curvature and it can be shown to be given by 
\bea
\Lambda=\left(\pi-\frac{\theta_d}{2}\right) \sum_{\alpha' \in {\cal S}_{d, d-2}} \epsilon_{\alpha'}.
\eea
We note that the expression for ${E}$ in Eq.  $(\ref{ER})$ differs from the Regge action \cite{Regge,Rovelli,Dittrich} by an overall sign, and by the fact that Eq. $(\ref{Eh})$ contains the energy of the $(d-2)$-faces while in the Regge action their role is played  by the volume of the $(d-2)$-faces.
Additionally, it is possible to define the {\em total energy of the boundary } $\hat{E}$ of the NGF as given by the sum of the energies of the $(d-1)$-faces at the boundary, i.e. 
\bea
\hat{E}=\sum_{\alpha \in {\cal S}_{d, d-1}}\epsilon_{\alpha} (1-n_{\alpha}),
\eea
with $\hat{E}$ and $E$ being related by 
\bea
E+\hat{E}=\sum_{\alpha \in {\cal S}_{d, d-1}}\epsilon_{\alpha}.
\eea
The total energy $\hat{E}$ of the the boundary can be written as
\bea
\hat{E}=\frac{B_d}{\theta_d} \left(\sum_{\alpha' \in {\cal S}_{d, d-2}}\epsilon_{\alpha'}R_{\alpha'}-\hat{\Lambda}\right)
\label{Eh}
\eea
with $\hat{\Lambda}$ being independent of the curvature and given by 
\bea
\hat{\Lambda}=(\pi-\theta_d) \sum_{\alpha' \in {\cal S}_{d, d-2}} \epsilon_{\alpha'}.
\eea
We note that the expression for $\hat{E}$ in Eq.  $(\ref{Eh})$ differs from the Regge action \cite{Regge,Rovelli,Dittrich} by the fact that Eq. $(\ref{Eh})$ contains the energy of the $(d-2)$-faces while in the Regge action their role is played  by the volume of the $(d-2)$-faces.

\section{ The generalized degree distributions at  $\beta=0$}
\label{seccriticald}
\subsection{The dependence of the  generalized degree distribution on dimensions $d,\delta$ and  flavor $s$ }
The NGFs  display a number of  critical dimensions marking changes in the structure of these networks as their dimension $d$ changes. These structural changes are   revealed by the  statistical properties associated with the distribution of the generalized degree $k_{d,\delta}$ of their $\delta-$faces with  $0 \leq\delta<d$.  
To show this, here we focus  on the effect of the dimensions $d$ and $\delta$ on the distribution $P_{k,\delta}^{[s]}(k)$ of the generalized degrees $k_{d,\delta}$ of NGF of flavor $s$. For simplicity,  our study will focus first on the simpler case $\beta=0,$ where the energies of the nodes play no role in the NGF dynamics. 
Using the master equation approach  \cite{RMP,Doro_Book,Newman_Book} we  show that   depending on the dimensions $d$ and $\delta$, and on the flavor $s$,  the generalized degrees   $k_{d,\delta}$ can follow either   binomial or exponential or power-law distributions.
The power-law distributions are characterized by the asymptotic behavior for  large generalized degree $k_{d,\delta}=k\gg 1$ given by  
\bea
P_{k,\delta}^{[s]}(k)\simeq k^{-\gamma_{d,\delta}^{[s]}}. 
\label{powerlaw}
\eea
Our results on the generalized degree distribution of NGF of different flavor $s$, dimension $d$  and $\beta=0$ are summarized in Table $\ref{table1}$.

\begin{table}
\center
\label{table1}
\caption{\label{table1} Distribution of generalized degrees of faces of dimension $\delta$ in a $d$-dimensional NGF of flavor $s$ at $\beta=0$.  For $d\geq d_c^{[\delta,s]}=2(\delta+1)-s$ the power-law distributions are scale-free, i.e.  the second moment of the distribution diverges. }
\footnotesize
\begin{tabular}{@{}llll}
\hline
flavor &$s=-1$&$s=0$&$s=1$\\
\hline
$\delta=d-1$&Binomial&Exponential&Power-law\\
\hline
$\delta=d-2$&Exponential&Power-law& Power-law\\
\hline
$\delta\leq d-3$&Power-law&Power-law& Power-law\\
\hline
\end{tabular}\\
\end{table}

Additionally,  power-law distributions can be characterized either by a power-law exponent $\gamma_{d,\delta}^{[s]}>3$ or  $\gamma_{d,\delta}^{[s]}\leq 3$ indicating, in the second case, a divergent second moment $ \Avg{k_{d,\delta}^2}$ of the generalized degree distribution $P_{d,\delta}^{[s]}(k)$. 
The critical dimension $d_c^{[\delta,s]}$ is the smallest dimension $d$ of the NGF of flavor $s$ for which the generalized degree distribution $P_{d,\delta}^{[s]}(k)$ is scale-free. 
  For obtaining the exact asymptotic expression for the generalized degree distribution  $P_{d,\delta}^{[s]}(k)$ of generalized degree $k_{d,\delta}=k$ in NGF of flavor $s$ with  $s=-1,0,1$ we use the master equation approach  \cite{Doro_Book,RMP,Newman_Book} .
 Here we discuss in detail the results in the cases $s=-1,0,1$. For  details of the calculation we refer the reader to the  Supplementary Material  \cite{SM}. \\
 \begin{figure*}
\begin{center}
{\includegraphics[width=2.0\columnwidth]{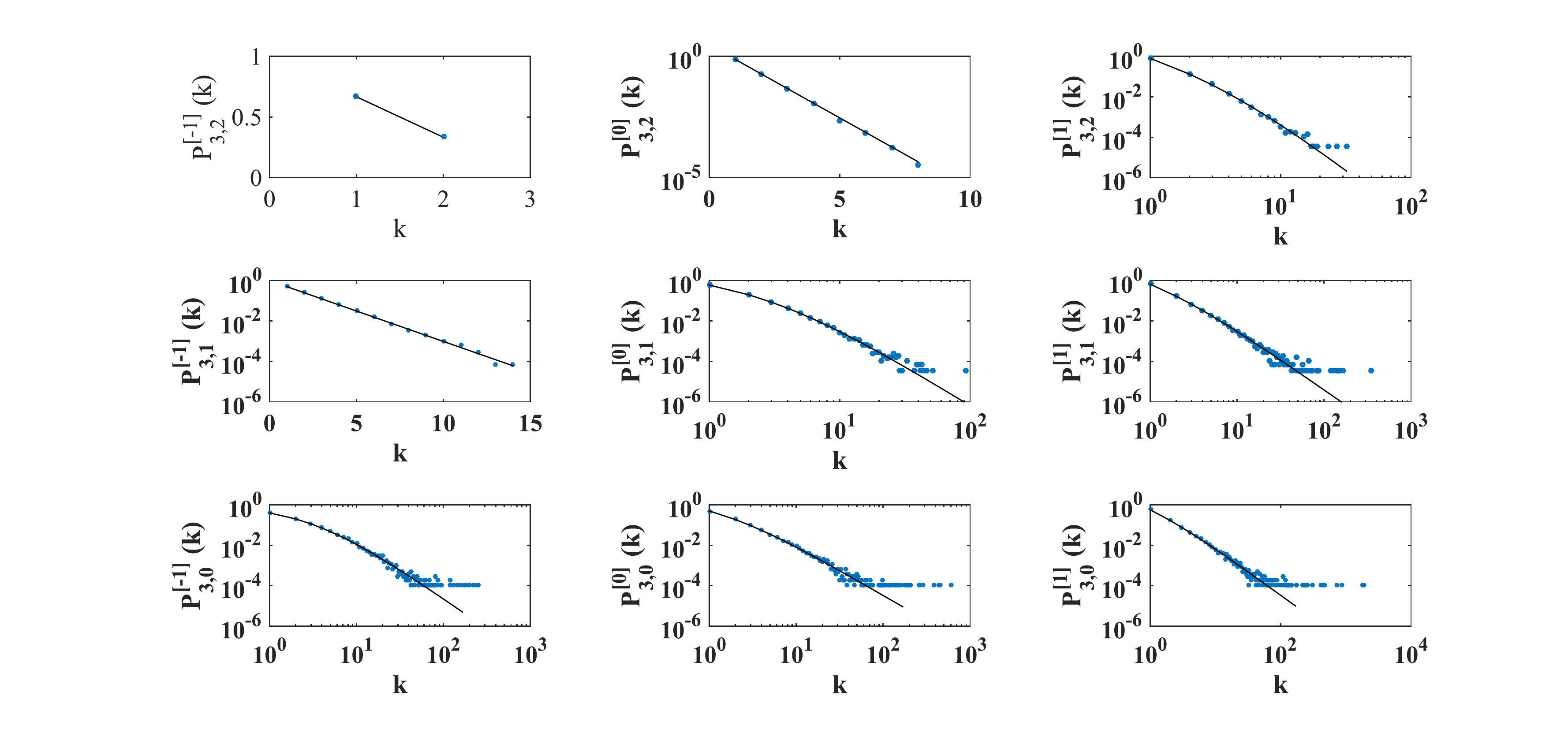}}
\end{center}
\caption{(Color online) The generalized degree distribution $P_{d,\delta}^{[s]}(k)$ of generalized degrees $k_{d,\delta}=k$ in NGF of dimension $d=3$ with value of flavor $s=-1,0,1$ and $\beta=0$. The simulation results   indicated with blue circles are shown for networks of $N=10^4$ nodes. These results perfectly match the theoretical predictions of Eqs. $(\ref{Pks-1}),$ $(\ref{Pks0})$ and $(\ref{Pks1})$ indicated here with solid black lines. }
\label{fig3}
\end{figure*}
\subsection{ Generalized degree distribution  $P_{d,\delta}^{[-1]}(k)$ for  $s=-1,\beta=0$ }
In the case $s=-1$ NGF generates manifolds  also called CQNM  \cite{CQNM}. 
At $\beta=0$ the  generalized degree follows a binomial distribution for faces of dimension $\delta=d-1$,  an exponential distribution for faces of dimension $\delta=d-2$, and a power-law distribution for faces of dimension $\delta\leq d-3$ (see Table $\ref{table1}$).  In particular, the distributions $P_{d,\delta}^{[-1]}(k)$ of generalized degrees $k_{d,\delta}$   are given by 
\bea
P_{d,d-1}^{[-1]}(k)&=&\left\{\begin{array}{lll}\frac{d-1}{d} &\mbox{for} & k=1 \\
\frac{1}{d} &\mbox{for} & k=2\end{array}\right. \ , \nonumber \\
P_{d,d-2}^{[-1]}(k)&=&\left(\frac{2}{d+1}\right)^{k}\frac{d-1}{2},\nonumber \\
\label{Pkex}
P_{d,\delta}^{[-1]}(k)&=&\frac{d-1}{d-\delta-2}\frac{\Gamma[1+(d+1)/(d-\delta-2)]}{\Gamma[1+2/(d-\delta-2)]}\nonumber \\
&&\hspace{-20mm}\times\frac{\Gamma[k+2/(d-\delta-2)]}{\Gamma[k+1+(d+1)/(d-\delta-2)]}, \ \mbox{for }    \delta\leq d-3. 
\label{Pks-1}
\eea
These distributions perfectly match the simulation results as shown in Figure $\ref{fig3}.$
For $\delta\leq d-3$ and  for large values of $k$, the distribution $P_{d,\delta}^{[-1]}(k)$   can be fitted by a power-law given by Eq. $(\ref{powerlaw})$ with power-law exponent $\gamma_{d,\delta}^{[-1]}$ given by 
\bea
\gamma_{d,\delta}^{[-1]}=1+\frac{d-1}{d-\delta-2}. 
\eea
This exponent is lower than 3, i.e.  $\gamma_{d,\delta}^{[-1]}\leq 3$ indicating a scale-free distribution of generalized degrees above the critical dimension, i.e.  for  $d\geq d_{c}^{[\delta,-1]}$ where 
\bea
d_c^{[\delta,-1]}=2\delta+3. 
\label{duno}
\eea

Therefore for   NGF with flavor $s=-1$ and $\beta=0$ the generalized degree of faces of dimension $\delta=d-2$ follows an exponential distribution. This result  implies that in this case   the Regge curvature $R$  given by Eq. $(\ref{RG})$ is following an exponential distribution, too.

\subsection{ Generalized degree distribution  $P_{d,\delta}^{[0]}(k)$ for  $s=0,\beta=0$}
In the case $s=0$, the  generalized degree of $(d-1)-$faces follows an exponential distribution, while the generalized degree of faces of dimension $\delta\leq d-2$ follows a power-law distribution (see Table $\ref{table1}$).  Specifically, the distribution $P_{d,\delta}^{[0]}(k)$ of generalized degree $k_{d,\delta}$ is given by
\bea
&&P_{d,d-1}^{[0]}(k)=\left(\frac{1}{d+1}\right)^{k}d \nonumber \\
&&P_{d,\delta}^{[0]}(k)=\frac{d}{d-\delta-1}\frac{\Gamma[1+(d+1)/(d-\delta-1)]}{\Gamma[1+1/(d-\delta-1)]} \nonumber \\
&&\hspace{-2mm}\times\frac{\Gamma[k+1/(d-\delta-1)]}{\Gamma[k+1+(d+1)/(d-\delta-1)]},\  \mbox{for} \    \delta\leq d-2.  
\label{Pks0}
\eea
These distributions perfectly match the simulation results as shown in Figure $\ref{fig3}.$
For $\delta\leq d-2$ and   for large values of $k$ the distribution $P_{d,\delta}^{[0]}(k)$ can be fitted by a power-law given by Eq. $(\ref{powerlaw})$ with power-law exponent $\gamma_{d,\delta}^{[0]}$ given by 
\bea
\gamma_{d,\delta}^{[0]}=1+\frac{d}{d-\delta-1}. 
\eea
This exponent is lower than 3, i.e.  $\gamma_{d,\delta}^{[0]}\leq 3$ indicating a scale-free distribution of generalized degrees above the critical dimension, i.e. for  $d\geq d_{c}^{[\delta,0]}$ where 
\bea
d_c^{[\delta,0]}=2\delta+2. 
\label{ddue}
\eea
\subsection{ Generalized degree distribution  $P_{d,\delta}^{[1]}(k)$ for  $s=1,\beta=0$}
In the case $s=1$ the generalized degree distribution $P_{d,\delta}^{[-1]}$ is power-law (see Table $\ref{table1}$) for any dimension $\delta\leq d-1$ and is given by 
\bea
P_{d,\delta}^{[1]}(k)&=&\frac{d+1}{d-\delta}{\Gamma[1+(d+1)/(d-\delta)]} \nonumber \\
&&\times \frac{\Gamma[k]}{\Gamma[k+1+(d+1)/(d-\delta)]}. \label{Pks1}
\eea
These distributions perfectly match the simulation results as shown in Figure $\ref{fig3}.$
For any $\delta\leq d-1$   for large values of $k$ the distribution $P_{d,\delta}^{[1]}(k)$   can be fitted by a power-law given by Eq. $(\ref{powerlaw})$ with power-law exponent $\gamma_{d,\delta}^{[1]}$ given by 
\bea
\gamma_{d,\delta}^{[1]}=1+\frac{d+1}{d-\delta}. 
\eea
This exponent is lower than 3, i.e.  $\gamma_{d,\delta}^{[1]}\leq 3$ indicating a scale-free distribution of generalized degrees above the critical dimension, i.e. for  $d\geq d_{c}^{[\delta,1]}$ where 
\bea
d_c^{[\delta,1]}=2\delta+1. 
\label{dtre}
\eea
\subsection{ The critical dimensions $d_c^{[\delta,s]}$ }

Summarizing the results of the previous paragraphs,  NGFs of flavor $s$ follow a regular pattern, with the flavor $s$ having the effect of shifting the statistical properties of generalized degree $k_{d,\delta}$ as indicated in  Table~\ref{table1}. 
The critical dimension for having a scale-free distribution of generalized degree for faces of dimension $\delta$ in NGF of dimension $d$  at $\beta=0$ is given by 
\bea
d_c^{[\delta,s]}=2(\delta+1)-s,
\eea 
which is a simple expression which summarizes the Eqs.  $(\ref{duno})-(\ref{ddue})-(\ref{dtre})$. \\
Therefore  the generalized degree $k_{d,\delta}$ of NGF of flavor $s$ is scale-free for every dimension $d$ of the NGF satisfying 
 \bea
d \geq d_c^{[\delta,s]}=2(\delta+1)-s. 
\eea

Since in NGF the generalized degree of node $\alpha$,  $k_{d,0}(\alpha)$,  is related to its degree   $K(\alpha)$ by the simple relation
\bea
K(\alpha)=k_{d,0}(\alpha)+d-1,
\eea
the critical dimension $d_c^{[0,s]}$  indicates also the smallest dimension $d$  of the NGF for which the NGF  has a scale-free degree distribution.  
Therefore the  NGFs at $\beta=0$ are scale-free networks as long as the dimension $d$ is greater than the critical dimension $d_c^{[0,s]}$, i.e. 
\bea
d \geq d_c^{[0,s]}=2-s. 
\eea
Therefore for $s=-1$ NGF at $\beta=0$ are scale-free for $d\geq d_c^{[0,-1]}=3, $ while for $s=0$ they are scale-free for any $d\geq d_c^{[0,0]}=2$, and for $s=1$ they are scale-free for any dimension $d\geq d_c^{[0,-1]}=1$. 

This interesting result implies that an explicit preferential attachment rule is not necessary for generating scale-free NGF in dimension $d>1$.  In fact both NGF with flavor $s=0$ and $s=-1$ do not have an explicit preferential attachment rule, but they can generate scale-free networks respectively for $d\geq 2$ and $d\geq 3$.  
This apparent contradiction with the results obtained by the seminal Barab\'asi-Albert model   \cite{BA} is solved by observing that  NGFs of dimension $d>1$ and flavor $s\neq -1$ that are scale-free, although they do not evolve according to  an explicit preferential attachment rule,    follow an effective preferential attachment rule emergent from their dynamics (see Supplementary Material  \cite{SM} for details ). 

\section{Quantum network states}
To each  NGF of flavor $s$, evolved up to time $t$, we can associate a quantum network state $\ket{\psi^{[s]}(t)}$ belonging to the Hilbert space ${\cal H}_{tot}^{[s]}$ by  following a similar procedure as the one used in precedent works  \cite{graphity_rg,graphity1,graphity2,PRE,CQNM}.  An Hilbert space ${\cal H}_{tot}^{[s]}$ is associated to a simplicial complex of $N$  nodes formed by gluing together $d$-dimensional simplices along $(d-1)$-faces. 
The Hilbert space ${\cal H}_{tot}^{[s]}$ is the tensorial product of the Hilbert spaces ${\cal H}_{node}$ associated to the nodes of the NGF and of two Hilbert spaces ${\cal H}_{d,d-1}$ and $\tilde{\cal H}_{d,d-1}^{[s]}$ associated to each of the possible $(d-1)-$faces of the NGF, i.e.  
\bea
{\cal H}_{tot}^{[s]}=\bigotimes^N {\cal H}_{node} \bigotimes^{P}{\cal H}_{d,d-1} \bigotimes^{P} \tilde{{\cal H}}_{d,d-1}^{[s]},
\eea
with $P=\left(\begin{array}{c}N \nonumber \\
d\end{array}\right)$ indicating  the maximum number of $(d-1)$-faces in a network of   $N$ nodes. 
The  Hilbert space ${\cal H}_{node}$  is  the one of   a fermionic oscillator of energy $\epsilon_i$, with basis 
$\{\ket{o_i,\epsilon}\}$, with $o_i=0,1$.   We indicate with  $b_i^{\dag}(\epsilon),b_i(\epsilon)$ respectively the fermionic creation and annihilation operators acting on this space. 
The Hilbert space  ${\cal H}_{d,d-1}$   associated to a $(d-1)$-face  $\alpha$ is the Hilbert space of a  fermionic oscillator with basis
 $\{\ket{a_{\alpha}}\}$, with $a_{\alpha}=0,1$.   We indicate with  $c_{\alpha}^{\dag},c_{\alpha}$ respectively the fermionic creation and annihilation operators acting on this space. 
 Finally the  Hilbert space  $\tilde{\cal H}_{d,d-1}^{[s]}$   associated to a $(d-1)$-face $\alpha$ has a different definition depending on the {\em flavor} $s$ of the NGF. 
 For $s=-1$,  $\tilde{\cal H}_{d,d-1}^{[-1]}$ is the Hilbert space of a  fermionic oscillator with basis
 $\{\ket{n_{\alpha}}\}$, with $n_{\alpha}=0,1$.  
 For $s=1$,  $\tilde{\cal H}_{d,d-1}^{[1]}$ is the Hilbert space of a  bosonic oscillator with basis
 $\{\ket{n_{\alpha}}\}$, with $n_{\alpha}=0,1,2,3,. . . $. 
  For $s=0$,  $\tilde{\cal H}_{d,d-1}^{[0]}$ is the Hilbert space   with basis
 $\{\ket{n_{\alpha}}\}$, with $n_{\alpha}=0,1,2,3,. . . $. 
 For $s=1,$ and $s=-1$ we indicate with  $h_{\alpha}^{\dag,[s]},h_{\alpha}^{[s]}$ the fermionic/bosonic  creation and annihilation operators acting respectively on the space   $\tilde{\cal H}_{d,d-1}^{[-1]}$ and   $\tilde{\cal H}_{d,d-1}^{[1]}$.  
 For $s=0$  we indicate with $h_{\alpha}^{\dag,[0]},h_{\alpha}^{[0]}$ the operators with commutation relations
 \bea
 [h_{\alpha}^{\dag,[0]},h_{\alpha}^{[0]}]=b
 \eea
 with the operator $b$ having elements 
 \bea
 b_{mn}=\bra{m}b\ket{n}=\delta_{m,n}\delta_{m,0},
 \eea
 such that 
 \bea
 h^{\dag,[0]}\ket{n}=\ket{n+1}
 \eea
 and 
 \bea
 \begin{array}{cclcc}
 h^{[0]}\ket{n}&=&\ket{n-1} & \mbox{for}& n>0\nonumber \\
 h^{[0]}\ket{0}&=&0&&.
 \end{array}
 \eea
 \\
 Having introduced the Hilbert space ${\cal H}_{tot}^{[s]}$, we can decompose any quantum network state $\ket{\phi}\in {\cal H}_{tot}^{[s]}$  as
\bea
\ket{\phi}&=&\sum_{\{o_i,\epsilon_i,a_{\alpha},n_{\alpha}\}} C(\{o_i,\epsilon_i,a_{\alpha},n_{\alpha}\})\nonumber \\
&&\times \bigotimes_{i=1}^N \ket{o_i,\epsilon_i} \bigotimes_{\alpha\in{\cal Q}_{d,d-1}(N)}\left(\ket{a_{\alpha}} \otimes \ket{n_{\alpha}}\right),
\eea
where with ${\cal Q}_{d,d-1}(N)$ we indicate all the possible $(d-1)$-faces of a  network of $N$ nodes. \\

 The node states  $\ket{o_i,\epsilon}$ are  mapped respectively to the presence  ($\ket{o_i=1,\epsilon}$) or the absence ($\ket{o_i=0,\epsilon}$) of a node $i$ of energy $\epsilon_i=\epsilon$ in the simplicial complex. 
 The state $\ket{a_{\alpha}=1}$ is mapped to the presence of the $(d-1)-$face  $\alpha\in {\cal S}_{d,d-1}$ in the network while the quantum state $\ket{a_{\alpha}=0}$ is mapped 
 to the absence of such a face.  
 Moreover, when $a_{\alpha}=1$, 
 the quantum number  $n_{\alpha}$ is mapped to the generalized degree of the  face $\alpha$  minus one $k_{d,d-1}(\alpha)-1$. 
Note that for $s=-1$ the Hilbert space  $\tilde{\cal H}_{d,d-1}^{[s]}$ is the one of a fermionic oscillator therefore allowing only $n_{\alpha}=0,1$ corresponding to generalized degrees $k_{d,d-1}(\alpha)=1,2$. \\
As already proposed  in the literature  \cite{CQNM,PRE,graphity_rg},
here we assume that the quantum network state follows a Markovian evolution.  In particular we assume that at time $t=1$ the state is given by 
\bea
\ket{\psi^{[s]}(1)}&=&\frac{1}{\sqrt{\hat{\cal Z}^{[s]}(1)}}\sum_{\{\epsilon_i\}_{i=1,. . d+1}}\prod_{i=1}^{d+1}\sqrt{g(\epsilon_i)}b_i^{\dag}(\epsilon_i)\nonumber \\
&&\times \prod_{\alpha\in{\cal Q}_{d,d-1}(d+1)}c^{\dag}_{\alpha}\ket{0},
\eea 
where ${\cal Z}^{[s]}(1)$ is fixed by the normalization condition $\Avg{\psi^{[s]}(1)|\psi^{[s]}(1)}=1$. 
The  quantum network state  is updated at each time $t>1$ according to the transition matrix $T_t^{[s]}$, i.e. 
\bea
\ket{\psi^{[s]}(t)}=T_t^{[s]}\ket{\psi(t-1)}
\label{markov}
\eea
with $T_t^{[s]}$ given by 
\bea
T_t^{[s]}&=&\sqrt{\frac{\hat{\cal Z}^{[s]}(t-1)}{\hat{\cal Z}^{[s]}(t)}}\sum_{\epsilon_{t+d}}\sqrt{g(\epsilon_{t+d})}b^{\dag}_{t+d}(\epsilon_{t+d}) \nonumber \\
&&\hspace{-10mm}\times \sum_{\alpha\in {\cal Q}_{d,d-1}(t+d-1)}e^{-\beta \epsilon_{\alpha}/2}\left[\prod_{\alpha' \in {\cal F}(t+d,\alpha)}c^{\dag}_{\alpha'}\right]h^{\dag,[s]}_{\alpha}c^{\dag}_{\alpha}c_{\alpha},\nonumber
\label{Udef}
\eea
where  ${\cal F}(i,\alpha)$ indicates the set of all the $(d-1)$-faces $\alpha'$ formed by the node $i$ and a subset of the nodes in $\alpha\in {\cal Q}_{d,d-1}(N)$,  ${\cal Z}^{[s]}(t)$ is fixed by the normalization condition
\bea
\langle \psi^{[s]}(t)|\psi^{[s]}(t)\rangle=1. 
\label{normdef}
\eea
The quantity $\hat{\cal Z}^{[s]}(t)$  is a path integral over NGF evolutions determined by the sequences $\{\epsilon_{i}\}_{i\leq t+d},\{\alpha_{t'}\}_{t'\leq t}$. 
In fact, using the normalization condition  in Eq.  $(\ref{normdef})$ and the evolution of the quantum network state given by Eqs.  $(\ref{markov})$, $(\ref{Udef})$ we get 
\bea
\hat{\cal Z}^{[s]}={\cal Z}^{[s]}
\eea
 where ${\cal Z}^{[s]}$ defined in Eq.  $(\ref{calZ})$ describes the temporal evolution of NGF, and therefore 
\bea
\hat{\cal Z}^{[s]}=\sum_{\{\alpha(t')\}_{t' \leq t}} e^{-\beta E}\prod_{\alpha\in Q_{d,d-1}(t)}(1+sn_{\alpha}(t))!
\eea
This implies that the set of all  classical evolutions of the CQNM fully determines the properties of the quantum network state evolving through the Markovian dynamics given by Eq.  $(\ref{markov})$. 

\section{ Quantum statistics  in Network Geometry with Flavor $s$}

\subsection{ Fermi-Dirac, Boltzmann and Bose-Einstein statistics describe the properties of the generalized degree of $\delta$-faces }
For $\beta>0$, as long as $\beta$ is sufficiently low, we can define self-consistently the chemical potentials $\mu_{d,\delta}^{[s]}$ and express the distributions $P_{k,\delta}^{[s]}(k)$ of the generalized degrees $k_{d,\delta}$ as convolution of binomial, exponential or power-law distributions  corresponding to the generalized degrees of  $\delta-$faces of energy $\epsilon$. These distributions  depend on the chemical potentials  $\mu_{d,\delta}^{[s]}$.
When we average the generalized degrees of $\delta-$faces of energy $\epsilon$  and  subtract one, i.e. we evaluate $\Avg{k_{d, \delta}-1|\epsilon,s}$, we observe that these quantities obey either the Fermi-Dirac, the Boltzmann or the Bose-Einstein statistics,  depending on the dimensions $d$ and $\delta$ and on the flavor $s$ of the NGF, where the Fermi-Dirac $n_F(\epsilon,\mu)$, the Boltzmann $n_B(\epsilon,\mu)$ and the Bose-Einstein statistics are given  \cite{Kardar} by the expressions
\bea
n_F(\epsilon,\mu)&=&\frac{1}{e^{\beta(\epsilon-\mu)}+1},\nonumber \\
n_Z(\epsilon,\mu)&=&e^{-\beta(\epsilon-\mu)},\nonumber \\
n_B(\epsilon,\mu)&=&\frac{1}{e^{\beta(\epsilon-\mu)}-1}. 
\label{ns}
\eea
The results are summarized  in Table $\ref{table2}$ and simulation results are compared with the theoretical expectations in Figure $\ref{figure4}.$ 

\begin{table}
\label{table2}
\center
\caption{\label{table2} The average  ${\Avg{k_{d, \delta}-1|\epsilon,s}}$  of the generalized degrees $k_{d,\delta}$ of $\delta$-faces with  energy $\epsilon$ minus one in a $d$-dimensional NGF of flavor $s$ follows either the Fermi-Dirac, the Boltzmann or the Bose-Einstein statistics  depending on the values of the dimensions $d$ and $\delta$.  }
\footnotesize
\begin{tabular}{@{}llll}
\hline
flavor &$s=-1$&$s=0$&$s=1$\\
\hline
$\delta=d-1$&Fermi-Dirac &Boltzmann&Bose-Einstein\\
\hline
$\delta=d-2$&Boltzmann&Bose-Einstein& Bose-Einstein\\
\hline
$\delta\leq d-3$&Bose-Einstein&Bose-Einstein& Bose-Einstein\\
\hline
\end{tabular}
\end{table}
We note here that the average  of $k_{d,d-1}(\alpha)-1=n_{\alpha}$ obeys the Fermi-Dirac statistics for $s=-1$, the Boltzmann statistics for $s=0$ and the Bose-Einstein statistics for $s=1$. 
This is particularly surprising because it shows that the statistical properties of NGF are intertwined with the properties of quantum network states in which $n_{\alpha}$ is mapped to a quantum number which is fermionic in the case $s=-1$ and bosonic in the case $s=1$.   Therefore, statistically,  on the NGF $n_{\alpha}$  follows the Fermi-Dirac statistics for $s=-1$ and the Bose-Einstein statistics for $s=1$ even if the NGF does not follow quantum equilibrium statistical mechanics. 
In order to show this result, let us give the  results of the master-equation approach for the generalized degree distribution $P_{d,\delta}^{[s]}(k)$ for $\beta>0$ (for the details of the derivation see the Supplementary Material  \cite{SM}). 
We will distinguish the cases in which the flavor $s$ takes value $s=-1,0,1$. 

\subsection{ Generalized degree distribution  $P_{d,\delta}^{[-1]}(k)$ for  $s=-1,\beta>0$}
\begin{figure*}
\begin{center}
\hspace*{-20mm}{\includegraphics[width=2.5\columnwidth]{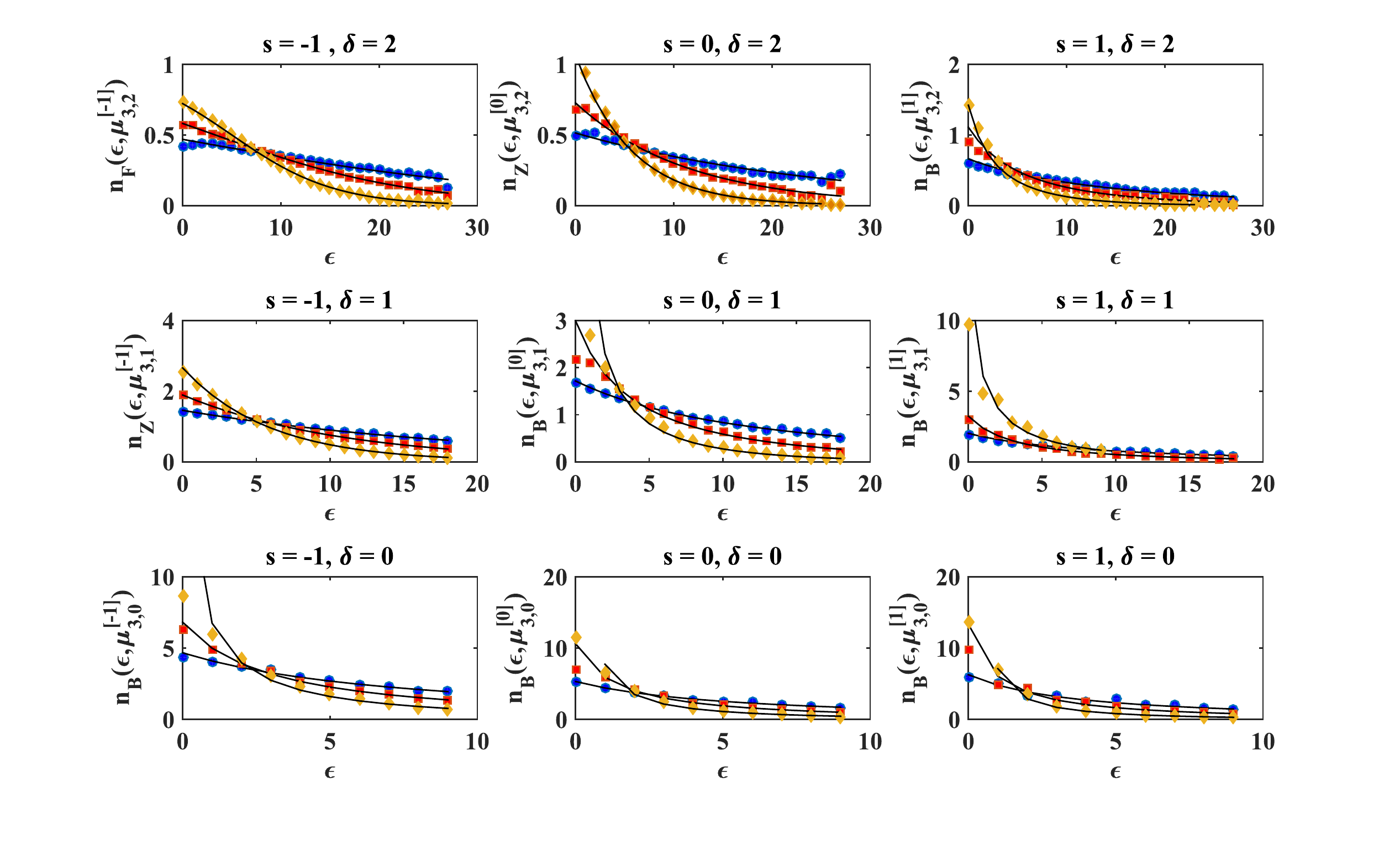}}
\end{center}
\caption{(Color online)The average $\Avg{k_{d,\delta}-1|\epsilon, s}$ for NGF of flavor $s$ in dimension $d=3$ follows either the Fermi-Dirac statistics $n_F(\epsilon,\mu)$, the Boltzmann statistics $n_Z(\epsilon,\mu)$ or the Bose-Einstein statistics $n_B(\epsilon,\mu)$ given by Eqs. $(\ref{ns})$ depending on the value of $\delta$ and $s$ as predicted by Eqs. $(\ref{nB-1})$, $(\ref{nB0})$ and $(\ref{nB1})$. Here the simulation results for NGF of dimension $d=3$ formed by $N=10^4$ nodes for $\beta=0.05,0.1,0.2$ (indicated respectively by blue circles, red squares and orange diamonds)  averaged $20$ times are compared with the theoretical expectations (indicated with solid black lines). The energies of the nodes take integer values and their uniform distribution is given by $g(\epsilon)=1/10$ for $0\leq \epsilon<10$.}
\label{figure4}
\end{figure*}

As long as the NGF is not a chain, i.e. $d>1$, and as long as we consider  sufficiently low values of the inverse temperature $\beta$, we can define a set of self-consistent quantities that we call {\em the chemical potentials } $\mu_{d,\delta}^{[-1]}$.
The generalized degrees $k_{d,\delta}=k$ of NGF with $d>1$ follow the distribution $P_{d,\delta}^{[-1]}(k)$ that depends on the chemical potential $\mu_{d,\delta}^{[-1]}$, and is given by a binomial distribution defined only for $k=1,2$ (for $\delta=d-1$),  by a convolution of exponentials (for $\delta=d-2$), or by a convolution of power-law distributions (for $\delta\leq d-3$)  \cite{CQNM}. In fact the exact asymptotic expression of the distribution  $P_{d,\delta}^{[-1]}(k)$ of the generalized degree $k_{d,\delta}=k$ obtained with the master equation approach is given by 
\bea
&&\hspace*{-3mm}P_{d,d-1}^{[-1]}(1)=
\sum_{\epsilon}\rho_{d, d-1}^{[-1]}(\epsilon)\left(1-\frac{1}{\exp{\left[\beta\left(\epsilon-\mu_{d, d-1}^{[-1]}\right)\right]}+1}\right),  \nonumber \\
&&P_{d,d-1}^{[-1]}(2)=\sum_{\epsilon}\rho_{d, d-1}^{[-1]}(\epsilon)\frac{1}{\exp{\left[\beta\left(\epsilon-\mu_{d, d-1}^{[-1]}\right)\right]}+1},  \nonumber \\
&&P_{d,d-2}^{[-1]}(k)=\sum_{\epsilon}\rho_{d, d-2}^{[-1]}(\epsilon)\frac{\exp\left[\beta\left(\epsilon-\mu_{d, d-2}^{[-1]}\right)\right]}{\left(\exp{\left[\beta\left(\epsilon-\mu_{d, d-2}^{[-1]}\right)\right]}+1\right)^{k}},\nonumber \\
&&P_{d,\delta}^{[-1]}(k)=\sum_{\epsilon}\rho_{d, \delta}^{[-1]}(\epsilon)\nonumber \\
&&\times\frac{ \exp{\left[\beta\left(\epsilon-\mu_{d, \delta}^{[-1]}\right)\right]}\Gamma\left[k+2/(d-\delta-2)\right]}{\Gamma\left[k+1+2/(d-\delta-2)+\exp{\left[\beta\left(\epsilon-\mu_{d, \delta}^{[-1]}\right)\right]}\right]} \nonumber \\
&&\times \frac{\Gamma\left[1+2/(d-\delta-2)+\exp{\left[\beta\left(\epsilon-\mu_{d, \delta}^{[-1]}\right)\right]}\right]}{\Gamma\left[1+2/(d-\delta-2)\right]},  \nonumber
\label{Pksb-1}
\eea
where  $\rho_{d, \delta}^{[-1]}(\epsilon)$ indicates the probability that a $\delta$-face has energy $\epsilon$, the dimension $d$ is greater than one, i.e. $d>1$, and  the last expression is valid for values of $\delta$ satisfying $0\leq \delta\leq d-3$.
The average of the generalized degree minus one, performed over $\delta-$faces of energy $\epsilon$ in dimension $d>1$, is given by the Fermi-Dirac statistics for $\delta=d-1$, the Boltzmann statistics for $\delta=d-2$ and the Bose-Einstein statistics for $\delta\leq d-3$  \cite{CQNM}
\bea
{\Avg{k_{d, d-1}-1|\epsilon,s=-1}}&=&n_F\left(\epsilon, \mu_{d, d-1}^{[-1]}\right),   \\
{\Avg{k_{d, d-2}-1|\epsilon,s=-1}}&=&n_Z\left(\epsilon, \mu_{d, d-2}^{[-1]}\right),    \nonumber\\
{\Avg{k_{d, \delta}-1|\epsilon,s=-1}}&=&A_{d,\delta}^{[-1]} n_B\left(\epsilon, \mu_{d, \delta}^{[-1]}\right),   \nonumber
\label{nB-1}
\eea
where the last expression is valid for $\delta\leq d-3,$  and where    $n_F(\epsilon,\mu),n_Z(\epsilon,\mu)$ and $n_B(\epsilon,\mu)$ are given by Eqs. $(\ref{ns})$, while $A_{d,\delta}^{[-1]}$ is given by 
\bea
A_{d,\delta}^{[-1]}=\frac{(d-\delta)}{(d-\delta-2)}. 
\eea
These relations perfectly match the simulation results for sufficiently low value of the inverse temperature $\beta$ (see Figure $\ref{figure4}$).
The self-consistent value of the chemical potential can be found by imposing the following geometrical relations satisfied by the generalized degrees of the NGF of every flavor $s$,
\bea
\lim_{t\to \infty}\frac{\sum_{\alpha\in {\cal S}_{d,\delta}(t)}k_{d,\delta}(\alpha)}{N_{d,\delta}(t)}=\frac{d+1}{\delta+1}.
\label{cond}
\eea
Imposing such condition is equivalent to fixing the normalization conditions for $n_F\left(\epsilon,\mu_{d,d-1}^{[-1]}\right),n_Z\left(\epsilon,\mu_{d,d-2}^{[-1]}\right),$ and $n_B\left(\epsilon,\mu_{d,\delta}^{[-1]}\right)$. These conditions are given by 
\bea
\sum_{\epsilon} \rho_{d,d-1}(\epsilon)n_F\left(\epsilon,\mu_{d,d-1}^{[-1]}\right)&=&\frac{1}{d},\nonumber \\
\sum_{\epsilon} \rho_{d,d-2}(\epsilon)n_Z\left(\epsilon,\mu_{d,d-2}^{[-1]}\right)&=&\frac{2}{d-1},\nonumber \\
\sum_{\epsilon} \rho_{d,\delta}(\epsilon)n_B\left(\epsilon,\mu_{d,\delta}^{[-1]}\right)&=&\frac{d-\delta-2}{\delta+1}.
\eea
The case $d=1$ is an exception because it is the only case in which the area $A$ of the NGF is not growing in time, in fact we have $A=2$ for every value of $t$. This property of the NGF of flavor $s=-1$ in dimension $d=1$ makes this case significantly different from the other cases, but fortunately this NGF has a much  simpler dynamics, since it is a  chain. 

\subsection{ Generalized degree distribution  $P_{d,\delta}^{[0]}(k)$ for  $s=0,\beta=0$ }
For NGF of flavor $s=0$, using the master equation approach together with the  self-consistent derivation, we can derive the distribution $P_{d,\delta}^{[0]}(k)$ of generalized degrees $k_{d,\delta}=k$. Therefore we define self-consistently the chemical potentials $\mu_{d,\delta}^{[0]}$, and express the distribution  $P_{d,\delta}^{[0]}(k)$ as a convolution of exponentials or a convolution of power-law distributions depending on the dimension $d$ and $\delta$. These distributions are given by
\bea
&&P_{d,d-1}^{[0]}(k)=\sum_{\epsilon}\rho_{d, d-1}^{[0]}(\epsilon)\frac{e^{\beta\left(\epsilon-\mu_{d, d-2}^{[0]}\right)}}{\left(e^{\beta\left(\epsilon-\mu_{d, d-2}^{[0]}\right)}+1\right)^{k}},  \\
&&P_{d,\delta}^{[0]}(k)=\sum_{\epsilon}\rho_{d, \delta}^{[0]}(\epsilon)\nonumber\\
&&\times\frac{ \exp{\left[\beta(\epsilon-\mu_{d, \delta}^{[0]})\right]} \Gamma\left[k+1/(d-\delta-1)\right]}{\Gamma\left[k+1+1/(d-\delta-1)+\exp{\left[\beta\left(\epsilon-\mu_{d, \delta}^{[0]}\right)\right]}\right]}\nonumber \\
&& \times \frac{\Gamma\left[1+1/(d-\delta-1)+\exp{\left[\beta\left(\epsilon-\mu_{d, \delta}^{[0]}\right)\right]}\right]}{\Gamma\left[1+1/(d-\delta-1)\right]},  \nonumber
\label{Pksb0}
\eea
where  $\rho_{d, \delta}^{[0]}(\epsilon)$ indicates the probability that a $\delta$-face has energy $\epsilon$, and where the last equation is valid for values of $\delta$ satisfying $0\leq\delta \leq d-2$.
Therefore the $(d-1)-$faces have generalized degree distribution $P_{d,d-1}^{[0]}(k)$ that is given by a convolution of exponentials, while the $\delta-$faces with $\delta\leq d-2$ have a generalized degree distribution $P_{d,\delta}^{[0]}(k)$ that is given by a convolution of power-laws.
When considering the average ${\Avg{k_{d, \delta}-1|\epsilon,s=0}}$, we observe that for  $\delta=d-1$ this quantity  is a Boltzmann distribution and for every $\delta \leq d-2$ is a Bose-Einstein distribution, i.e.
\bea
{\Avg{k_{d, d-1}-1|\epsilon,s=0}}&=&n_Z\left(\epsilon, \mu_{d, d-2}^{[0]}\right),    \\
{\Avg{k_{d, \delta}-1|\epsilon,s=0}}&=&A_{d,\delta}^{[0]} n_B\left(\epsilon, \mu_{d, \delta}^{[0]}\right), \  \mbox{for} \ \  \delta\leq d-2.  	\nonumber
\label{nB0}
\eea
with $n_Z(\epsilon,\mu)$ and $n_B(\epsilon,\mu)$ given by Eqs. $(\ref{ns})$
 and $A_{d,\delta}^{[0]}$ given by 
 \bea
A_{d,\delta}^{[0]}=\frac{(d-\delta)}{(d-\delta-1)}. 
\eea
The chemical potential $\mu_{d,\delta}^{[0]}$ can  then be found imposing the condition in Eq. $(\ref{cond})$ that all NGF must satisfy. Therefore,   the self-consistent equations that the chemical potentials must satisfy are 
\bea
\sum_{\epsilon} \rho_{d,d-1}(\epsilon)n_Z\left(\epsilon,\mu_{d,d-1}^{[-1]}\right)&=&\frac{1}{d},\nonumber \\
\sum_{\epsilon} \rho_{d,\delta}(\epsilon)n_B\left(\epsilon,\mu_{d,\delta}^{[-1]}\right)&=&\frac{d-\delta-1}{\delta+1},  \  \mbox{for }    \delta\leq d-2.\nonumber \\
\eea

\subsection{ Generalized degree distribution  $P_{d,\delta}^{[1]}(k)$ for  $s=1,\beta=0$ }
The NGF of flavor $s=1$, at sufficiently low inverse temperature $\beta$, has the generalized degrees $k_{d,\delta}=k$ with distribution $P_{d,\delta}^{[1]}(k)$ dependent on the chemical potential $\mu_{d,\delta}^{[1]}$. The generalized degree distributions $P_{d,\delta}^{[1]}(k)$ can be found using the master equation approach, and they are given by 
\bea
P_{d,\delta}^{[1]}(k)&=&\sum_{\epsilon}\rho_{d, \delta}^{[1]}(\epsilon)\frac{ \exp{\left[\beta\left(\epsilon-\mu_{d, \delta}^{[1]}\right)\right]} \Gamma\left[k\right]}{\Gamma\left[k+1+\exp{\left[\beta\left(\epsilon-\mu_{d, \delta}^{[1]}\right)\right]}\right]}\nonumber \\
&&\times\Gamma\left[1+\exp{\left[\beta\left(\epsilon-\mu_{d, \delta}^{[1]}\right)\right]}\right],
\label{Pksb1}
\eea
where $\rho_{d, \delta}^{[1]}(\epsilon)$ indicates the probability that a $\delta$-face has energy $\epsilon$.
In this case, if we perform the average ${\Avg{k_{d, \delta}-1|\epsilon,s=1}}$ over all $\delta-$faces with energy $\epsilon,$ we always get the Bose-Einstein distribution, independently  of $0\leq \delta< d$, i.e. we obtain
\bea
{\Avg{k_{d, \delta}-1|\epsilon,s=1}}&=& n_B\left(\epsilon, \mu_{d, \delta}^{[1]}\right), 
\label{nB1}
\eea
with $n_B(\epsilon,\mu)$ given by Eq. $(\ref{ns})$.
The chemical potentials $\mu^{[1]}_{d,\delta}$ must satisfy Eq. $(\ref{cond})$. Therefore they can be found self-consistently by solving 
\bea
\sum_{\epsilon} \rho_{d,\delta}(\epsilon)n_B\left(\epsilon,\mu_{d,\delta}^{[1]}\right)=\frac{d-\delta}{\delta+1}.
\eea

\subsection{ The low temperature regime}
In the regime of low temperatures, i.e. high enough values of $\beta$, it is possible to observe a breakdown of the self-consistent hypothesis made for solving the generalized degree distribution and the self-consistent equations might not have a solution. In the NGF of $d=1$ and flavor $s=1$ there is a well-defined phase transition in which one node grabs a finite fraction of all the links. This phase transition is also called Bose-Einstein condensation in complex networks and has been characterized in Ref.  \cite{Bose}. In general NGF of higher dimensions and also different flavors might show  phase transitions modifying the generalized degree distribution of different $\delta$-faces as shown for the case $d=2$ and flavors  $s=-1$ and $s=1$ in Ref.  \cite{PRE}.
A full investigation of the nature of the possible phase transitions occurring in NGF is beyond the scope of this paper.
\section{Conclusions}
In conclusion here we have presented the model of Network Geometry with Flavor $s$. This is a model for growing simplicial complexes in dimension $d$.
Simplicial complexes are very useful generalizations of networks and can be used to model interactions involving more than just two nodes, as the one occurring for example  in collaboration networks, or in protein-interaction networks. Moreover simplicial complexes of dimension $d$ are useful structures to discretize a geometrical $d$-dimensional space, and for this reason they are widely used in quantum gravity.\\
Network Geometry with flavor $s$ evolves by a non-equilibrium dynamics that enforces an indefinite growth of these geometrical structures. Moreover these networks are formed by simplices having heterogeneous properties modeled by assigning  an energy to them that determines their evolution.
The statistical mechanics of the NGF allows to characterize the thermodynamic properties of these networks and to relate these networks to complexity theory on the one side and to quantum geometry on the other side.\\
The thermodynamic properties of NGF reveal that these networks  obey the area law and the change in their entropy $S$ depends on the change of their area $A$.
From the point of view of network theory we observe that characterizing NGF of dimensionality $d>1$  allows for a significant generalization of previous results, showing that an explicit preferential attachment is not necessary for obtaining scale-free networks in the case of NGF of $d>1$.
Finally the significant interplay between  the NGF and their quantum mechanical description in terms of quantum network states is revealed by the statistical properties of the generalized degrees of $\delta$-faces, whose average follows either the Fermi-Dirac, the Boltzmann or the Bose-Einstein statistics depending on the dimensions $d,\delta$ and on the flavor $s$.\\
Overall we have proposed the  theoretical framework of NGF for describing the non-equilibrium dynamics of  simplicial complexes. Our framework  generates a large variety of network geometries, from chains and higher dimensional manifolds to scale-free networks with communities and small-world properties. 
Interestingly, NGF with flavor $s=-1,0,1$ displays a   strikingly regular pattern in their structural properties.
We believe that these results extend our understanding of growing complex networks to simplicial complexes of larger dimensionality and can be used in network theory to model network-like structures where nodes are connected by interactions involving more than two nodes. Finally we hope that this work, showing the rich interplay between NGF and their quantum mechanical description,  will stimulate the  cross-fertilization between network theory and quantum gravity.

\clearpage
\renewcommand\theequation{{S-\arabic{equation}}}
\renewcommand\thetable{{S-\Roman{table}}}
\renewcommand\thefigure{{S-\arabic{figure}}}
\setcounter{equation}{0}
\setcounter{figure}{0}
\setcounter{section}{0}

\onecolumngrid

\section*{\Large SUPPLEMENTARY INFORMATION }

\section*{INTRODUCTION}
 Network Geometries with Flavor $s=-1,0,1$ (NGFs) are simplicial complexes of dimension $d$  formed by gluing $d$-simplices along $(d-1)$-faces.
The NGFs evolve according to a non-equilibrium dynamics that enforces the simplicial complex to grow continuously by the subsequent addition  of $d$-dimensional simplices. 
In this Supplementary Material we will first provide some useful definition of important  properties of the NGFs (Sec. $\ref{2}$),  then we will define the dynamical evolution of NGF (Sec. \ref{3}). In the subsequent sections we will provide details of the analytic results reported in the main body of the paper and provide the codes for the simulation of NGF in dimension $d=1,2,3$.
In particular in Sec. \ref{3b} we will provide the details of the Eq. (21) and Eq. (28) used to derive the thermodynamic properties of NGF in the main text  and  
in Sec. $\ref{4}$ we discuss the generalized degree distribution for $\beta=0$ and $\beta>0$, and  in Sec. $\ref{6}$ we will provide the codes for generating NGF of flavor $s=-1,0,1$ in dimensions $d=1,2,3$.

\section*{GENERALIZED DEGREE AND ENERGY OF THE $\delta-$FACES} 
\label{2}

Here we provide some useful definitions of important structural properties of the $\delta$-faces of the NGFs.
Let us indicate  with ${\cal Q}_{d,\delta}(N)$ the set of all possible $\delta$-dimensional faces (also called $\delta-$faces) with $\delta\leq d$ of a  $d$-dimensional NGF formed by $N$ nodes. Moreover we will  indicate  with ${\cal S}_{d,\delta}$ the set of all $\delta$-dimensional faces (also called $\delta-$faces) with $\delta<d$ belonging to the $d$-dimensional NGF of $N$ nodes.
The {\it generalized degrees} $k_{d,\delta}(\alpha)$ of the $\delta$-face $\alpha\in {\cal S}_{d,\delta}$  in a $d$-dimensional NGF is the number of $d-$dimensional simplices incident to it.
Given  the adjacency tensor ${\bf a}$ with the generic element  $a_{\alpha'}$ with $\alpha'\in{\cal Q}_{d,d}(N)$ taking the values 
\bea
a_{\alpha'}=\left\{\begin{array}{ll}1& \mbox{if }  \alpha'\in {\cal Q}_{d,d}(N)\\ 0 & \mbox{otherwise} 
\end{array}\right.,
\eea  the generalized degree  of a $\delta$-face $\alpha$ is given by 
\bea
k_{d,\delta}(\alpha)=\sum_{\alpha' | \alpha \subset \alpha'}a_{\alpha'}.
\label{kdde2}
\eea
For example, in a NGF of dimension $d=2$, the generalized degree $k_{2,1}(\alpha)$ is the number of triangles incident to a link $\alpha$ while the generalized degree $k_{2,0}(\alpha)$ indicates the number of triangles incident to a node $\alpha$.
Similarly in a NGF of dimension $d=3$, the generalized degrees $k_{3,2}$, $k_{3,1}$ and $k_{3,0}$ indicate  the number of tetrahedra incident respectively to a triangular face, a link or a node.
A useful quantity to associate to each $(d-1)$-face $\alpha\in {\cal S}_{d,d-1}$ is $n_{\alpha}$ given by  
\bea
n_{\alpha}=k_{d,d-1}(\alpha)-1,
\eea
 indicating the generalized degree of the face minus one, i.e. how many $d$-dimensional simplices have been glued to the $(d-1)-$face $\alpha$ during the NGF evolution.
Moreover, to each node $i$ we assign an {\it energy }  $\epsilon_i$ drawn  from a distribution $g(\epsilon)$ and quenched  during the evolution of the network.
To every  $\delta$-face  $\alpha\in {\cal S}_{d,\delta}$ we associate an {\it energy} $\epsilon_{\alpha}$ given by the sum of the energy of the nodes that belong to $\alpha$, 
\bea
\epsilon_{\alpha}=\sum_{i \subset \alpha}\epsilon_i\ .
\label{e2}
\eea
The energies of $\delta-$faces characterize their heterogeneous properties, which are not captured by the generalized degree.
Here we will  always consider the case in which the energies of the nodes take only integer values, although the extension to models having real energy values is straightforward.

\section*{EVOLUTION OF THE NGF}
\label{3}
The NGFs evolve by a non-equilibrium dynamics depending on the energy of their $\delta-$faces which  enforces an indefinite growth of the NGF.
Here we give the algorithm determining the NGF evolution.\\

At time $t=1$ the NGF of dimension $d$ and flavor  $s=-1,0,1$ is formed by a single $d$-dimensional simplex.\\
At  each time $t>1$ we add a simplex of dimension $d$ to a $(d-1)$-face $\alpha\in{\cal S}_{d,d-1}$ chosen with probability $\Pi_{\alpha}^{[s]}$ given by  
\bea
\Pi_{\alpha}^{[s]}&=&\frac{1}{Z^{[s]}}e^{-\beta \epsilon_{\alpha}}(1+s n_{\alpha}),
\label{P1}
\eea
where $\beta\geq 0$ is a parameter of the model called {\it inverse temperature} and $Z^{[s]}$ is a normalization sum given by 
\bea
Z^{[s]}=\sum_{\alpha\in {\cal S}_{d,d-1}}e^{-\beta \epsilon_{\alpha}}(1+sn_{\alpha}),
\eea
and $s$ is the flavor of the NGF.
Having chosen the $(d-1)$-face $\alpha$, we glue to it a new $d$-dimensional complex containing  all the nodes of the face $\alpha$  plus the new node $i$. It follows that the new node $i$ is 
linked to each node $j$ belonging to $\alpha$.\\

The NGFs of flavor $s=-1$ are Complex Quantum Network Manifolds introduced in \cite{CQNM}  having generalized degrees of  their $(d-1)-$faces taking only values $k_{d,d-1}=k=1,2.$
The NGF of flavor $s=1$ includes an explicit preferential attachment rule since $\Pi_{\alpha}^{[1]}\propto k_{d,d-1}(\alpha)$, i.e. for each $(d-1)-$face the probability to attract the new $d-$dimensional simplex is proportional to the number of $d-$dimensional simplices already incident to it, i.e. to its generalized degree $k_{d,d-1}(\alpha)=1+n_{\alpha}$.
Therefore the NGF with flavor $s=1$ and dimension $d=1$ is the Bianconi-Barab\'asi model \cite{Fitness,Bose} and for $\beta=0$ the Barab\'asi-Albert model \cite{BA}.
The case $s=0$ and $d=2$ has been first proposed as a scale-free network in \cite{Doro_link}. The NGF in $d=2$ is also related to the recent papers \cite{Emergent,PRE}.

Since at time $t=1$ the number of nodes in the NGF is $N(1)=d+1$, and at each time we add a new additional node, the total number of nodes is $N(t)=t+d$.
The NGF evolution up to time $t$ is fully determined  by the sequences $\{\epsilon_{i}\}_{i\leq t+d},\{\alpha_{t'}\}_{t'\leq t}$, where $\epsilon_{t'+d}$ indicates the energy of the 
node added to the NGF at time $t'>1$, $\epsilon_i$ with $i\leq d+1$ indicates the energy of an initial node $i$ of the NGF, and   $\alpha_{t'}$ indicates  the $(d-1)$-face to which the new $d$-dimensional complex is added at time $t'$.

\section*{THERMODYNAMIC PROPERTIES OF THE NGFs} 
\label{3b}
The thermodynamic properties of the NGF are widely discussed in the main text where it is shown that the NGF follows a generalized area law.
Here we provide additional details of the derivation of two equations used in the main text, Eq. $(21)$ and Eq. $(28)$.

\subsection{Derivation of Eq. (21) of the main text}
Here we want to provide the detailed derivation  of Eq. (21) of the main text.
Given the quantity 
\bea
z_N^{[s]}=\frac{1}{N!}L^{[s]}(T)
\label{treb2}
\eea
 where $L^{[s]}(T)$ indicates the number of different NGF temporal evolutions giving rise to the same network $G_N$, we want to show (Eq. (21) of the main text),  that 
 \bea
z_N^{[s]}=C^{[s]}e^{\beta \nu^{[s]} N},
\label{zns2}
\eea
where $C^{[s]}$ is a subleading factor, and $\nu^{[s]}$ depends on the degree distribution of the dual of the NGF, and therefore depends on its flavor $s$. 
 It can easily be realized that $L^{[s]}(T)$ indicates  the number of different labelings of the tree $T$ that constitutes the dual network of the NGF. 
The introduced quantity $z_N^{[s]}$ can be calculated by  following the derivation given in Ref. \cite{Burda}, as long as the NGF is in a stationary state.
In fact it is possible to evaluate the scaling of $z_N^{[s]}$ by writing a recursive equation for $L^{[s]}(T)$  where  the tree $T$ is  given by a  root node connected to $p$ subtrees $T_1,T_2\ldots T_p$ formed respectively by  $N_1,N_2\ldots N_p$ nodes. The recursive equation is given by 
\bea
L^{[s]}(T)=\frac{(N-1)!}{N_1!N_2! \ldots N_p!}\delta_{\sum_i N_i+1,N}\prod_{i=1}^pL^{[s]}(T_i).
\label{lst12}
\eea
Here, differently from the case analyzed in \cite{Burda}, the different branches $T_1,T_2,\ldots, T_p$ of the tree $T$ are not exchangeable since the tree $T$ is a dual of a labelled NGF where the labels indicate the different energies of the nodes. 
In oder to use the recursive Eq. $(\ref{lst12})$, we consider the generating function $G(\hat{\mu})$ given by  
\bea
G^{[s]}(\hat{\mu})=\sum_{N}z_N^{[s]}e^{-\hat{\mu}N}.
\eea 
The recursive Eq. $(\ref{lst12})$ can be written in terms of the generating function as 
\bea
\frac{dG^{[s]}(\hat{\mu})}{d\hat{\mu}}=-e^{-\hat{\mu}}\sum_{p=1}^{\infty}\pi^{[s]}(p+1)[G^{[s]}(\hat{\mu})]^{p}=-e^{-\hat{\mu}}F^{[s]}[G^{[s]}(\hat{\mu})],
\eea
where $\pi^{[s]}(p)$ is the degree distribution of the dual of the NGF of flavor $s$.
Integrating this differential equation we obtain
\bea
e^{-\hat{\mu}(G^{[s]})}=H^{[s]}(G^{[s]})=C+\int_{G_0}^{G^{[s]}} \frac{dz}{F^{[s]}(z)},
\eea
where $C$ and $G_0$ are constants.
For any NGF different from the chain (flavor $s=-1$ and $d=1$), as long as we consider sufficiently low values of $\beta$, we obtain stationary NGF characterized by a stationary degree distribution $\pi^{[s]}(p)$ of the dual network that is not trivial (i.e. $\pi^{[s]}(p)$ is not equal to zero for all $p\neq 1,2$).
Under this very general conditions, the function $H^{[s]}(G^{[s]})$ is a positive monotonically increasing function of $G^{[s]}$, bounded from above. Hence $\hat{\mu}$ is bounded from below and $G^{[s]}(\hat{\mu})$ has a singularity at some $\hat{\mu}=\beta \nu^{[s]}$, with $\nu^{[s]}$ given by 
\bea
\beta \nu^{[s]}=-\ln H^{[s]}(\overline{x})
\eea
where $\overline{x}$ is the radius of convergence of $H^{[s]}(x)$.
The singularity of the generating function $G^{[s]}(\hat{\mu})$ for $\hat{\mu}=\beta \nu^{[s]}$ dictates the leading behavior of $z_N^{[s]}$, given by  Eq. $(\ref{zns2})$ for large values of $N$.

\subsection{Derivation of Eq. (28) of the main text}
Here our aim is to provide the detailed derivation of the combinatorial Eq. (28) of the main text, given by
\bea
 \sum_{\alpha \in {\cal S}_{d, d-1}}\epsilon_{\alpha} k_{d,d-1}({\alpha})=B_d \sum_{\alpha' \in {\cal S}_{d, d-2}} \epsilon_{\alpha'}k_{d,d-2}(\alpha'),
 \label{282}
\eea
with $B_d={2}/{(d-1)}$.
This equation can be easily derived using the definition of generalized degree $k_{d,\delta}(\alpha)$ and energy $\epsilon_{\alpha}$ for the generic $\delta$-face $\alpha$, given respectively by Eqs. ($\ref{kdde2}$) and ($\ref{e2}$).
Using Eqs. $(\ref{kdde2})$ and $(\ref{e2})$ and assuming that the $(d-1)$-face $\alpha$ includes the nodes $i_1,i_2,\ldots i_d$  while  $\alpha'=\{i_1,i_2,\ldots i_{d+1}\}$ is a $d$-dimensional simplex comprising $\alpha$, (i.e. such that $\alpha\subset \alpha'$), we have 
\bea
 \sum_{\alpha \in {\cal S}_{d, d-1}}\epsilon_{\alpha} k_{d,d-1}({\alpha})=\frac{1}{d!}\sum_{\{i_1,i_2,\ldots,i_d\}}\sum_{p=1}^{d}\epsilon_{i_p}\sum_{i_{d+1}}a_{i_1,i_2,\ldots i_{d+1}}
 \label{id}
\eea
where $\{i_1,i_2\ldots i_d\}$ indicates a generic ordered sequence of $d$ labels of the nodes.
The $d!$ in the right side of Eq. $(\ref{id})$ takes into account all the equivalent permutations of the $d$ label of the nodes in the sequence  $\{i_1,i_2\ldots i_d\}$.
The expression can be further simplified by noticing that 
\bea
\sum_{p=1}^{d}\epsilon_{i_p}=\frac{1}{d-1}\sum_{\alpha'\in {\cal D}_{d,d-2}(\alpha)}\epsilon_{\alpha'},
\label{ep2}
\eea
where $\epsilon_{\alpha'}$ is the energy of any of the $d$  possible  $(d-2)$-faces $\alpha'\subset \alpha$, (i.e.
$\alpha' \in {\cal D}_{d,d-2}(\alpha)$ where ${\cal D}_{d,d-2}(\alpha)$ indicates all the $(d-2)$-faces that are subsets of $\alpha$).
For example, for $d=3$ we have 
\bea
\epsilon_1+\epsilon_2+\epsilon_3=\frac{1}{2}\left(\epsilon_{(1,2)}+\epsilon_{(2,3)}+\epsilon_{(1,3)}\right)
\eea
where the energy $\epsilon_{(i,j)}$ of the link $\alpha=(i,j)$ is given by $\epsilon_{(i,j)}=\epsilon_i+\epsilon_j$.
Using  Eq. $(\ref{ep2})$ and changing the order of the sums in Eq. $(\ref{id})$ we get 
 \bea 
 \sum_{\alpha \in {\cal S}_{d, d-1}}\epsilon_{\alpha} k_{d,d-1}({\alpha})=\frac{1}{(d-1)!}\frac{1}{d-1}\sum_{\{i_1,i_2,\ldots,i_{d-1}\}}\epsilon_{(i_1,i_2,\ldots, i_{d-1})}\sum_{i_d,i_{d+1}}a_{i_1,i_2,\ldots i_{d+1}}
 \eea
 Given that 
 \bea
 k_{d,d-2}(\alpha')=k_{d,d-1}(i_1,i_2,\ldots i_{d-1})=\frac{1}{2}\sum_{i_d,i_{d+1}}a_{i_1,i_2,\ldots i_{d+1}},
 \eea
  we have
 \bea
 \sum_{\alpha \in {\cal S}_{d, d-1}}\epsilon_{\alpha} k_{d,d-1}({\alpha}) &=&\frac{2}{d-1}\frac{1}{(d-1)!}\sum_{\{i_1,i_2,\ldots,i_{d-1}\}}\epsilon_{(i_1,i_2,\ldots, i_{d-1})}k_{d,d-2}(i_1,i_2,\ldots, i_{d-1})\nonumber \\
 &=&\frac{2}{d-1}\sum_{\alpha' \in {\cal S}_{d, d-2}} \epsilon_{\alpha'}k_{d,d-2}(\alpha').
 \label{idb2}
 \eea
 This final expression is equivalent to Eq. $(\ref{282})$ (and  Eq. (28) in the main text).
 
\section*{DISTRIBUTION OF THE GENERALIZED DEGREES}
\label{4}
In this section we provide the details for deriving the generalized degree distribution for NGF of flavor $s=-1,0,1$ for  $\beta=0$ and $\beta>0$.
For deriving our results at $\beta=0$ we use the master equation approach \cite{Doro_Book,RMP,Newman_Book} that provides exact asymptotic solutions for the distribution. In the case $\beta>0$ we combine the master equation approach with the self-consistent approach introduced in the context of the Bianconi-Barab\'asi model \cite{Fitness,Bose} yielding exact results for the generalized degree distribution as long as the self-consistent hypothesis is satisfied, i.e. for low enough values of the inverse temperature $\beta$.
The NGFs with flavor $s=-1$ are the Complex Quantum Network  Manifolds introduced in Ref. \cite{CQNM}. Nevertheless, for completeness, we report here  the details of the derivation of the generalized degree distribution also for $s=-1$.
 
\subsection{Distribution of Generalized Degrees for $\beta=0$}

The  distribution $P_{d,\delta}^{[s]}(k)$ of the generalized degrees $k_{d,\delta}=k$ in NGF with flavor $s$ can be obtained for $\beta=0$ using the master-equation approach \cite{Doro_Book}. Here we give the details for the derivation of the results presented in the main paper, distinguishing between the three cases $s=-1,0,1$.
\subsubsection{Case $s=-1$}
For NGF with flavor $s=-1$ the generalized degree of $(d-1)-$faces can only take values $k_{d,d-1}=1,2$. We will call the $(d-1)-$faces with generalized degree $k_{d,d-1}(\alpha)=1$ unsaturated, and the $(d-1)-$ faces $\alpha$ with generalized degree $k_{d,d-1}(\alpha)=2$ saturated.
The evolution of the NGF with flavor $s=-1$ described in Sec. $\ref{3}$ allows  the new $d-$dimensional simplex to attach exclusively to unsaturated $(d-1)-$faces.  
In the case $d=1$ the NGF with flavor $s=-1$ in dimension $d$  are $d$-dimensional manifolds. In the case $d=1$ the NGF with flavor $s=-1$ is a chain.
Here we focus first on the case $d>1$ and at the end of the paragraph we will discuss the case $d=1$.\\
The explicit expression of $\Pi^{[-1]}_{\alpha}$ is easily derived.
In fact the  number of unsaturated $(d-1)$-faces in the NGF of flavor $s=-1$ evolved up to time $t$ is given by  $(d-1)t$ since  at each time we add $d$ unsaturated  $(d-1)-$faces belonging to the new $d$-dimensional simplex, while the face where the simplex is attached to changes from being unsaturated to being saturated.
Since any new $d$-dimensional simplex can be glued only to unsaturated $(d-1)-$faces, the probability $\Pi^{[-1]}_{\alpha}$ that a new $d$-dimensional simplex is attached to a $(d-1)$-face $\alpha$ is given by 
\bea
\Pi^{[-1]}_{\alpha}=\left\{\begin{array}{ccc}\frac{1}{(d-1)t}& \mbox{for} &k_{d,d-1}(\alpha)=1,\nonumber \\
0&\mbox{for}& k_{d,d-1}(\alpha)=2. \end{array}\right.
\label{pia0}
\eea 
Now we  observe that each $\delta$-face, with  $\delta<d-1$, which has generalized  degree $k_{d,\delta}(\alpha)=k$, is incident to 
\bea
2+(d-\delta-2)k
\label{2d}
\eea
  unsaturated $(d-1)$-faces. \\
In fact, is is easy to check that a $\delta-$face  with generalized degree  $k_{d,\delta}=1$   is incident to $d-\delta$ unsaturated $(d-1)$-faces. 
Moreover, at each time we add to a $\delta$-face a new $d$-dimensional simplex, a number $d-\delta-1$ of unsaturated $(d-1)-$faces are added to the $\delta$-face  while a previously unsaturated $(d-1)$-face incident to it  becomes saturated.
Therefore the number of unsaturated $(d-1)$-faces incident to a $\delta$-face of generalized degree $k_{d,\delta}=k$ follows 
Eq. $(\ref{2d})$.
This result allows us to evaluate the average number $\tilde{m}_{d,\delta}^{[-1]}(k)$ of $\delta$-faces of generalized degree $k_{d,\delta}=k$ that increase their generalized degree by one.
For  $\delta=d-1$ and large times $t\gg1$  it is given by 
\bea
\tilde{m}_{d, d-1}^{[-1]}(k)=\frac{1}{(d-1)t}\delta_{k, 1}
\label{pia011}
\eea 
where $\delta_{x, y}$ indicates the Kronecker delta. 
For  $\delta<d-1$, instead,  it is given by 
\bea
\tilde{m}_{d, \delta}^{[-1]}(k)=\frac{2+(d-\delta-2)k}{(d-1)t}.
\label{pia111}
\eea 
From Eq. $(\ref{pia111})$ it follows that, as long as $\delta<d-2$, the generalized degree follows an {\it effective preferential attachment} mechanism  \cite{BA,RMP,Newman_Book,Doro_Book}. In fact each $\delta$-face will be incident to an additional $d$-dimensional simplex with a probability that depends linearly on its generalized degree, indicating how many $d-$dimensional simplices are already incident to the $\delta-$face. Therefore, even if the evolution of the NGF with flavor $s=-1$ does not contain an explicit preferential attachment mechanism, this mechanism emerges from its dynamical rules. 

Using Eqs. $(\ref{pia011})-(\ref{pia111})$ and the master equation approach \cite{RMP,Doro_Book,Newman_Book} , it is possible to derive the exact distribution for the generalized degrees.
We indicate with $N_{d,\delta}^{t,[-1]}(k)$ the average  number of $\delta$-faces that at time $t$ have generalized degree $k_{d,\delta}=k$ during the temporal evolution of a $d$-dimensional CQNM. The master equation \cite{RMP,Doro_Book,Newman_Book}  for $N_{d,\delta}^t(k)$ reads
\bea
N^{t+1, [-1]}_{d,\delta}(k)-N^{t, [-1]}_{d,\delta}(k)=\tilde{m}_{d,\delta}^{[-1]}(k-1)N_{d,\delta}^{t,[-1]}(k-1)(1-\delta_{k,1})-\tilde{m}_{d,\delta}^{[-1]}(k)N^{t,[-1]}_{d,\delta}(k)+m_{d,\delta}\delta_{k,1}
\eea 
with $k\geq 1$.
Here  $m_{d,\delta}=\left(\begin{array}{c}d\nonumber \\ \delta\end{array}\right)$ is the number of  $\delta$-faces added at each time $t$ to the CQNM.
The master equation is solved by observing that  for large times $t\gg1 $ we have $N_{d,\delta}(k)\simeq m_{d,\delta}tP_{d,\delta}^{[-1]}(k)$ where $P_{d,\delta}^{[-1]}(k)$ is the generalized degree distribution.
For $\delta=d-1$ we obtain the  bimodal distribution
\bea
P_{d,d-1}^{[-1]}(k)=\left\{\begin{array}{lll}\frac{d-1}{d}, &\mbox{for} & k=1\nonumber \\
\frac{1}{d} &\mbox{for} & k=2\end{array}\right.\ .
\eea
For $0\leq \delta=d-2$ instead, we find an exponential distribution, i.e.
\bea
\begin{array}{ccccc}
P_{d,d-2}^{[-1]}(k)&=&\left(\frac{2}{d+1}\right)^{k}\frac{d-1}{2},& \mbox{for} & k\geq 1 \ .
\end{array}
\label{Pkex}
\eea
Finally for $0\leq\delta<d-2$ we have the distribution 
\bea
\begin{array}{ccccc}
P_{d,\delta}^{[-1]}(k)&=&\frac{d-1}{d-\delta-2}\frac{\Gamma[1+(d+1)/(d-\delta-2)]}{\Gamma[1+2/(d-\delta-2)]}\frac{\Gamma[k+2/(d-\delta-2)]}{\Gamma[k+1+(d+1)/(d-\delta-2)]},& \mbox{for} &  k\geq 1, 0\leq \delta\leq d-3.
\end{array}
\label{Pksfm1}
\eea
From Eq. $(\ref{Pksfm1})$ it follows that for  $0\leq\delta<d-2$ and $k\gg1$ the generalized degree distribution follows a power-law with exponent $\gamma_{d,\delta}^{[-1]}$, i.e.
\bea 
P_{d,\delta}^{[-1]}(k)\simeq Ck^{-\gamma_{d,\delta}^{[-1]}}& \mbox{for} &0\leq \delta\leq d-3,
\eea
and
\bea
\gamma_{d,\delta}^{[-1]}=1+\frac{d-1}{d-\delta-2}.
\label{gammaddm1}
\eea
Therefore the generalized degree  distribution $P_{d,\delta}^{[-1]}(k)$ given by Eq. $(\ref{Pksfm1})$ is scale-free, i.e. it has diverging second moment $\Avg{\left(k_{d,\delta}\right)^2}$, as long as  $\gamma_{d,\delta}^{[-1]}\in(2,3]$.
This implies that the  generalized degree distribution $P_{d,\delta}^{[-1]}(k)$ is scale free for NGF of dimension $d$ satisfying,
\bea
d\geq d_c^{[\delta,-1]}=2\delta+3.
\eea
\subsubsection{Case $s=0$}
In the case $s=0$ every new $d-$dimensional simplex can be attached to an arbitrary $(d-1)$-face of the NGF.  Therefore the generalized degree $k_{d,d-1}=k$ can take any value $k\geq 1$.
Since any new $d$-dimensional simplex can be glued  to any  $(d-1)-$face, the probability $\Pi^{[0]}_{\alpha}$ that a new $d$-dimensional simplex is attached to a $(d-1)$-face $\alpha$ is given by 
\bea
\Pi^{[0]}_{\alpha}=\frac{1}{d t},
\label{pia00}
\eea 
for $t\gg 1$.
In fact the number of $(d-1)$-faces at time $t$ is equal to $d t$, because each new $d$-dimensional simplex adds  a number $d$ of $(d-1)$-faces to the NGF.
Let us now observe that each $\delta$-face, which has generalized  degree $k_{d,\delta}(\alpha)=k$, is incident to 
\bea
1+(d-\delta-1)k
\label{2d0}
\eea
$(d-1)$-faces. \\
In fact, is is easy to check that a $\delta-$face  with generalized degree  $k_{d,\delta}=k=1$   is incident to $d-\delta$  $(d-1)$-faces.  Moreover, at each time  a new $d$-dimensional simplex is glued  to a $\delta$-face $\alpha$, adding a  number $d-\delta-1$ of  $(d-1)-$faces incident to it.
Therefore the number of $(d-1)$-faces incident to a $\delta$-face of generalized degree $k_{d,\delta}=k$ follows 
Eq. $(\ref{2d0})$.
This implies that the average number $\tilde{m}_{d,\delta}^{[0]}(k)$ of $\delta$-faces of generalized degree $k_{d,\delta}=k$ that increase their generalized degree by one  is given by 
\bea
\tilde{m}_{d, \delta}^{[0]}(k)=\frac{1+(d-\delta-1)k}{d\ t}.
\label{pia100}
\eea 
From Eq. $(\ref{pia100})$  follows that, as long as $\delta<d-1$, the generalized degree follows an {\it effective preferential attachment} mechanism  \cite{BA,RMP,Newman_Book,Doro_Book}. In fact each $\delta$-face will be incident to an additional $d$-dimensional simplex with a probability that  depends linearly on its generalized degree, indicating how many $d-$dimensional simplices are already incident to the $\delta-$face. Therefore, even if the evolution of the NGF with flavor $s=0$ does not contain an explicit preferential attachment mechanism, the preferential attachment emerges from its dynamical rules.

Using  the master equation approach \cite{RMP,Newman_Book,Doro_Book}, it is possible to derive the exact distribution for the generalized degrees.
We indicate with $N_{d,\delta}^{t,[0]}(k)$ the average  number of $\delta$-faces that at time $t$ have generalized degree $k_{d,\delta}=k$. The master equation \cite{RMP,Newman_Book,Doro_Book} for $N_{d,\delta}^{t,[0]}(k)$ reads
\bea
N^{t+1,[0]}_{d,\delta}(k)-N^{t,[0]}_{d,\delta}(k)=\tilde{m}_{d,\delta}^{[0]}(k-1)N_{d,\delta}^{t,[0]}(k-1)(1-\delta_{k,1})-\tilde{m}_{d,\delta}^{[0]}(k)N^{t,[0]}_{d,\delta}(k)+m_{d,\delta}\delta_{k,1}
\eea 
with $k\geq 1$.
Here  $m_{d,\delta}=\left(\begin{array}{c}d\nonumber \\ \delta\end{array}\right)$ is the number of  $\delta$-faces added at each time $t$ to the NGF.
The master equation is solved by observing that  for large times $t\gg1 $ we have $N_{d,\delta}(k)\simeq m_{d,\delta}tP_{d,\delta}(k)$ where $P_{d,\delta}(k)$ is the generalized degree distribution.
For $\delta=d-1$ we obtain the  exponential distribution
\bea
\begin{array}{ccccc}
P_{d,d-1}^{[0]}(k)&=&\left(\frac{1}{d+1}\right)^{k}d & \mbox{for} & k\geq 1 \ .
\end{array}
\label{Pkex0}
\eea
Instead, for $0\leq\delta<d-1$ we obtain  the distribution 
\bea
\begin{array}{ccccc}
P_{d,\delta}^{[0]}(k)&=&\frac{d}{d-\delta-1}\frac{\Gamma[1+(d+1)/(d-\delta-1)]}{\Gamma[1+1/(d-\delta-1)]}\frac{\Gamma[k+1/(d-\delta-1)]}{\Gamma[k+1+(d+1)/(d-\delta-1)]},& \mbox{for} &  k\geq 1, 0\leq \delta \leq d-2.
\end{array}
\label{Pksf0}
\eea
From Eq. $(\ref{Pksf0})$ it follows that for  $0\leq\delta<d-1$ and $k\gg1$ the generalized degree distribution follows a power-law with exponent $\gamma_{d,\delta}^{[0]}$, i.e.
\bea 
P_{d,\delta}^{[0]}(k)\simeq Ck^{-\gamma_{d,\delta}^{[0]}}& \mbox{for} &0\leq \delta \leq d-2,
\eea
and
\bea
\gamma_{d,\delta}^{[0]}=1+\frac{d}{d-\delta-1}.
\label{gammadd}
\eea
Therefore the generalized degree  distribution $P_{d,\delta}^{[0]}(k)$ given by Eq. $(\ref{Pksf0})$ is scale-free, i.e. it has diverging second moment $\Avg{\left(k_{d,\delta}\right)^2}$, as long as  $\gamma_{d,\delta}^{[0]}\in(2,3]$.
This implies that the  generalized degree distribution $P_{d,\delta}^{[0]}(k)$ is scale-free for NGF of dimension $d$ satisfying
\bea
d\geq d_c^{[\delta,0]}=2\delta+2.
\eea

\subsubsection{Case $s=1$}
In NGF with flavor $s=1$ and $\beta=0$ each  $(d-1)-$face  $\alpha\in{\cal S}_{d,d-1}$ has a probability to be selected  proportional to  its generalized degree $k_{d,d-1}(\alpha)$.
In fact the probability $\Pi_{\alpha}^{[1]}$ defined in  Eq. $(\ref{P1})$ includes an explicit preferential attachment mechanism \cite{BA,RMP, Newman_Book,Doro_Book} as is given by 
\bea
\Pi_{\alpha}^{[1]}=\frac{k_{d,d-1}(\alpha)}{Z^{[1]}}.
\eea
For $t\gg1$ the sum of all generalized degrees is given by 
\bea
Z^{[1]}=\sum_{\alpha \in{\cal S}_{d,d-1}}k_{d,d-1}({\alpha})= (d+1)t,
\eea
 since each  $d-$dimensional simplex augments  the generalized degree of its $(d+1)$ $(d-1)-$faces by one.
Therefore we have 
\bea
\Pi_{\alpha}^{[1]}=\frac{k_{d,d-1}(\alpha)}{(d+1)t}.
\eea
Here we want to show that the average number $\tilde{m}_{d,\delta}^{[1]}(k)$ of $\delta-$faces with generalized degree $k_{d,\delta}=k$ that increase their generalized degree by one at a generic time $t>1$  is given by 
\bea
\tilde{m}_{d,\delta}^{[1]}(k)=\frac{(d-\delta)k}{(d+1)t}.
\label{p1b}
\eea
In fact the probability $\pi_{\alpha}$ to attach a new $d$-dimensional simplex to a  $\delta$-face $\alpha$ with $0\leq \delta\leq d-1$ is given by 
\bea
\pi_{\alpha}=\sum_{\alpha'\in{\cal S}_{d,d-1}|\alpha\subset \alpha'}\Pi_{\alpha'}^{[1]}=\sum_{\alpha'\in{\cal S}_{d,d-1}|\alpha\subset \alpha'}\frac{k_{d,d-1}(\alpha')}{(d+1)t}.
\eea
Now we observe that every $d$-dimensional simplex attached to the $\delta$-face $\alpha$ increases the generalized degree of all the $(d-1)-$faces $\alpha'$ incident to it  by one. The number of the  $(d-1)$-faces $\alpha'$ incident to the $\delta$-face $\alpha$ is given by $\left(\begin{array}{c} d-\delta\\ d-\delta-1\end{array}\right)=(d-\delta)$.
Therefore 
\bea
\pi_{\alpha}=\frac{(d-\delta) k_{d,\delta}(\alpha) }{(d+1)t}.
\eea
Therefore t $\tilde{m}_{d,\delta}^{[1]}(k)$   is given by Eq. $(\ref{p1b})$  at a generic time $t>1$ .
Using  the master equation approach \cite{RMP,Doro_Book,Newman_Book} , it is possible to derive the exact distribution for the generalized degrees.
We indicate with $N_{d,\delta}^{t, [1]}(k)$ the average  number of $\delta$-faces that at time $t$ have generalized degree $k_{d,\delta}=k$. The master equation\cite{RMP,Doro_Book,Newman_Book}  for $N_{d,\delta}^{t,[1]}(k)$ reads
\bea
N^{t+1, {[1]}}_{d,\delta}(k)-N^{t, [1]}_{d,\delta}(k)=\tilde{m}_{d,\delta}^{[1]}(k-1)N_{d,\delta}^{t, [1]}(k-1)(1-\delta_{k,1})-\tilde{m}_{d,\delta}^{[1]}(k)N^{t,[1]}_{d,\delta}(k)+m_{d,\delta}\delta_{k,1}
\eea 
with $k\geq 1$.
Here  $m_{d,\delta}=\left(\begin{array}{c}d\nonumber \\ \delta\end{array}\right)$ is the number of  $\delta$-faces added at each time $t$ to the NGF.
The master equation is solved by observing that  for large times $t\gg1 $ we have $N_{d,\delta}^{t,[1]}(k)\simeq m_{d,\delta}tP_{d,\delta}^{[1]}(k)$ where $P_{d,\delta}^{[1]}(k)$ is the generalized degree distribution.
We find 
\bea
\begin{array}{ccccc}
P_{d,\delta}^{[1]}(k)&=&\frac{d+1}{d-\delta}{\Gamma[1+(d+1)/(d-\delta)]}\frac{\Gamma[k]}{\Gamma[k+1+(d+1)/(d-\delta)]},& \mbox{for} &  k\geq 1.
\end{array}
\label{Pksf1}
\eea
From Eq. $(\ref{Pksf1})$ it follows that for   $k\gg1$ the generalized degree distribution follows a power-law with exponent $\gamma_{d,\delta}$, i.e.
\bea 
P_{d,\delta}^{[1]}(k)\simeq Ck^{-\gamma_{d,\delta}^{[1]}},
\eea
and
\bea
\gamma_{d,\delta}^{[1]}=1+\frac{d+1}{d-\delta}.
\label{gammadd}
\eea
Therefore the generalized degree  distribution $P_{d,\delta}^{[1]}(k)$ given by Eq. $(\ref{Pksf1})$ is scale-free, i.e. it has diverging second moment $\Avg{\left(k_{d,\delta}\right)^2}$, as long as  $\gamma_{d,\delta}^{[1]}\in(2,3]$.
This implies that the  generalized degree distribution is scale free for 
\bea
d\geq d_c^{[\delta,1]}={2 \delta+1}.
\eea

\subsection{Distribution of Generalized Degrees for $\beta>0$}

For $\beta>0$, as long as $\beta$ is sufficiently low, we can define self-consistently the chemical potentials $\mu_{d,\delta}^{[s]}$ and derive, using the master equation approach \cite{RMP,Doro_Book,Newman_Book}, the distributions $P_{k,\delta}^{[s]}(k)$ of the generalized degrees $k_{d,\delta}$ as convolution of binomials, exponentials or power-law distributions  corresponding to the generalized degrees of  $\delta-$faces of energy $\epsilon$. These distributions will depend on the chemical potentials  $\mu_{d,\delta}^{[s]}$.
When we average the generalized degrees of $\delta-$faces of energy $\epsilon$,  and we remove one, i.e. we evaluate $\Avg{k_{d, \delta}-1|\epsilon,s}$, we observe that these quantities obey either the Fermi-Dirac, the Boltzmann or the Bose-Einstein statistics,  depending on the dimensions $d$ and $\delta$ and on the flavor $s$ of the NGF. The Fermi-Dirac distribution $n_F(\epsilon,\mu)$, the Boltzmann distribution $n_Z(\epsilon,\mu)$ and the Bose-Einstein distribution $n_B(\epsilon,\mu)$ are given \cite{Kardar} by 
\bea
n_F(\epsilon,\mu)&=&\frac{1}{e^{\beta(\epsilon-\mu)}+1},\nonumber \\
n_Z(\epsilon,\mu)&=&e^{-\beta(\epsilon-\mu)},\nonumber \\
n_B(\epsilon,\mu)&=&\frac{1}{e^{\beta(\epsilon-\mu)}-1}. 
\label{ns}
\eea
In the following we will  consider the cases $s=-1,0,1$ separately.
\subsubsection{Case $s=-1$}   
Let us derive the distribution $P_{d,\delta}^{[-1]}(k)$ of the generalized degrees $k_{d,\delta}$ for every $0\leq \delta\leq d-1$ in a NGF with flavor $s=-1$ and dimension $d$.\\
The average number $N_{d,d-1}^{t,[-1]}(k|\epsilon)$ of $(d-1)$-faces of energy $\epsilon$ that at  time $t$ have  generalized degrees $k_{d,\delta}=k$ in a NGF with flavor $s=-1$ follows the master equation given by 
\bea
{ N_{d,d-1}^{t+1,[-1]}(k=2|\epsilon)}-N_{d,d-1}^{t, [-1]}(k=2|\epsilon)&=&\frac{e^{-\beta \epsilon}}{Z^{[-1]}}N_{d,d-1}^{t, [-1]}({k=1}|\epsilon),\nonumber \\
&&\nonumber \\
{N_{d,d-1}^{t+1,[-1]}(k=1|\epsilon)}-N_{d,d-1}^{t, [-1]}(k=1|\epsilon)&=&-\frac{e^{-\beta \epsilon}}{Z^{[-1]}}N_{d,d-1}^{t, [-1]}(k=1|\epsilon)+m_{d,d-1}\rho_{d,d-1}^{[-1]}(\epsilon)
\label{Nm1n}
\eea
where  $\rho_{d,d-1}^{[-1]}(\epsilon)$ is the probability that a    $(d-1)$-face  added to the network at a generic time $t\gg1$ has energy $\epsilon$ and $m_{d,d-1}=\left(\begin{array}{c} d\nonumber \\ d-1 \end{array}\right)=d$ is the number of $\delta$-faces added to the network at each time $t$.
In order to solve this master equation we assume that the normalization constant $Z^{[-1]}\propto t$  has a finite limit, and we put
\bea
e^{-\beta \mu_{d,d-1}^{[-1]}}&=&\lim_{t\to \infty}\frac{Z^{[-1]}}{t}.
\label{selfF1}
\eea
For large times $t\gg1$, using the asymptotic expression of $Z^{[-1]}\simeq e^{-\beta \mu_{d,d-1}^{[-1]}}t$,   we can rewrite  Eqs. $(\ref{Nm1n})$ as
\bea
{ N_{d,d-1}^{t+1,[-1]}(k=2|\epsilon)}-N_{d,d-1}^{t, [-1]}(k=2|\epsilon)&=&\frac{e^{-\beta \left(\epsilon-\mu_{d,d-1}^{[-1]}\right)}}{t}N_{d,d-1}^{t, [-1]}({k=1}|\epsilon),\nonumber \\
&&\nonumber \\
{N_{d,d-1}^{t+1,[-1]}(k=1|\epsilon)}-N_{d,d-1}^{t, [-1]}(k=1|\epsilon)&=&-\frac{e^{-\beta \left(\epsilon-\mu_{d,d-1}^{[-1]}\right)}}{t}N_{d,d-1}^{t, [-1]}(k=1|\epsilon)+m_{d,d-1}\rho_{d,d-1}^{[-1]}(\epsilon).
\label{Nm1}
\eea
Imposing that for large times $t\gg1$, $N_{d,d-1}^{t,[-1]}(k|\epsilon)\simeq m_{d,d-1}t\rho_{d,d-1}(\epsilon) P_{d,d-1}^{[-1]}(k|\epsilon)$, where $P_{d,d-1}^{[-1]}(k|\epsilon)$ is the distribution of generalized degrees $k_{d,d-1}=k$ of nodes with energy $\epsilon$, we can solve  Eqs. $(\ref{Nm1})$ for $P_{d,d-1}^{[-1]}(k|\epsilon)$, obtaining
\bea
P_{d,d-1}^{[-1]}(k|\epsilon)&=&\left\{\begin{array}{lcc}
\left(1-\frac{1}{\exp{\left[\beta\left(\epsilon-\mu_{d, d-1}^{[-1]}\right)\right]}+1}\right),  &\mbox{for }& k=1,\nonumber \\
\\
\frac{1}{\exp{\left[\beta\left(\epsilon-\mu_{d, d-1}^{[-1]}\right)\right]}+1}, &\mbox{for }& k=2.\end{array}\right.
\label{Nsm1}
\eea
Using a similar approach one can write the master equation for the average number $N_{d,d-2}^{t,[-1]}(k|\epsilon)$ of $(d-2)$-faces of energy $\epsilon$ that at  time $t$ have  generalized degrees $k_{d,d-1}=k$ in a NGF with flavor $s=-1$,  as 
\bea
\hspace*{-0mm}{N_{d,d-2}^{t+1,[-1]}(k|\epsilon)}-N_{d,d-2}^{t,[-1]}(k | \epsilon)&=&\frac{e^{-\beta (\epsilon-{\mu}_{d,d-2}^{[-1]})}}{t}N_{d,d-2}^{t,[-1]}(k-1|\epsilon)[1-\delta_{k,1}]-\frac{e^{-\beta (\epsilon-{\mu}_{d,d-2}^{[-1]})}}{t}N_{d,d-2}^{t,[-1]}(k|\epsilon)\nonumber \\
&&+m_{d,d-2}\rho_{d,d-2}^{[-1]}(\epsilon)\delta_{k,1},
\label{N2F}
\eea
where $m_{d,d-2}=d(d-1)/2$ is the number of $(d-2)-$faces added at each time $t$ to the NGF, $\rho_{d,d-2}^{[-1]}(\epsilon)$ is the probability that such faces have energy $\epsilon$,   and the chemical potential $\mu_{d,d-2}^{[-1]}$ is defined self-consistently by 
\bea
\hspace*{-10mm}e^{\beta {\mu}_{d,d-2}^{[-1]}}=e^{\beta \mu_{d,d-1}^{[-1]}}\lim_{t \to \infty} \Avg{\frac{\sum_{\alpha \in{\cal Q}_{d,d-2}(t)}\sum_{\alpha'\in {\cal Q}_{d,d-1}(t)| \alpha\subset \alpha'}e^{-\beta (\epsilon_{\alpha'}-\epsilon_{\alpha})}\delta(k_{d,d-1}({\alpha'}),1)\delta(k_{d,d-2}(\alpha),k)}{\sum_{\alpha \in{\cal Q}_{d,d-2}(t)}\delta(k_{d,d-2}(\alpha),k)}}_{k}.
\label{selfF2}
\eea
In Eq. $(\ref{selfF2})$ we indicate by $\Avg{\ldots}_k$ the average over different values of $k$. Assuming that $N_{d,d-2}^{t, [-1]}(k|\epsilon)\simeq t m_{d,d-2}\rho_{d,d-2}^{[-1]}(\epsilon) P_{d,d-2}^{[-1]}(k|\epsilon)$ for $t\gg1$,  we can solve Eqs. $(\ref{N2F})$  finding that the distribution $P_{d,d-2}^{[-1]}(k|\epsilon)$ that $\delta$-faces of energy $\epsilon$ have generalized degree $k_{d,\delta}=k$ is given by
\bea
P_{d,d-2}^{[-1]}(k|\epsilon)&=&\frac{\exp\left[\beta\left(\epsilon-\mu_{d, d-2}^{[-1]}\right)\right]}{\left(\exp{\left[\beta\left(\epsilon-\mu_{d, d-2}^{[-1]}\right)\right]}+1\right)^{k}}.
\eea
Finally we can derive the expression for the distribution  $P_{d,d-2}^{[-1]}(k|\epsilon)$ of generalized degrees $k_{d,\delta}$  for $\delta$-faces with energy $\epsilon$ and  $0\leq \delta \leq d-3$.
The chemical potentials $\mu_{d,\delta}^{[-1]}$ for  $0\leq \delta \leq d-3$ are defined self-consistently as
\bea
\hspace*{-10mm}
e^{\beta {\mu}_{d,\delta}^{[-1]}}=e^{\beta \mu_{d,d-1}^{[-1]}} \lim_{t \to \infty} \Avg{\frac{\sum_{\alpha \in{\cal Q}_{d,\delta}(t)}\sum_{\alpha'\in {\cal Q}_{d,d-1}(t)| \alpha\subset \alpha'}e^{-\beta (\epsilon_{\alpha'}-\epsilon_{\alpha})}\delta(k_{d,d-1}({\alpha'}),1)\delta(k_{d,\delta}(\alpha),k)}{\sum_{\alpha \in{\cal Q}_{d,\delta}(t)}\left[k+2/(d-2-\delta)\right]\delta(k_{d,\delta}(\alpha),k)}}_{k}.\label{selfdelta}
\eea
Assuming that the chemical potential $\mu_{d,\delta}^{[-1]}$ exists and is finite, the master equations \cite{RMP,Doro_Book,Newman_Book}  for the average number $N^t_{d,\delta}(k|\epsilon)$ of $\delta$-faces with energy $\epsilon$ and generalized degree $k\geq 1$ read  
\bea
{N_{d,\delta}^{t+1,[-1]}(k|\epsilon)}-N_{d,\delta}^{t ,[-]}(k | \epsilon)&=&\frac{e^{-\beta (\epsilon-{\mu}_{d,\delta})}[k-1+2/(d-\delta-2)]}{t}N_{d,\delta}^{t, [-1]}(k-1|\epsilon)[1-\delta_{k,1}]\nonumber \\
&&-\frac{e^{-\beta (\epsilon-{\mu}_{d,\delta})}[k+2/(d-\delta-2)]}{t}N_{d,\delta}^{t, [-1]}(k|\epsilon)+m_{d,\delta}\rho_{d,\delta}^{[-1]}(\epsilon)\delta_{k,1},
\label{N3kF}
\eea
where $m_{d,\delta}=\left(\begin{array}{c}d\nonumber \\\delta\end{array}\right)$ is the number of $\delta-$faces added at each time $t$ to the NGF, $\rho_{d,\delta}(\epsilon)$ is the probability that such faces have energy $\epsilon$, and $\delta_{x,y}$ indicates the Kronecker delta.
Since for large times $t\gg1$  the average number of  $\delta$-faces with generalized degree $k_{d,\delta}=k$ scales like $N_{d,\delta}^{t, [-1]}(k|\epsilon)\simeq m_{d,\delta}P_{d,\delta}^{[-1]}(k|\epsilon)$ we can derive $P_{d,\delta}^{[-1]}(k|\epsilon)$  for $0\leq \delta\leq d-3$  
 \bea
 P_{d,\delta}^{[-1]}(k|\epsilon)&=&\frac{ \exp{\left[\beta\left(\epsilon-\mu_{d, \delta}^{[-1]}\right)\right]}\Gamma\left[k+2/(d-\delta-2)\right]}{\Gamma\left[k+1+2/(d-\delta-2)+\exp{\left[\beta\left(\epsilon-\mu_{d, \delta}^{[-1]}\right)\right]}\right]}  \frac{\Gamma\left[1+2/(d-\delta-2)+\exp{\left[\beta\left(\epsilon-\mu_{d, \delta}^{[-1]}\right)\right]}\right]}{\Gamma\left[1+2/(d-\delta-2)\right]}.  
\label{PkF3-1}
 \eea
 In order to obtain the  distributions $P_{d,\delta}^{[-1]}(k)$ of generalized degree $k_{d,\delta}=k$ of $\delta-$faces  we use the relation 
 \bea
  P_{d,\delta}^{[-1]}(k)=\sum_{\epsilon}\rho_{d, \delta}^{-1}(\epsilon) P_{d,\delta}^{[-1]}(k|\epsilon),
  \label{conv}
 \eea
finding
\bea
\hspace*{-3mm}P_{d,d-1}^{[-1]}(1)&=&\left\{\begin{array}{lcc}
\sum_{\epsilon}\rho_{d, d-1}^{[-1]}(\epsilon)\left(1-\frac{1}{\exp{\left[\beta\left(\epsilon-\mu_{d, d-1}^{[-1]}\right)\right]}+1}\right)  &\mbox{for}& k=1,\nonumber \\
\sum_{\epsilon}\rho_{d, d-1}^{[-1]}(\epsilon)\frac{1}{\exp{\left[\beta\left(\epsilon-\mu_{d, d-1}^{[-1]}\right)\right]}+1} &\mbox{for}& k=2,\end{array}\right.\nonumber \\
P_{d,d-2}^{[-1]}(k)&=&\sum_{\epsilon}\rho_{d, d-2}^{[-1]}(\epsilon)\frac{\exp\left[\beta\left(\epsilon-\mu_{d, d-2}^{[-1]}\right)\right]}{\left(\exp{\left[\beta\left(\epsilon-\mu_{d, d-2}^{[-1]}\right)\right]}+1\right)^{k}},\nonumber \\
P_{d,\delta}^{[-1]}(k)&=&\sum_{\epsilon}\rho_{d, d-2}^{[-1]}(\epsilon)\frac{ \exp{\left[\beta\left(\epsilon-\mu_{d, \delta}^{[-1]}\right)\right]}\Gamma\left[k+2/(d-\delta-2)\right]}{\Gamma\left[k+1+2/(d-\delta-2)+\exp{\left[\beta\left(\epsilon-\mu_{d, \delta}^{[-1]}\right)\right]}\right]} \nonumber \\
&& \times \frac{\Gamma\left[1+2/(d-\delta-2)+\exp{\left[\beta\left(\epsilon-\mu_{d, \delta}^{[-1]}\right)\right]}\right]}{\Gamma\left[1+2/(d-\delta-2)\right]},  \ \mbox{for} \ 0\leq \delta\leq d-3.
\label{Pksb-1}
\eea
The average of the generalized degree minus one, performed over $\delta-$faces of energy $\epsilon$ is given by the Fermi-Dirac statistics for $\delta=d-1$, the Boltzmann statistics for $\delta=d-2$ and the Bose-Einstein statistics for $\delta\leq d-3$ \cite{CQNM}. In particular, using Eqs. $(\ref{Pksb-1})$ we obtain
\bea
{\Avg{k_{d, d-1}-1|\epsilon,s=-1}}&=&n_F\left(\epsilon, \mu_{d, d-1}^{[-1]}\right),   \\
{\Avg{k_{d, d-2}-1|\epsilon,s=-1}}&=&n_Z\left(\epsilon, \mu_{d, d-2}^{[-1]}\right),    \nonumber\\
{\Avg{k_{d, \delta}-1|\epsilon,s=-1}}&=&A_{d,\delta}^{[-1]} n_B\left(\epsilon, \mu_{d, \delta}^{[-1]}\right),   \nonumber
\label{nB-1}
\eea
where the last expression is valid for $\delta\leq d-3,$ and where    $n_F(\epsilon,\mu),n_Z(\epsilon,\mu)$ and $n_B(\epsilon,\mu)$ are given by Eqs. $(\ref{ns})$, while $A_{d,\delta}^{[-1]}$ is given by 
\bea
A_{d,\delta}^{[-1]}=\frac{(d-\delta)}{(d-\delta-2)}. 
\eea
The self-consistent value of the chemical potentials $\mu_{d,\delta}^{[-1]}$ can be found by imposing the following geometrical relations satisfied by the generalized degrees of the NGF of any flavor $s=-1,0,1$, 
\bea
\lim_{t\to \infty}\frac{\sum_{\alpha\in {\cal S}_{d,\delta}(t)}k_{d,\delta}(\alpha)}{N_{d,\delta}(t)}=\frac{d+1}{\delta+1}.
\label{cond}
\eea
Imposing such condition is equivalent to fixing the normalization conditions for $n_F\left(\epsilon,\mu_{d,d-1}^{[-1]}\right),n_Z\left(\epsilon,\mu_{d,d-2}^{[-1]}\right),$ and $n_B\left(\epsilon,\mu_{d,\delta}^{[-1]}\right)$. These conditions are given by 
\bea
\sum_{\epsilon} \rho_{d,d-1}(\epsilon)n_F\left(\epsilon,\mu_{d,d-1}^{[-1]}\right)&=&\frac{1}{d},\nonumber \\
\sum_{\epsilon} \rho_{d,d-2}(\epsilon)n_Z\left(\epsilon,\mu_{d,d-2}^{[-1]}\right)&=&\frac{2}{d-1},\nonumber \\
\sum_{\epsilon} \rho_{d,\delta}(\epsilon)n_B\left(\epsilon,\mu_{d,\delta}^{[-1]}\right)&=&\frac{d-\delta-2}{\delta+1}.\nonumber \\
\eea

\subsubsection{Case $s=0$}
Let us derive the distribution $P_{k,\delta}^{[-1]}(k)$ of the generalized degrees $k_{d,\delta}$ for every $0\leq \delta\leq d-1$ in a NGF with flavor $s=0$.\\
The average number $N_{d,d-1}^{t,[0]}(k|\epsilon)$ of $(d-1)$-faces of energy $\epsilon$ that at  time $t$ have  generalized degrees $k_{d,\delta}=k$ in a NGF with flavor $s=0$ follows the master equation given by 
\bea
\hspace*{-0mm}{N_{d,d-1}^{t+1 [0]}(k|\epsilon)}-N_{d,\delta}^{t}(k | \epsilon)&=&e^{-\beta \left(\epsilon-{\mu}_{d,d-1}^{[0]}\right)}\frac{1}{t}N_{d,\delta}^{t [0]}(k-1|\epsilon)[1-\delta_{k,1}]-e^{-\beta \left(\epsilon-{\mu}_{d,d-1}^{[0]}\right)}\frac{1}{t}N_{d,\delta}^{t, [0]}(k|\epsilon)\nonumber \\
&&+m_{d,\delta}\rho_{d,\delta}^{[0]}(\epsilon)\delta_{k,1},
\label{N0a}
\eea
where $\rho_{d, d-1}^{[0]}(\epsilon)$ indicates the probability that a $(d-1)$-face has energy $\epsilon$ and the chemical potential $\mu_{d,d-1}^{[0]}$ is defined self-consistently as
\bea
e^{-\beta \mu_{d,d-1}^{[0]}}&=&\lim_{t\to \infty}\frac{Z^{[0]}}{t}.
\label{selfF0a}
\eea
For large times $t\gg1$ the average number $N_{d,d-1}^{t, [0]}(k|\epsilon)$ of $(d-1)-$faces with degree $k$ and energy $\epsilon$ in NGF of flavor $s=0$ satisfies $N_{d,d-1}^{t,[0]}(k|\epsilon)\simeq tm_{d,\delta}\rho_{d,\delta}^{[0]}(\epsilon)P_{d,d-1}^{[0]}(k|\epsilon)$, where $P_{d,d-1}^{[0]}(k|\epsilon)$ indicates the probability distribution that a $\delta$-face of energy $\epsilon$ has generalized degree $k_{d,\delta}=k$. Therefore we can solve Eq. $(\ref{N0b})$ for $P_{d,d-1}^{[0]}(k|\epsilon)$, obtaining
\bea
P_{d,d-1}^{[0]}(k|\epsilon)&=&\frac{e^{\beta\left(\epsilon-\mu_{d, d-2}^{[0]}\right)}}{\left(e^{\beta\left(\epsilon-\mu_{d, d-2}^{[0]}\right)}+1\right)^{k}}.  
\eea
The average number $N_{d,\delta}^{t,[0]}(k|\epsilon)$ of $\delta$-faces of energy $\epsilon$ with $0\leq \delta \leq d-2 $ that at  time $t$ have  generalized degrees $k_{d,\delta}=k$ in a NGF with flavor $s=0$ follows the master equation given by 
\bea
\hspace*{-0mm}{N_{d,\delta}^{t+1, [0]}(k|\epsilon)}-N_{d,\delta}^{t,[0]}(k | \epsilon)&=&e^{-\beta \left(\epsilon-{\mu}_{d,\delta}^{[0]}\right)}\frac{[(k-1)+1/(d-\delta-1)]}{t}N_{d,\delta}^{t,[0]}(k-1|\epsilon)[1-\delta_{k,1}]\nonumber \\
&&-e^{-\beta \left(\epsilon-{\mu}_{d,\delta}^{[0]}\right)}\frac{[k+1/(d-\delta-1)]}{t}N_{d,\delta}^{t, [0]}(k|\epsilon)+m_{d,\delta}\rho_{d,\delta}^{[0]}(\epsilon)\delta_{k,1},
\label{N0b}
\eea
where $\rho_{d, \delta}^{[0]}(\epsilon)$ indicates the probability that a $\delta$-face has energy $\epsilon$ and the chemical potentials $\mu_{d,\delta}^{[0]}$ for $0\leq \delta \leq d-2$ are self-confidently defined as
\bea
e^{-\beta \mu_{d,\delta}^{[0]}}&=&e^{\beta \mu_{d,d-1}^{[0]}}\Avg{\frac{\sum_{\alpha\in {\cal S}_{d,\delta}}\sum_{\alpha'\in {\cal S}_{d,d-1}|\alpha\subset\alpha'}e^{-\beta(\epsilon_{\alpha'}-\epsilon_{\alpha})}\delta(k_{d,\delta}(\alpha),k)}{\sum_{\alpha''\in{\cal S}_{d,\delta}}[k+1/(d-\delta-1)] \delta(k_{d,\delta}(\alpha''),k)}}.
\label{selfF0b}
\eea
For large times $t\gg1$ the average number $N_{d,\delta}^{t, [0]}(k|\epsilon)$ of $\delta-$faces with generalized degree $k_{d,\delta}=k$ and energy $\epsilon$ in NGF of flavor $s=0$ satisfies $N_{d,\delta}^{t,[0]}(k|\epsilon)\simeq tm_{d,\delta}\rho_{d,\delta}^{[0]}(\epsilon)P_{d,\delta}^{[0]}(k|\epsilon)$. Therefore we can solve Eq. $(\ref{N0b})$ for $P_{d,\delta}^{[0]}(k|\epsilon)$, obtaining 
\bea
P_{d,\delta}^{[0]}(k|\epsilon)&=&\frac{ \exp{\left[\beta(\epsilon-\mu_{d, \delta}^{[0]})\right]} \Gamma\left[k+1/(d-\delta-1)\right]}{\Gamma\left[k+1+1/(d-\delta-1)+\exp{\left[\beta\left(\epsilon-\mu_{d, \delta}^{[0]}\right)\right]}\right]} \frac{\Gamma\left[1+1/(d-\delta-1)+\exp{\left[\beta\left(\epsilon-\mu_{d, \delta}^{[0]}\right)\right]}\right]}{\Gamma\left[1+1/(d-\delta-1)\right]}.  \nonumber
\eea
Finally, using Eq. $(\ref{conv})$ we can  obtain  the distribution $P_{d,\delta}^{[0]}(k)$ of generalized degrees $k_{d,\delta}=k$. These are given by  
\bea
P_{d,d-1}^{[0]}(k)&=&\sum_{\epsilon}\rho_{d, d-1}^{[0]}(\epsilon)\frac{e^{\beta\left(\epsilon-\mu_{d, d-2}^{[0]}\right)}}{\left(e^{\beta\left(\epsilon-\mu_{d, d-2}^{[0]}\right)}+1\right)^{k}},   \\
P_{d,\delta}^{[0]}(k)&=&\sum_{\epsilon}\rho_{d, \delta}^{[0]}(\epsilon)\frac{ \exp{\left[\beta(\epsilon-\mu_{d, \delta}^{[0]})\right]} \Gamma\left[k+1/(d-\delta-1)\right]}{\Gamma\left[k+1+1/(d-\delta-1)+\exp{\left[\beta\left(\epsilon-\mu_{d, \delta}^{[0]}\right)\right]}\right]} \frac{\Gamma\left[1+1/(d-\delta-1)+\exp{\left[\beta\left(\epsilon-\mu_{d, \delta}^{[0]}\right)\right]}\right]}{\Gamma\left[1+1/(d-\delta-1)\right]},  \nonumber
\label{P0s}
\eea
where the last equation is valid for $0\leq \delta \leq d-2.$
Therefore the $(d-1)-$faces have generalized degree distribution $P_{d,d-1}^{[0]}(k)$ that is given by a convolution of exponentials, while the $\delta-$faces with $\delta\leq d-2$ have a generalized degree distribution $P_{d,\delta}^{[0]}(k)$ that is given by a convolution of power-laws.
When considering the average ${\Avg{k_{d, \delta}-1|\epsilon,s=0}}$, we observe that this quantity for  $\delta=d-1$ is a Boltzmann distribution and for every $\delta \leq d-2$ is a Bose-Einstein distribution, i.e.
\bea
{\Avg{k_{d, d-1}-1|\epsilon,s=0}}&=&n_Z\left(\epsilon, \mu_{d, d-2}^{[0]}\right),    \\
{\Avg{k_{d, \delta}-1|\epsilon,s=0}}&=&A_{d,\delta}^{[0]} n_B\left(\epsilon, \mu_{d, \delta}^{[0]}\right), \  \mbox{for} \ \  \delta\leq d-2, 	\nonumber
\label{nB0}
\eea
with $n_Z(\epsilon,\mu)$ and $n_B(\epsilon,\mu)$ given by Eqs. $(\ref{ns})$
 and $A_{d,\delta}^{[0]}$ given by 
 \bea
A_{d,\delta}^{[0]}=\frac{(d-\delta)}{(d-\delta-1)}. 
\eea
The chemical potential $\mu_{d,\delta}^{[0]}$ can  then be found imposing the condition in Eq. $(\ref{cond})$ that all NGF must satisfy. Therefore,   the self-consistent equations that the chemical potentials must satisfy are 
\bea
\sum_{\epsilon} \rho_{d,d-1}(\epsilon)n_Z\left(\epsilon,\mu_{d,d-1}^{[-1]}\right)&=&\frac{1}{d},\nonumber \\
\sum_{\epsilon} \rho_{d,\delta}(\epsilon)n_B\left(\epsilon,\mu_{d,\delta}^{[-1]}\right)&=&\frac{d-\delta-1}{\delta+1},  \  \mbox{for }    \delta\leq d-2.
\eea

\subsubsection{Case $s=1$}
In this section we derive the distribution $P_{d,\delta}^{[1]}(k)$ of generalized degrees $k_{d,\delta}=k$ in NGF with flavor $s=1$.\\
The master equation for the average number ${N_{d,\delta}^{t+1,[1]}(k|\epsilon)}$ of $\delta-$faces with generalized degree $k_{d,\delta}=k$ and energies $\epsilon$ is given by 
\bea
\hspace*{-0mm}{N_{d,\delta}^{t+1,[1]}(k|\epsilon)}-N_{d,\delta}^{t,[1]}(k | \epsilon)&=&\frac{e^{-\beta \left(\epsilon-{\mu}_{d,\delta}^{[1]}\right)}(k-1)}{t}N_{d,\delta}^{t,[1]}(k-1|\epsilon)[1-\delta_{k,1}]\nonumber \\
&&-\frac{e^{-\beta \left(\epsilon-{\mu}_{d,\delta}^{[1]}\right)}k}{t}N_{d,\delta}^{t, [1]}(k|\epsilon)+m_{d,\delta}\rho_{d,\delta}^{[1]}(\epsilon)\delta_{k,1},
\label{NZk1}
\eea
where $\rho_{d, \delta}^{[1]}(\epsilon)$ indicates the probability that a $\delta$-face has energy $\epsilon$ and the chemical potentials $\mu_{d,\delta}^{[1]}$ are defined self-consistently respectively for the cases $\delta=d-1$ and $0\leq \delta\leq d-2$ as,
\bea
e^{-\beta \mu_{d,d-1}^{[1]}}&=&\lim_{t\to \infty}\frac{Z^{[1]}}{t},
\label{selfF1a} \\
e^{-\beta \mu_{d,\delta}^{[1]}}&=&e^{\beta \mu_{d,d-1}^{[1]}}\Avg{\frac{\sum_{\alpha\in {\cal S}_{d,\delta}}\sum_{\alpha'\in {\cal S}_{d,d-1}|\alpha\subset\alpha'}e^{-\beta(\epsilon_{\alpha'}-\epsilon_{\alpha})}\delta(k_{d,\delta}(\alpha),k)}{\sum_{\alpha''\in{\cal S}_{d,\delta}}k \delta(k_{d,\delta}(\alpha''),k)}}.
\label{selfF1b}
\eea
Since asymptotically in time we observe that  $N_{d,\delta}^{[k]}(k|\epsilon)\simeq m_{d,\delta}\rho_{d,\delta}^{[1]}(\epsilon)P^{[1]}_{d,\delta}(k|\epsilon)$, where $P^{[1]}_{d,\delta}(k|\epsilon)$ is the probability distribution that a $\delta$-face of energy $\epsilon$ has generalized degree $k_{d,\delta}=k,$ we obtain 
\bea
P_{d,\delta}^{[1]}(k|\epsilon)&=&\frac{ \exp{\left[\beta\left(\epsilon-\mu_{d, \delta}^{[1]})\right)\right]} \Gamma\left[k\right]}{\Gamma\left[k+1+\exp{\left[\beta\left(\epsilon-\mu_{d, \delta}^{[1]}\right)\right]}\right]}\Gamma\left[1+\exp{\left[\beta\left(\epsilon-\mu_{d, \delta}^{[1]}\right)\right]}\right]. 
\label{Pksb1e}
\eea

Therefore, using Eq. $(\ref{conv})$ we obtain that the probability $P_{d,\delta}^{[1]}(k)$ that a $\delta$-face has generalized degree $k_{d,\delta}=k$ is given by 
\bea
P_{d,\delta}^{[1]}(k)&=&\sum_{\epsilon}\rho_{d, \delta}^{[1]}(\epsilon)\frac{ \exp{\left[\beta\left(\epsilon-\mu_{d, \delta}^{[1]})\right)\right]} \Gamma\left[k\right]}{\Gamma\left[k+1+\exp{\left[\beta\left(\epsilon-\mu_{d, \delta}^{[1]}\right)\right]}\right]}\Gamma\left[1+\exp{\left[\beta\left(\epsilon-\mu_{d, \delta}^{[1]}\right)\right]}\right]. 
\label{Pksb1}
\eea
In this case, if we perform the average ${\Avg{k_{d, \delta}-1|\epsilon,s=1}}$ over all $\delta-$faces with energy $\epsilon,$ we always get the Bose-Einstein distribution, independently of the value of $0\leq \delta< d$, i.e. we obtain
\bea
{\Avg{k_{d, \delta}-1|\epsilon,s=1}}&=& n_B\left(\epsilon, \mu_{d, \delta}^{[1]}\right), 
\label{nB1}
\eea
with $n_B(\epsilon,\mu)$ given by Eq. $(\ref{ns})$.
The chemical potentials $\mu^{[1]}_{d,\delta}$ must satisfy Eq. $(\ref{cond})$. Therefore they can be found self-consistently by solving 
\bea
\sum_{\epsilon} \rho_{d,\delta}(\epsilon)n_B\left(\epsilon,\mu_{d,\delta}^{[1]}\right)=\frac{d-\delta}{\delta+1}.
\eea
\section*{CODES FOR GENERATING   NGF}
\label{6}
In this section we provide three MATLAB codes for generating NGF in dimension $d=1,2,3$.
\subsection{Code for $d=1$}

\begin{lstlisting}

function [a,kn] = NGF_d1(N,s,beta,figure)
%%%%%%%%%%%%%%%%%%%%%%%%%%%%%%%%%%%%%%%%%%%%%%%%%%%%%%%%%%%%%%%%%%%%%%%%%%%
% If you use this code, please cite 
% G. Bianconi and C. Rahmede 
% "Network geometry with flavour: from complexity to quantum geometry"
%%%%%%%%%%%%%%%%%%%%%%%%%%%%%%%%%%%%%%%%%%%%%%%%%%%%%%%%%%%%%%%%%%%%%%%%%%%
% Code that generates NGF in dimension d=2 and flavour s=-1,0,1.

% a adjacency matrix
% kn vector of  generalized degrees k_{1,0} (the degree) of the nodes

% This code uses 
% N maximal number of nodes in the NGF
% Flavour of the NGF  s=-1,0,1
% Inverse temperature: beta>0 or beta=0
% figure=1 will print the edge list of the network in file 
% "NGF_edgelist_d1_s%d.edges"
% figure=0 will not print the edge list of the network
% energy of the nodes epsilon is uniform from 0-9

%%%%%%%%%%%%%%%%%%%%%%%%%%%%%%%%%%%%%%%%%%%%%%%%%%%%%%%%%%%%%%%%%%%%%%%%%%%

% Initialization
a=sparse(N,N);
a_occ=zeros(1,N);

%%%%%%%%%%%%%%%%%%%%%%%%%%%%%%%%%%%%%%%%%%%%%%%%%%%%%%%%%%%%%%%%%%%%%%%%%%%
% Assign energies to the nodes
% If using Poisson and power-law you must define
% the parameters mu, or kappa
% Examples:
% mu=10;
% kappa=1;
for i=1:N 
    epsilon(i)=floor(10*rand(1));
    % Alternative energy distributions
    % epsilon(i)=random('Poisson',mu); 
    % poisson distribution with average mu
    % epsilon(i)=rand(1)^(1/(kappa+1)); 
    % power-law distribution with exponent kappa
   
end
%%%%%%%%%%%%%%%%%%%%%%%%%%%%%%%%%%%%%%%%%%%%%%%%%%%%%%%%%%%%%%%%%%%%%%%%%%%     
% Initial condition: at time t=1 a single link (1,2)

        a(1,2)=exp(-beta*(epsilon(1)+epsilon(2)));
        a(2,1)=exp(-beta*(epsilon(1)+epsilon(2)));
        k(1)=1;  
        k(2)=1;  
        a_occ(1)=1;
        a_occ(2)=1;
        
%%%%%%%%%%%%%%%%%%%%%%%%%%%%%%%%%%%%%%%%%%%%%%%%%%%%%%%%%%%%%%%%%%%%%%%%%%%        
   % Addition of new links at time t=in-1 the node in is added to the
   % network geometry with flavour
   
  for in=2+1:N,
    % Choose the node to which attach a new link

    V=exp(-beta*epsilon).*a_occ;
    norm=sum(V);   
    x=rand(1)*norm;
    if (norm>0)
	    for nj1=1:in-1,
		    x=x-V(nj1);
		 if x<0,
		     j=nj1;
		     break;
		 end
        end
  end
   % Attach the new link between node in and node j
        a(in,j)=exp(-beta*epsilon(in)-beta*epsilon(j));
        a(j,in)=a(in,j);
        a_occ(in)=1;
        a_occ(j)=a_occ(j)+s;	  
end
%%%%%%%%%%%%%%%%%%%%%%%%%%%%%%%%%%%%%%%%%%%%%%%%%%%%%%%%%%%%%%%%%%%%%%%%%%%
% Generalized degree (degree) of the nodes
kn=sum(a>0);
a=a>0;

%%%%%%%%%%%%%%%%%%%%%%%%%%%%%%%%%%%%%%%%%%%%%%%%%%%%%%%%%%%%%%%%%%%%%%%
% Print network file
if figure==1
    [I,J,A]=find(tril(a));
    filename=sprintf('NGF_edgelist_d1_s%d.edges',s);
    fid=fopen(filename,'w');
    for it=1:max(size(A)),
        fprintf(fid, '%d  %d  \n', I(it), J(it));
    end
    fclose(fid);
end

end

\end{lstlisting}
\subsection{Code for $d=2$}
\begin{lstlisting}

function [a,kn,kl] = NGF_d2(N,s,beta,figure)
%%%%%%%%%%%%%%%%%%%%%%%%%%%%%%%%%%%%%%%%%%%%%%%%%%%%%%%%%%%%%%%%%%%%%%%%%%%
% If you use this code, please cite 
% G. Bianconi and C. Rahmede 
% "Network geometry with flavour: from complexity to quantum geometry"
%%%%%%%%%%%%%%%%%%%%%%%%%%%%%%%%%%%%%%%%%%%%%%%%%%%%%%%%%%%%%%%%%%%%%%%%%%%
% Code that generates NGF in dimension d=2 and flavour s=-1,0,1.

% a adjacency matrix
% kn vector of  generalized degrees k_{2,0}  of the nodes
% kl vector of  generalized degrees k_{2,1} of links  

% This code uses 
% N maximal number of nodes in the NGF
% Flavour of the NGF  s=-1,0,1
% Inverse temperature: beta>0 or beta=0
% figure=1 will print the edge list of the network in file 
% "NGF_edgelist_d2_s%d.edges"
% figure=0 will not print the edge list of the network
% energy of the nodes epsilon is uniform from 0-9

%%%%%%%%%%%%%%%%%%%%%%%%%%%%%%%%%%%%%%%%%%%%%%%%%%%%%%%%%%%%%%%%%%%%%%%%
% Inizialization
a=sparse(N,N);
a_occ=sparse(N,N);
a_occ2=sparse(N,N);

%%%%%%%%%%%%%%%%%%%%%%%%%%%%%%%%%%%%%%%%%%%%%%%%%%%%%%%%%%%%%%%%%%%%%%%%%%%
% Assign energies to the nodes
% If using Poisson and power-law you must define 
% the parameters mu, or kappa
% Examples:
% mu=10;
% kappa=1
for i=1:N
     epsilon(i)=floor(10*rand(1));
end
%%%%%%%%%%%%%%%%%%%%%%%%%%%%%%%%%%%%%%%%%%%%%%%%%%%%%%%%%%%%%%%%%%%%%%%%%%%
% Initial condition at time t=1 including a single triangle between nodes
% 1,2,3
L=0;
for i1=1:3,
    for i2=(i1+1):3,     
        L=L+1;
        a(i1,i2)=exp(-beta*(epsilon(i1)+epsilon(i2)));
        a(i2,i1)=exp(-beta*(epsilon(i1)+epsilon(i2)));
        a_occ(i1,i2)=1;  
        a_occ(i2,i1)=1;  
        a_occ2(i1,i2)=1;
        a_occ2(i2,i1)=1;
    end
end

%%%%%%%%%%%%%%%%%%%%%%%%%%%%%%%%%%%%%%%%%%%%%%%%%%%%%%%%%%%%%%%%%%%%%%%%%%%
% At each time t=in-2 we attach a new triangle

for in=(3+1):N,
    % Choose edge (l1,l2) to which we will attach the new triangle
    
    [I,J,V]=find(tril(a.*(a_occ)));
   
    norm=sum(V);   
    x=rand(1)*norm;
    if (norm>0)
	    for nj1=1:numel(V),
		    x=x-V(nj1);
            if x<0,
                nj=nj1;
                break;
            end
        end
	    l1=I(nj);
	    l2=J(nj);
	    
	     a_occ(l1,l2)=a_occ(l1,l2)+s;
	     a_occ(l2,l1)=a_occ(l2,l1)+s;
	     a_occ2(l1,l2)=a_occ2(l1,l2)+1;
         a_occ2(l2,l1)=a_occ2(l2,l1)+1;
         
         % Attach the new node in to the node l1;
         L=L+1;
	     a(in,l1)=exp(-beta*(epsilon(l1)+epsilon(in)));
         a(l1,in)=exp(-beta*(epsilon(l1)+epsilon(in)));
         a_occ(in,l1)=1;
         a_occ(l1,in)=1;
         a_occ2(in,l1)=1;
         a_occ2(l1,in)=1;
        
         % Attach the new node in to the node l2;
         L=L+1;
         a(in,l2)=exp(-beta*(epsilon(l2)+epsilon(in)));
         a(l2,in)=exp(-beta*(epsilon(l2)+epsilon(in)));
         a_occ(in,l2)=1;
         a_occ(l2,in)=1;
         a_occ2(in,l2)=1;
         a_occ2(l2,in)=1;
    end    
end
%%%%%%%%%%%%%%%%%%%%%%%%%%%%%%%%%%%%%%%%%%%%%%%%%%%%%%%%%%%%%%%%%%%%%%%%%%%
% Generalized degrees 
k=sum(a>0);

[I,J,kl]=find(tril(a_occ2));
kn=k-ones(size(k));
a=a>0;
%%%%%%%%%%%%%%%%%%%%%%%%%%%%%%%%%%%%%%%%%%%%%%%%%%%%%%%%%%%%%%%%%%%%%%%%%%%   
% Print network file

if figure==1
    [I,J,A]=find(tril(a));
    filename=sprintf('NGF_edgelist_d2_s%d.edges',s);
    fid=fopen(filename,'w');
    for it=1:max(size(A)),
        fprintf(fid, '%d  %d  \n', I(it), J(it));
    end
    fclose(fid);
end


end


\end{lstlisting}
\subsection{Code for $d=3$}
\begin{lstlisting}

function [a,kn,kl,kt] = NGF_d3(N,s,beta,figure)
%%%%%%%%%%%%%%%%%%%%%%%%%%%%%%%%%%%%%%%%%%%%%%%%%%%%%%%%%%%%%%%%%%%%%%%%%%%
% If you use this code, please cite 
% G. Bianconi and C. Rahmede 
% "Network geometry with flavour: from complexity to quantum geometry"
%%%%%%%%%%%%%%%%%%%%%%%%%%%%%%%%%%%%%%%%%%%%%%%%%%%%%%%%%%%%%%%%%%%%%%%%%%%
% Code that generates NGF in dimension d=3 and flavour s=-1,0,1.

% a adjacency matrix
% kn vector of  generalized degrees k_{3,0}  of the nodes
% kl vector of  generalized degrees k_{3,1} of links  
% kt vector of  generalized degrees k_{3,2} of triangles  

% This code uses 
% N maximal number of nodes in the NGF
% Flavour of the NGF  s=-1,0,1
% Inverse temperature: beta>0 or beta=0
% figure=1 will print the edge list of the network in file 
%"NGF_edgelist_d3_s%d.edges"
% figure=0 will not print the edge list of the network
% energy of the nodes epsilon is uniform from 0-9

%%%%%%%%%%%%%%%%%%%%%%%%%%%%%%%%%%%%%%%%%%%%%%%%%%%%%%%%%%%%%%%%%%%%%%%%%%%
% Initialization
a=sparse(N,N);

nt=0;   
%%%%%%%%%%%%%%%%%%%%%%%%%%%%%%%%%%%%%%%%%%%%%%%%%%%%%%%%%%%%%%%%%%%%%%%%%%%
% Assign energies to the nodes
% If using Poisson and power-law you must define
% the parameters mu, or kappa
% Examples:
% mu=10;
% kappa=1;
for i=1:N,
     epsilon(i)=floor(10*rand(1));
% Alternative energy distributions
    %epsilon(i)=random('Poisson',mu); 
    %poisson distribution with average mu
    %epsilon(i)=rand(1)^(1/(kappa+1)); 
    %power-law distribution with exponent kappa
   
end

%%%%%%%%%%%%%%%%%%%%%%%%%%%%%%%%%%%%%%%%%%%%%%%%%%%%%%%%%%%%%%%%%%%%%%%%%%%     
% Initial condition: at time t=1 a single tedrahedron (1,2,3,4)

for i1=1:4,
    for i2=(i1+1):4,        
        a(i1,i2)=1;
        a(i2,i1)=1; 
       for i3=(i2+1):4,           
           nt=nt+1;
           tri(nt,1)=i1;
           tri(nt,2)=i2;
           tri(nt,3)=i3;
           at(nt)=exp(-beta*(epsilon(i1)+epsilon(i2)+epsilon(i3)));
           a_occ(nt)=1;
           a_occ3(nt)=1;
       end
    end    
end


%%%%%%%%%%%%%%%%%%%%%%%%%%%%%%%%%%%%%%%%%%%%%%%%%%%%%%%%%%%%%%%%%%%%%%%%%%%
% At each time t=in-3 we attach a new tetrahedron

for in=4+1:N,
% Choose triangular face to which to attach the new tetrahedron 

    [I,J,V]=find(at.*a_occ);

    norm=sum(V);
    x=rand(1)*norm;
    for nj1=1:numel(V),
            x=x-V(nj1);
         if x<0,
             nj=J(nj1);
             break;
         end
    end
 
    l(1)=tri(nj,1);
    l(2)=tri(nj,2);
    l(3)=tri(nj,3);

     a_occ(nj)=a_occ(nj)+s;
     a_occ3(nj)=a_occ3(nj)+1;
    %Add the tethaedron
    for n=1:3,
    a(in,l(n))=1;
    a(l(n),in)=1;  
    end
    for n1=1:3,
        for n2=n1+1:3,
            a(l(n1),l(n2))=a(l(n1),l(n2))+1;
            a(l(n2),l(n1))=a(l(n2),l(n1))+1;
        end
    end
    for n=1:3,
        for n2=n+1:3,           
            nt=nt+1;
            tri(nt,1)=l(n);
            tri(nt,2)=l(n2);
            tri(nt,3)=in;   
            at(nt)=exp(-beta*(epsilon(l(n))+epsilon(l(n2))+epsilon(in)));
            a_occ(nt)=1;
            a_occ3(nt)=1;
        end
    end
end


%%%%%%%%%%%%%%%%%%%%%%%%%%%%%%%%%%%%%%%%%%%%%%%%%%%%%%%%%%%%%%%%%%%%%%%%%%%
% Generalized degrees
kn=sum(a>0);
kn=kn-2;
[I2,J2,A2]=find(tril(a));
kl=A2;
kt=a_occ3;
a=a>0;
%%%%%%%%%%%%%%%%%%%%%%%%%%%%%%%%%%%%%%%%%%%%%%%%%%%%%%%%%%%%%%%%%%%%%%%%%%%   
% Print network file
[I2,J2,A2]=find(tril(a>0));
if (figure==1)
   filename=sprintf('NGF_edgelist_d3_s%d.edges',s);
   fid=fopen(filename,'w');
   for it=1:numel(A2),
       fprintf(fid, ' %d %d  \n', I2(it), J2(it));
   end
   fclose(fid);
end

\end{lstlisting}

\begin{thebibliography}{99}

\bibitem{interdisciplinary}
G. Bianconi, EPL{\bf 111} 56001 (2015).

\bibitem{Kleinberg}
R. Kleinberg,  
In INFOCOM 2007. 26th IEEE International Conference on Computer Communications. IEEE,  1902,  (2007).
\bibitem{Boguna_navigability}
M.  Bogu\~n\'a,  D.    Krioukov,  and K. C.   Claffy,  
{ Nature Physics} {\bf 5},  74  (2008).
\bibitem{Boguna_Internet}
M.  Bogu\~n\'a,  F.   Papadopoulos,  and D.   Krioukov, 
{ Nature Commun.} {\bf 1},  62  (2010).
\bibitem{Aste_filtering}
M. Tumminello,  T. Aste, T. Di Matteo, and R. N. Mantegna,
PNAS {\bf 102},  10421 (2005).
\bibitem{Reka}
R. Albert, B. DasGupta, and N. Mobasheri,
Phys. Rev. E {\bf 89}, 032811 (2014).
\bibitem{Vaccarino2}
G.  Petri,  P.   Expert,   F.  Turkheimer,  R.    Carhart-Harris,  D.    Nutt,  P.J.   Hellyer,   and F.  Vaccarino, 
{ Journal of The Royal Society Interface} {\bf 11},    20140873 (2014).
\bibitem{Mason}
D. Taylor,  F. Klimm, H. A. Harrington, M. Kramar, K. Mischaikow, M. A. Porter, and P. J. Mucha,
Nature Commun.  {\bf 6},7723 (2015).
\bibitem{Caldarelli}
M. Borassi, A. Chessa and G. Caldarelli,Phys. Rev. E {\bf 92}, 032812(2015).

\bibitem{Mukherjee}
J. Steenbergen, C. Klivans, and S. Mukherjee,
Advances in Applied Mathematics {\bf 56},  56 (2014).

\bibitem{Aste}
T. Aste,   T.   Di Matteo,   and S: T.   Hyde,   
{ Physica A: Statistical Mechanics and its Applications} {\bf 346},   20 (2005).

\bibitem{Hyperbolic}
D. Krioukov, F. Papadopoulos, M. Kitsak, A. Vahdat, and M. Bogu\~n\'a, Phys. Rev. E 82, 036106 (2010). 


\bibitem{Boguna_growing}
 F. Papadopoulos,   M.   Kitsak,  M.A.   Serrano,  M.    Bogu\~n\'a,   and D.   Krioukov,   
  { Nature} {\bf 489},   537 (2012).
  \bibitem{Saniee}
O.  Narayan,  and I.  Saniee,  
{ Phys. Rev. E} {\bf 84},  066108 (2011).
  
\bibitem{det}
S. N. Dorogovtsev,A. V. Goltsev, and J. F. F. Mendes. 
Phys. Rev. E {\bf 65}  066122 (2002).

\bibitem{Apollonian1}
J. S.  Andrade Jr,  H. J. Herrmann, R. F. S. Andrade, and L. R. da Silva
Phys. Rev. Lett. {\bf 94},  018702 (2005).
\bibitem{Apollonian4}
Z. Zhang, L. Rong, L. and F. Comellas, 
Phstatysica A {\bf 364}, 610 (2006).
\bibitem{Apollonian2}
T. Zhou,  G. Yan, and B.-H. Wang,
Phys. Rev. E {\bf 71},  046141 (2005).
\bibitem{Apollonian3}
Z. Zhang,  F. Comellas, G. Fertin, and L. Rong,
Journal of physics A {\bf 39}, 1811 (2006).
\bibitem{Aste2}
T. Aste, R. Gramatica, and T. Di Matteo, Phys. Rev. E {\bf 86}, 036109 (2012).




\bibitem{Emergent}
Z. Wu, G. Menichetti, C. Rahmede and G. Bianconi, Scientific Reports {\bf 5}, 10073 (2015).
\bibitem{PRE}
G. Bianconi, C. Rahmede, Z. Wu, Phys. Rev. E  92, 022815 (2015).
\bibitem{CQNM}
G. Bianconi, C. Rahmede, Scientific Reports, 5, 13979 (2015).
\bibitem{Franzosi1}
R. Franzosi, D. Felice, S. Mancini, and M. Pettini
EPL (Europhysics Letters) {\bf 111},  20001 (2015).



\bibitem{Yau1}
Y. Lin,   L.   Lu,   and S.-T. Yau,   
{\ Tohoku Mathematical Journal} {\bf 63},    605 (2011).
\bibitem{Yau2}
 Y. Lin,   and S.-T. Yau,   
 {\it Math. Res. Lett} {\bf 17} 343 (2010).
\bibitem{Jost}
F. J.  Bauer,    J. Jost,   and S. Liu,   
arXiv preprint arXiv:1105.3803 (2011).
\bibitem{Ollivier}
Y. Ollivier,  
{ Journal of Functional Analysis} {\bf 256},   810 (2009).



\bibitem{Gromov}
M.  Gromov,    {\it Hyperbolic groups}. (Springer,   New York,   1987).

\bibitem{Majid}
S. Majid, 
Journal of Geometry and Physics {\bf 69}, 74 (2013).

\bibitem{Bombelli}
  L.~Bombelli, J.~Lee, D.~Meyer and R.~D.~Sorkin,
  Phys.\ Rev.\ Lett.\  {\bf 59} (1987) 521.
  \bibitem{Dowker}
  F. Dowker, J. Henson, and R. D. Sorkin
  Modern Physics Letters A {\bf 19}, 1829, (2004).
  \bibitem{CDT1}
J. Ambjorn,    J.   Jurkiewicz,   and R. Loll,
 {\it Phys. Rev. D} {\bf 72},   064014 (2005).
\bibitem{CDT2}
J. Ambjorn,   J.   Jurkiewicz,  and R.  Loll 
{\it Phys. Rev. Lett.} {\bf 93},    131301 (2004).

\bibitem{Burda1}
P. Bialas,  Z. Burda, and D. Johnston,
 Nuclear Physics B {\bf 542},  413 (1999).
 \bibitem{Burda2}
 P. Bialas, Z. Burda, A. Krzywicki, and B. Petersson,
 Nuclear Physics B {\bf 472}, 293 (1996).
  \bibitem{Oriti}
D. Oriti, Reports on Progress in Physics {\bf 64}, 1703 (2001).
\bibitem{Oriti_PRL}
S. Gielen, D. Oriti, and L. Sindoni,
Phys. Rev. Lett. 111, 031301 (2013).
\bibitem{Smolin1}
C. Rovelli, and L. Smolin,
Nucl. Phys. B {\bf 442},  593 (1995).
\bibitem{Smolin2}
C. Rovelli, and L. Smolin,
Nuclear Physics B {\bf 331}, 80 (1990).
 \bibitem{Rovelli}
C. Rovelli and F. Vidotto, {\it Covariant Loop Quatum Gravity}, (Cambridge University Press,Cambridge, 2015).

\bibitem{Energetic1}
M. Cort\^es, L. Smolin,  Phys. Rev. D {\bf 90}, 084007 (2014).
\bibitem{Energetic2}
M. Cort\^es, L. Smolin,  Physical Review D,  {\bf 90},  044035 (2014).
\bibitem{Trugenberger}
C. A. Trugenberger, 
Phys. Rev. D {\bf 92},  084014 (2015).
\bibitem{graphity_rg}
F. Antonsen, 
{  International journal of theoretical physics}, {\bf 33}, 11895 (1994).
\bibitem{graphity1}
T. Konopka,  F. Markopoulou,  S. Severini,   
{\it Phys. Rev. D} {\bf 77},    104029 (2008).
\bibitem{graphity2}
A. Hamma, F. Markopoulou, S. Lloyd, F. Caravelli, S. Severini, and K. Markstr\"om,
Phys. Rev. D {\bf 81}, 104032 (2010).


  \bibitem{Cosmology}
 D. Krioukov,	
M. Kitsak,	
R. S. Sinkovits,	
D. Rideout,	
D. Meyer, M. Bogu\~n\'a,
    Scientific Reports {\bf 2},    793 (2012).
\bibitem{Evans}
J. R. Clough and T. Evans, preprint arXiv:1408.1274 (2014).


\bibitem{RMP} R. Albert and A.-L. Barabasi,
   {Rev. Mod. Phys.} \textbf{74}, 47 (2002).

\bibitem{Doro_Book}
S.N. Dorogovtsev and J. F.F. Mendes {\it Evolution of networks: From biological nets to the Internet and WWW} (Oxford University Press,Oxford, 2003).


\bibitem{Newman_Book}
Newman,   M.  E.  J.  {\it Networks: An introduction. } (Oxford University Press, Oxford, 2010). 

 
\bibitem{Dima_SC}
K. Zuev, O. Eisenberg, D. Krioukov,
arXiv:1502.05032 (2015).

\bibitem{Farber1}
A. Costa and M. Farber, 
arxiv:1412.5805 (2014).

\bibitem{Farber2}
 D. Cohen,
A. Costa,
M. Farber,
T. Kappeler 
Discrete and Computational Geometry, {\bf 47}, 117 (2012). 
\bibitem{Kahle}
M. Kahle, %
Topology of random simplicial complexes: a survey 
AMS Contemp. Math {\bf 620}, 201 (2014).

\bibitem{Newman1}
G. Ghoshal, V. Zlati\'c, G. Caldarelli, and M. E. J. Newman,
Phys.Rev. E {\bf 79},  066118 (2009).
\bibitem{Newman2}
V. Zlati\'c, G. Ghoshal, and G. Caldarelli,
Phys. Rev. E {\bf 80},  036118 (2009).


\bibitem{WS}
 D. J. Watts,   and S.  Strogatz, 
 { Nature} {\bf 393},    440 (1998).
\bibitem{BA}
A. -L. Barab\'asi,   and R.   Albert,   
{ Science} {\bf 286},   509 (1999).

\bibitem{Santo}
S. Fortunato,  
 {\it Phys. Rep.} {\bf 486},  75 (2010).

\bibitem{Fitness}
G. Bianconi, A.-L. Barab\'asi, EPL {\bf 54}, 436 (2001).
     \bibitem{Bose}
 G. Bianconi and A.-L. Barab\'asi, Phys. Rev. Lett. 86, 5632  (2001).
 \bibitem{Chayes}
C. Borgs, J. Chayes, C. Daskalakis, and S. Roch,
in {\em Proceedings of the thirty-ninth annual ACM symposium on Theory of computing}, 135 (2007).
\bibitem{Weight}
G. Bianconi, 
EPL {\bf 71}, 1029 (2005). 

\bibitem{Doro_link}
S. N. Dorogovtsev, J. F. F. Mendes and A. N. Samukhin,
Phys. Rev. E {\bf 63} 062101 (2001).
\bibitem{crit}
S: N: Dorogovtsev, A. V. Goltsev, and J. F.F. Mendes,
Rev. of Mod. Phys. {\bf 80}, 1275, (2008).

\bibitem{Dynamics}
A. Barrat, M. Barthelemy, and A. Vespignani,
{\em  Dynamical processes on complex networks}
(Cambridge University Press, Cambridge, 2008).
\bibitem{Fermi}
G. Bianconi, Phys. Rev. E {\bf 66}, 036116  (2002).
\bibitem{Complex}
G. Bianconi, Phys. Rev. E {\bf 66}, 056123 (2002).
\bibitem{Multiplex}
G. Bianconi, Phys. Rev. E {\bf 91},  012810 (2015).
\bibitem{Garlaschelli}
D. Garlaschelli, M. I. Loffredo, Phys. Rev. Lett. {\bf 102}, 038701 (2009).

\bibitem{Jacobson}
T. Jacobson, 
Phys. Rev. Lett. {\bf 75}, 1260 (1995).
\bibitem{Chirco_liberati}
G. Chirco, and S. Liberati,
Phys. Rev. D {\bf 81},  024016 (2010).
\bibitem{Chirco_Rovelli}
G. Chirco,  H. M. Haggard, A. Riello, and C. Rovelli,
Phys. Rev. D {\bf 90},  044044 (2014).
\bibitem{SM}
See  Supplementary Material at 
 \bibitem{Doro_add}
S. N.  Dorogovtsev,  J. F. F. Mendes, and A. N. Samukhin,
Phys. Rev. Lett. {\bf 85}, {\bf 21}, 4633 (2000).

\bibitem{Burda}
P. Bialas,  Z. Burda, J. Jurkiewicz, and A. Krzywicki,
Physical Review E {\bf 67},   066106 (2003).
\bibitem{Ambjorn}
 J. Ambjorn,  B. Durhuus,  T. J\'onnson,  Phys. Lett. {\bf B244},403 (1990).
\bibitem{Regge}
T.Regge, 
Il Nuovo Cimento Series {\bf 10}, 558 (1961).
\bibitem{Dittrich}
B. Dittrich, and P. A. H\"ohn,
Classical and quantum gravity {\bf 29},  115009 (2012).
\bibitem{Kardar}
M. Kardar,{\it Statistical Physics of Particles} (Cambridge University Press, Cambridge 2007)

\end{thebibliography}
\end{document}